\documentclass[12pt,a4paper]{article}

\usepackage{amsmath,amsfonts,latexsym,amssymb}
\usepackage{verbatim,amsthm,curves,graphics}
\usepackage{mathrsfs}

\def\ben{\begin{equation}}
\def\een{\end{equation}}
\def\bena{\begin{eqnarray}}
\def\eena{\end{eqnarray}}
\def\bml{\begin{multline}}
\def\eml{\end{multline}}

\theoremstyle{definition}
\newtheorem{thm}{Theorem}[section]

\newtheorem{lemma}{Lemma}[section]
\newtheorem{prop}{Proposition}[section]

\begin{document}

\newcommand{\myI}{\stackrel{\circ}{I}}
\newcommand{\mr}{\mathbb{R}} 
\newcommand{\mz}{\mathbb{Z}} 
\newcommand{\mc}{\mathbb{C}} 
\newcommand{\mh}{\mathbb{H}} 
\newcommand{\eps}{{\mbox{\boldmath $\epsilon$}}}
\newcommand{\mn}{\mathbb{N}} 
\newcommand{\F}{{\mathcal F}_{\rm class}} 
\newcommand{\h}{{\bf h}} 
\newcommand{\pgator}{/\!\!\! S}
\newcommand{\rP}{\operatorname{P}}
\renewcommand{\i}{{\rm i}}
\newcommand{\Ihaves}{\frac{{\rm i}}{2}}
\renewcommand{\d}{\operatorname{d}}  
\newcommand{\rSO}{\operatorname{SO}}      
\newcommand{\rO}{\operatorname{O}}        
\newcommand{\rCliff}{\operatorname{Cliff}}
\newcommand{\WF}{\operatorname{WF}}       
\newcommand{\Pol}{{\rm WF}_{pol}}         
\renewcommand{\O}{\mathcal{O}}             
\newcommand{\cA}{\mathcal{A}}                
\newcommand{\cW}{\mathcal{W}}                
\newcommand{\A}{\mathscr{A}}
\newcommand{\T}{\mathscr{T}}             
\newcommand{\R}{\mathscr{R}}             
\newcommand{\cC}{\mathcal{C}}             
\newcommand{\cF}{\mathscr{F}}             
\newcommand{\E}{\mathcal{E}}             
\newcommand{\cH}{\mathscr{H}}
\renewcommand{\P}{\mathcal{P}}
\newcommand{\cM}{\mathscr{M}}
\newcommand{\cT}{\mathscr{T}}
\newcommand{\cN}{\mathscr{N}}
\newcommand{\cD}{\mathcal{D}}             
\newcommand{\cL}{\mathcal{L}}             
\newcommand{\W}{{\mathcal W}}             
\newcommand{\B}{{\mathcal B}}
\newcommand{\g}{{\bf g}}             
\renewcommand{\S}{\mathcal{S}}
\newcommand{\bU}{\mathcal{U}}
\newcommand{\bL}{\operatorname{OP}}
\newcommand{\glob}{{\textrm{\small global}}}
\newcommand{\loca}{{\textrm{\small local}}}
\newcommand{\singsupp}{\operatorname{singsupp}}
\newcommand{\dom}{\operatorname{dom}}
\newcommand{\clo}{\operatorname{clo}}
\newcommand{\supp}{\operatorname{supp}}
\newcommand{\rd}{{\rm d}}                 
\newcommand{\mslash}{/\!\!\!}             
\newcommand{\slom}{/\!\!\!\omega}         
\newcommand{\dirac}{/\!\!\!\nabla}        
\newcommand{\myid}{\leavevmode\hbox{\rm\small1\kern-3.8pt\normalsize1}}
\newcommand{\esssup}{\operatorname*{ess.sup}}
\newcommand{\ran}{\operatorname{ran}}
\newcommand{\sd}{{\rm sd}}
\newcommand{\reg}{\,{\rm l.n.o.}}
\newcommand{\V}{{\mathcal V}_{\rm class}}
\newcommand{\wk}{{\bf k}}
\newcommand{\ws}{{\bf s}}
\newcommand{\lno}{:\!}
\newcommand{\rno}{\!:}
\newcommand{\clim}{\operatorname*{coin.lim}}
\newcommand{\mydef}{\stackrel{\textrm{def}}{=}}
\newcommand{\rhoo}{\rho^0}
\newcommand{\partialo}{\stackrel{\circ}{\partial}}
\newcommand{\nablao}{\stackrel{\circ}{\nabla}}
\renewcommand{\min}{{\textrm{\small int}\,\mathscr{L}}}
\newcommand{\Exp}{\operatorname{Exp}}
\newcommand{\Ad}{\operatorname{Ad}}
\newcommand{\ret}{{\rm ret}} 
\newcommand{\adv}{{\rm adv}} 

\author{Stefan Hollands\thanks{\tt stefan@gr.uchicago.edu}
and Robert M. Wald\thanks{\tt rmwa@midway.uchicago.edu} \\ \it Enrico
Fermi Institute and Department of Physics \\ \it University of Chicago
\\ \it 5640 S.~Ellis Avenue, Chicago, IL~60637, USA}

\title{Conservation of the stress tensor in perturbative interacting quantum field theory in curved spacetimes}

\maketitle

\tableofcontents

\begin{abstract}
We propose additional conditions (beyond those considered in our
previous papers) that should be imposed on Wick products and
time-ordered products of a free quantum scalar field in curved
spacetime. These conditions arise from a simple ``Principle of
Perturbative Agreement'': For
interaction Lagrangians $L_1$ that are such that the 
interacting field theory can be
constructed exactly---as occurs when $L_1$ is a ``pure divergence'' or
when $L_1$ is at most quadratic in the field and contains no more than
two derivatives---then time-ordered products must be defined so that
the perturbative solution for interacting
fields obtained from the Bogoliubov formula agrees with the
exact solution. The conditions derived from this principle include a
version of the Leibniz rule (or ``action Ward identity'') and a
condition on time-ordered products that contain a factor of the free
field $\varphi$ or the free
stress-energy tensor $T_{ab}$. The main results of our paper are (1) a
proof that in spacetime dimensions greater than $2$, 
our new conditions can be consistently imposed in addition
to our previously considered conditions and (2) a proof that, if they
are imposed, then for {\em any} polynomial interaction Lagrangian
$L_1$ (with no restriction on the number of derivatives appearing in
$L_1$), the stress-energy tensor $\Theta_{ab}$ of the interacting
theory will be conserved. Our work thereby establishes
(in the context of perturbation theory) the conservation of
stress-energy for an arbitrary interacting scalar field in curved
spacetimes of dimension greater than $2$. Our approach requires us to 
view time-ordered products as
maps taking classical field expressions into the quantum field algebra
rather than as maps taking Wick polynomials of the quantum field into the
quantum field algebra.

\end{abstract}

\section{Introduction}

In \cite{hw1} and \cite{hw2}, we took an axiomatic approach toward
defining Wick powers and time-ordered products of a quantum scalar
field, $\varphi$, in curved spacetime. We provided a list of axioms that these
quantities are required to satisfy (see conditions T1-T9 of \cite{hw2}
or section 2 below) and then succeeded in proving both their
uniqueness (up to specified renormalization ambiguities) \cite{hw1}
and their existence \cite{hw1,hw2}.

Our previous analysis restricted attention to the case where the Wick
powers and the factors appearing in the time-ordered products do not
contain derivatives of the scalar field $\varphi$. In fact, however,
as we already noted in \cite{hw1,hw2}, our uniqueness and existence
results extend straightforwardly to the case where the Wick powers and
the factors appearing in the time-ordered products are arbitrary
polynomial expressions in $\varphi$ and its derivatives\footnote{Axiom
T9 was explicitly stated in \cite{hw2} only for the case of
expressions that do not contain derivatives. Its generalization to
expressions with derivatives is given in section 2 below.}.  We
excluded the explicit consideration of expressions containing
derivatives partly for simplicity but also
because it was clear to us that additional axioms should
be imposed on these quantities---and, consequently, stronger
uniqueness and existence theorems should be proven---but it was not
clear to us precisely what form these additional axioms
should take.  The main purpose of this paper is to provide these
additional axioms, to investigate some of their consequences---most
notably, conservation of the stress-energy of the interacting
field---and to prove the desired stronger existence and uniqueness
results for our new strengthened set of axioms.

Some simple examples should serve to illustrate the issues involved in
determining what additional conditions should be imposed. One obvious
possible requirement is the ``Leibniz
rule''. Consider, for example, the Wick monomials $\varphi^2$ and
$\varphi \nabla_a \varphi$ in $D=4$ spacetime dimensions. 
The uniqueness theorem of \cite{hw1}
applies to both of these expressions. It establishes that the first is
unique up to the addition of $c_1 R \myid$, where $c_1$ is an arbitrary
constant and $R$ denotes the scalar curvature. Similarly, the second
is unique up to the addition of $c_2 \nabla_a R \myid$, where $c_2$ is
an independent arbitrary constant. However, it would be natural to
require that 
\ben
\label{lr}
\nabla_a \varphi^2 = 2 \varphi \nabla_a \varphi 
\een 
where the left side denotes the distributional derivative of
$\varphi^2$.  If we wished to impose eq.~(\ref{lr}), then we would need to
strengthen our previous existence theorem to show that eq.~(\ref{lr}) can
be imposed in addition to our previous axioms. (This is easily done.)
Our above uniqueness result would then be strengthened in that we
would have $c_1=2c_2$, i.e, $c_1$ and $c_2$ would no longer be
independent. Note that the Leibniz rule eq.~(\ref{lr}) has an obvious
generalization to arbitrary Wick polynomials, but it is not so obvious, a
priori, what form the Leibniz rule should take on factors occurring in
time-ordered products.

A second ``obvious'' requirement that one might attempt to impose on
Wick polynomials and time-ordered products is that they respect the
equations of motion of the free field $\varphi$. Consider the case of
a massless Klein-Gordon field, so that $\nabla^a \nabla_a \varphi =
0$. Then it would seem natural to require the vanishing of any Wick
monomial containing a factor of $\nabla^a \nabla_a \varphi$---such as the
Wick monomials $\varphi \nabla^a \nabla_a \varphi$ and $(\nabla_b \varphi)
(\nabla^a \nabla_a \varphi)$. Similarly, it would be natural to require
the vanishing of any time-ordered product with the property 
that any of its arguments contains a factor of this form. However, it turns out
that---as we will explicitly prove in section 3 below---it is {\em
not} possible to impose this ``wave equation'' requirement together
with the Leibniz rule requirement of the previous paragraph. 

Should one impose the Leibniz rule or the free equations of motion (or
neither of them) on Wick polynomials or time-ordered products? If the
Leibniz rule is imposed, what form should it take for time-ordered
products? Should any conditions be imposed in addition to the Leibniz
rule or, alternatively, to the free equations of motion? In this
paper, we will take the view that these and other similar questions
should {\em not} be answered by attempting to make aesthetic
arguments concerning properties of Wick polynomials and
time-ordered products for the free field theory defined by the free
Lagrangian $L_0$. Rather, we will consider the properties of the
interacting quantum field theory defined by adding to $L_0$ an
interaction Lagrangian density $L_1$, 
which may contain an arbitrary (but finite) number of powers of 
$\varphi$ and its derivatives. As discussed in detail e.g., in section 3 of 
\cite{hw3} (see also subsection 4.1  below), 
an arbitrary interacting quantum field $\Phi_{L_1}$ 
(with $\Phi$ denoting an arbitrary polynomial in 
$\varphi$ and its derivatives) 
is defined perturbatively by the Bogoliubov formula, which expresses
$\Phi_{L_1}$ in terms of the free-field time-ordered products
with factors composed of $\Phi$ and $L_1$. The main basic idea 
of this paper is to
invoke the following simple principle, which we will refer to as the
``Principle of Perturbative Agreement'': {\em If the interaction
Lagrangian $L_1$ is such that the quantum field theory defined
by the full Lagrangian $L_0 + L_1$ can be solved exactly, then
the perturbative construction of the quantum field theory must agree
with the exact construction.}

There are two separate cases in which this principle yields nontrivial
conditions. The first is where the interaction Lagrangian corresponds
to a pure ``boundary term'', i.e., in differential forms notation, the
interaction Lagrangian is of the form $\d B$, where
$B$ is a smooth $(D-1)$-form of compact support depending polynomially 
on $\varphi$ and its derivatives.
Such an ``interaction'' produces an identically vanishing
contribution to the action, and the interacting quantum field theory
is therefore identical to the free theory. As we shall show in subsection
3.1, the imposition of the requirement that all perturbative corrections
vanish for any interaction Lagrangian of the form $\d B$
precisely yields the Leibniz rule for Wick polynomials and yields a
generalization of the Leibniz rule for time-ordered products. This
generalization states that, in effect, derivatives can be freely
commuted through the ``time ordering''. We will refer to this
condition as the generalized Leibniz rule 
and will label it as ``T10''.  Our condition T10 corresponds to 
the ``action Ward identity'' proposed in~\cite{stora, mwi} and 
proven recently in the context of flat spacetime theories 
in \cite{df2}.
In order for condition T10 to be
mathematically consistent, it is necessary that we adopt the
viewpoint of \cite{boas} and \cite{fd}---which we already 
adopted in \cite{hw3} for other
reasons---that time-ordered products are maps from {\em classical}
field expressions (on which the classical equations of motion 
are {\em not} imposed) into the quantum algebra of observables. This
viewpoint and the reasons that necessitate its adoption are explained
in detail in section 2. A proof that condition T10 can be consistently
imposed in addition to conditions T1-T9 is given in subsection 3.1.

The second case where the above principle yields nontrivial conditions
is where the interaction Lagrangian is at most quadratic in the field
and contains a total of at most two derivatives. This includes
interaction Lagrangians consisting of terms 
of the form $J \varphi$, $V \varphi^2$, 
and $h^{ab} \nabla_a \varphi \nabla_b
\varphi$, corresponding to the presence of an external classical
source, a spacetime variation of the mass, 
and a variation of the spacetime metric. 
In all of these cases,
the exact quantum field algebra of the theory with Lagrangian $L_0 +
L_1$ can be constructed directly, in a manner similar to the theory
with Lagrangian $L_0$. Our demand that perturbation theory reproduce
this construction yields new, nontrivial conditions on time-ordered
products (which are most conveniently formulated in terms of retarded
products). The general form of this requirement, which we label as
``T11'', is formulated in subsection 4.1. A useful infinitesimal version
of this condition
for the case of an external current interaction---which we label as 
condition T11a---is derived in subsection 4.2, and a corresponding
infinitesimal version for the case of a metric variation---which we label as 
condition T11b---is derived in subsection 4.3.

The consequences of our additional conditions are investigated in
section 5. The main results proven there---which also constitute some
of the main results of this paper---are that our conditions imply the
following: (i) The free stress-energy tensor, $T_{ab}$, in the free quantum
theory must be conserved. (ii) For an {\em arbitrary} 
polynomial interaction Lagrangian, $L_1$, (a)
the interacting quantum field $\varphi_{L_1}$ always satisfies the
interacting equations of motion and (b) the interacting stress-energy
tensor, $\Theta_{L_1}^{ab}$, of the interacting theory always is conserved. 
This is rather remarkable in that, a
priori, one might have expected properties (i) and (ii) to be entirely
independent of conditions T1-T11. Indeed, one might have expected that
if one required that (i) and (ii) be satisfied in perturbation theory,
one would obtain a further set of requirements on Wick polynomials and
time-ordered products. The fact that no additional conditions are
actually needed provides confirmation that T10 and T11 are the
appropriate conditions that are needed to supplement our original
conditions T1-T9. In effect, the analysis of section 5 shows the
following: Suppose that the definition of time-ordered products
satisfies T1-T10. Then, if the definition of time-ordered products is
further adjusted, if necessary, so that in perturbation theory the
quantum field satisfies the correct field equation in the presence of
an arbitrary classical current source $J$ (as required by T11a), then
the interacting field also will satisfy the correct field equation for
an arbitrary self-interaction. Furthermore, if, in perturbation
theory, the stress-energy tensor remains conserved in the presence of
an arbitrary metric variation (as is a consequence of T11b), it also will
remain conserved in the presence of an arbitrary self-interaction.

Finally, in section 6, we prove that condition T11a and---in spacetimes 
of dimension $D>2$---condition T11b can be
consistently imposed, in addition to conditions
T1-T10. The proof that condition T11a can be consistently imposed is
relatively straightforward, and is presented in subsection 6.1. The proof
that condition T11b also can be imposed when $D>2$ is much more complex
technically, and is presented in the seven sub-subsections of 6.2. Despite
its complexity, the proof is logically straightforward except for a
significant subtlety that is treated in sub-subsection 6.2.6. Here we find
that a potential obstruction to satisfying T11b arises from the requirement
that time-ordered-products containing more than one factor of the 
stress-energy tensor be symmetric in these factors. We show that this
potential obstruction does not actually occur
for the theory of a scalar field, as
treated here. However, this need not be the case for other fields, and, 
indeed, it presumably is the underlying cause of the inability to impose
stress-energy conservation in certain parity violating theories in curved
spacetimes of dimension $D=4k +2$, as found in~\cite{aw}. For scalar fields,
we are thereby able to show that
condition T11b can be consistently imposed in curved
spacetimes of dimension $D>2$. However, for $D=2$ a further difficulty arises
from the simple fact that the freedom to modify the definition of
$\varphi \nabla_a \nabla_b \varphi$ by the addition of an arbitrary local
curvature term does not give rise to a similar freedom to modify the
definition of $T_{ab}$, and we find that, as a consequence, condition 
T11b cannot be satisfied for a scalar field in $D=2$ dimensions.

It is our
view that conditions T1-T11
provide the complete characterization of Wick polynomials and
time-ordered products of a quantum scalar field in curved spacetime.

\paragraph{Notation and Conventions.} Our notation and conventions
generally follow those of our previous papers \cite{hw1}-\cite{hw3}.
The spacetime dimension is denoted as $D$, and
$(M,\g)$ always denotes an oriented, globally hyperbolic
spacetime. We denote by 
$\eps = \sqrt{-\g} \, \d x^0 \wedge \dots \wedge \d x^{D-1}$ 
the volume element (viewed as a $D$-form, or density of
weight 1) associated with $\g$. Abstract index notation is used
wherever it does not result in exceedingly many indices. However, abstract
index notation is generally not used for $\g=g_{ab}$ and $\eps=\epsilon_{ab
\dots c}$.

\section{The nature and properties of time-ordered products}

\subsection[The construction of the free quantum field algebra and the
nature \\
of time-ordered products]{The construction of the free quantum field algebra and the
nature of time-ordered products}

Consider a scalar field $\varphi$ on an arbitrary 
globally hyperbolic spacetime,
$(M, \g)$, with classical action
\ben
\label{s0}
S_0 = \int L_0 = -\frac{1}{2} \int (g^{ab} \nabla_a \varphi \nabla_b \varphi + m^2 \varphi^2 + \xi R\varphi^2) \eps. 
\een
The equations of motion derived from this action have unique fundamental advanced and retarded solutions
$\Delta^{\adv/ \ret}(x,y)$ satisfying
\ben
\label{propagators}
(\nabla^a \nabla_a - m^2 - \xi R)\Delta^{\adv /\ret} = \delta, 
\een
together with the support property
\ben
\supp \Delta^{\adv/ \ret} \subset \{(x, y) \in M \times M \mid x \in J^{-/+}(y) \},  
\een
where $J^{-/+}(S)$ is the causal past/future of a set $S$ in spacetime. Here 
we view the distribution kernel of $\Delta^{\adv/\ret}$ as undensitized, 
i.e., acting on test {\em densities} rather than {\em scalar} test 
functions\footnote{Consequently, 
the delta-distribution in eq.~\eqref{propagators} is also undensitized.},
i.e., we view $\Delta^{\adv/\ret}$ as a linear map from compactly supported, 
smooth {\em densities} to smooth {\em scalar} functions.

The quantum theory of the field $\varphi$ is defined by constructing a
suitable *-algebra of observables as follows:
We start with
the free *-algebra with identity $\myid$ generated by the formal expressions $\varphi(f)$ and $\varphi(h)^*$ where
$f, h$ are smooth compactly supported densities on $M$. Now factor this free *-algebra
by the following relations:
\begin{enumerate}
\item[(i)]
$\varphi(\alpha_1 f_1 + \alpha_2 f_2) =  \alpha_1 \varphi(f_1) + \alpha_2
\varphi(f_2)$, with $\alpha_1, \alpha_2
\in \mc$; 
\item[(ii)]
$\varphi(f)^* = \varphi(\bar{f})$; 
\item[(iii)]
$\varphi((\nabla^a\nabla_a - m^2  - \xi R)f) = 0$; and 
\item[(iv)]
$\varphi(f_1) \varphi(f_2) -\varphi(f_2) \varphi(f_1) = \i \Delta(f_1,
f_2) \myid$, where $\Delta$ denotes the causal propagator for the Klein-Gordon operator, 
\ben
\label{commutiv}
\Delta = \Delta^\adv - \Delta^\ret. 
\een
\end{enumerate}
We refer to the algebra, ${\mathcal A}(M, \g)$, defined by relations (i)--(iv)
as the CCR-algebra (for ``canonical commutation relations'').  Quantum
states on the CCR-algebra $\mathcal A$ are simply linear maps $\omega$
from $\mathcal A$ into $\mc$ that are normalized in the sense that
$\omega(\myid) = 1$ and that are positive in the sense that
$\omega(a^* a)$ is non-negative for any $a \in {\mathcal A}$. This
algebraic notion of a quantum state corresponds to the usual notion of
a state as a normalized vector in a Hilbert space as follows: Given a
representation, $\pi$, of $\mathcal A$ on a Hilbert space, $\mathcal
H$, (so that each $a \in {\mathcal A}$ is represented as a linear
operator $\pi(a)$ on $\mathcal H$), then any normalized vector state
$|\psi\rangle \in {\mathcal H}$ defines a state $\omega$ in the above
sense via taking expectation values, $\omega(a) = \langle \psi|\pi(a)|
\psi \rangle$. Conversely, given a state, $\omega$, the GNS
construction establishes that one always can find a Hilbert space,
$\mathcal H$, a representation, $\pi$ of $\mathcal A$ on $\mathcal
H$, and a vector $|\psi\rangle \in {\mathcal H}$ such that $\omega(a)
= \langle \psi|\pi(a)| \psi \rangle$.

By construction, the only observables contained in $\mathcal A$ are
the correlation functions of the quantum field $\varphi$. Even if we
were only interested in considering the free quantum field defined by
the action eq.~(\ref{s0}), there are observables of interest that are
not contained in $\mathcal A$, such as the stress-energy tensor of the
quantum field
\begin{multline}
\label{tabdef}
T^{ab} = 2 \eps^{-1} \frac{\delta L_0}{\delta g_{ab}} = \nabla^a \varphi \nabla^b \varphi - \frac{1}{2}
g^{ab} \nabla^c \varphi \nabla_c \varphi - \frac{1}{2}g^{ab} m^2 \varphi^2 \\
+ \xi[G^{ab} \varphi^2 - 2 \nabla^a(\varphi \nabla^b \varphi) + 2g^{ab} \nabla^c (\varphi \nabla_c \varphi)]. 
\end{multline}
We will refer to any polynomial expression, $\Phi$, in $\varphi$ and
its derivatives as a ``Wick polynomial''. All Wick polynomials, such as
$T_{ab}$, that involve quadratic or higher order powers of $\varphi$ are
intrinsically ill defined on account of the distributional character
of $\varphi$. It is natural, however, to try to interpret Wick
polynomials as arising from ``unsmeared'' elements of $\mathcal A$
that are then made well defined via some sort of ``regularization''
procedure. In Minkowski spacetime, a suitable regularization is
accomplished by ``normal ordering'', which can be interpreted in terms
of a subtraction of expectation values in the Minkowski vacuum
state. However, in curved spacetime, regularization via ``vacuum
subtraction'' is, in general, neither available (since there will, in
general, not exist a unique, preferred ``vacuum state'') nor
appropriate (since the resulting Wick polynomials will fail to be
local, covariant fields \cite{hw1}).

The necessity of going beyond observables in $\mathcal A$ becomes even
more clear if one attempts to construct the theory of a
self-interacting field (with a polynomial self-interaction) in terms
of a perturbation expansion off of a free field theory. First, the
interaction Lagrangian, $L_1$, itself will be a Wick polynomial and
thereby corresponds to an observable that does not lie in $\mathcal
A$. Second, the $n$th order perturbative corrections to
$\varphi$---or, more generally, the $n$th order perturbative
corrections to any Wick monomial $\Phi$---are formally given by the
Bogoliubov formula (see eq.~(\ref{pertseries}) below), 
which expresses the Wick
monomial $\Phi_{L_1}$, for the interacting field
as a sum of $\Phi$ and correction terms involving
the ``time-ordered products'' of expressions containing one factor of
$\Phi$ and $n$ factors of $L_1$. For the case of two Wick monomials,
$\Phi_1$ and $\Phi_2$, the time-ordered product is formally given by
\ben
\T(\Phi_1(x_1) \Phi_2(x_2)) = \vartheta(x_1^0 - x_2^0) \Phi_1(x_1) \Phi_2(x_2) +
\vartheta(x_2^0 - x_1^0) \Phi_2(x_2) \Phi_1(x_1)
\label{top}
\een 
where $\vartheta$ denotes the step function. (The formal generalization
of eq.~(\ref{top}) to time-ordered products with $n$-arguments is
straightforward.) However, even if Wick monomials have been suitably
defined, the time-ordered product (\ref{top}) is not well defined
since the Wick monomials also have a distributional character, and taking
their product with a step function is, in general, ill
defined. Nevertheless, in Minkowski spacetime, time-ordered products can be
defined by well known renormalization procedures.

Thus, the perturbative construction of the quantum field theory of an
interacting field requires the definition of Wick polynomials and
time-ordered products, both of which necessitate enlarging the algebra
of observables beyond the original CCR-algebra, $\mathcal A$. These
steps were successfully carried out in \cite{hw1,hw2}, based upon
prior results obtained in~\cite{bfk, bf}. The first key step is to construct
an algebra of observables, $\W(M,\g)$, which is large enough
to contain all Wick polynomials and time-ordered products.
To do so, consider the following
expressions in ${\mathcal A}(M,\g)$:
\bena
\label{a2}
W_n(u) &=& 
\int u(x_1, \dots, x_n) \, : \prod_i^n \varphi(x_i) :_\omega \nonumber\\
&\equiv& \int u(x_1, \dots, x_n) \frac{\delta^n}{\i^n \delta f(x_1) \cdots \delta f(x_n)}
e^{\i \varphi(f) + \frac{1}{2} \omega_2(f, f)} \Bigg|_{f=0}, \quad u \in C^\infty_0 
\eena
where $\omega_2$ is the two-point function of an arbitrarily chosen 
Hadamard state. Thus, $\omega_2$ is a distribution on 
$M \times M$ with antisymmetric part equal to $(\i/2)\Delta$, satisfying the 
spectrum condition given in eq.~\eqref{hadadef} and satisfying 
the Klein-Gordon equation in each entry, i.e., $(P \otimes 1)\omega_2 = 0 = 
(1 \otimes P) \omega_2$ where $P$ 
is the Klein-Gordon operator associated with $L_0$, 
\ben
P = \nabla^a \nabla_a - m^2 - \xi R.
\label{kgP}
\een
It follows from the above relations (i)--(iv) in the CCR-algebra
that $W_n(u)^* = W_n(\bar u)$, and that
\ben
\label{a4}
W_n(u) \cdot W_m(u') = \sum_{2k \le m+n} W_{n+m-2k}(u \otimes_k u'),
\een
where the ``$k$-times contracted tensor product'' $\otimes_k$ is defined by
\begin{multline}
\label{a5}
(u \otimes_{k} u')(x_1, \dots, x_{n+m-2k}) \mydef 
{\bf S} \frac{n!m!}{(n-k)!(m-k)!k!}
\int_{M^{2k}} 
u(y_{1}, \dots, y_{k}, x_1, \dots, x_{n-k})\times\\
u'(y_{k+1}, \dots, y_{k+i}, x_{n-k+1}, 
\dots, x_{n+m-2k})\prod_{i=1}^k \omega_2(y_{i}, y_{k+i}) \,
\eps(y_{i}) \eps(y_{k+i}) 
\end{multline} 
where $\bf S$ denotes symmetrization in $x_1, \dots, x_{n+m-2k}$. If either 
$m < k$ or $n < k$, then the contracted tensor product is defined to be 
zero. The above product formula can be recognized as Wick's theorem for 
normal ordered products. The enlarged algebra $\W(M, \g)$ is now 
obtained by allowing not only compactly supported smooth functions $u \in C^\infty_0$ 
as arguments of $W_n(u)$ but more generally any {\em distribution} $u$ in the space
\ben
\label{a3}
\E_n(M, \g) = \{u \in \cD'(M^{n}) \mid \WF(u) \cap (V^+)^n = \WF(u) 
\cap (V^-)^n = \emptyset \}.
\een
Here, $V^{+/-} \subset T^*M$ is the union of all future resp. past 
lightcones in the cotangent space
over $M$, and $\WF(u)$ is the wave front~\cite{horm} set of a distribution $u$. The key point is that 
Hadamard property of $\omega_2$ and the wave front set condition on the 
$u$ and $u'$ imposed in the definition of the spaces $\E'_n(M, \g)$ is 
necessary and sufficient in order to show that the distribution products appearing
in the contracted tensor product are well-defined and give a distribution in the desired class
$\E'_{m+n-2k}(M, \g)$. Note that the definition of the algebra 
$\W(M, \g)$ a priori depends on the choice of
$\omega_2$. However, it can be shown \cite{hw1}
that different choices give rise to 
*-isomorphic algebras. Thus, as an abstract algebra, $\W(M, \g)$ is 
independent of the choice of $\omega_2$.

Although the algebra $\W(M,\g)$ is ``large enough'' to contain all
Wick polynomials and time-ordered products, the above construction
does not determine which elements of $\W(M,\g)$ correspond to given
Wick polynomials or time-ordered products. (In particular, 
the normal-ordered quantities $W_n$, eq.~(\ref{a2}), with $u$ taken to be
a smooth function of one variable times a delta-function, clearly do
not provide an acceptable definition of Wick powers, since they fail
to define local, covariant fields \cite{hw1}.) In \cite{hw1, hw2}, an
axiomatic approach was then taken to determine which
elements of $\W$ correspond to given Wick polynomials and time-ordered
products. In other words, rather than attempting
to define Wick polynomials and time-ordered products by the adoption of
some particular regularization scheme, we provided a list of
properties that these quantities should satisfy. We proved the
existence of Wick polynomials and time-ordered products satisfying these
properties and also proved their uniqueness up to expected renormalization
ambiguities. As already discussed in the previous section, one of the
main purposes of the present paper is to supplement this list of
axioms with additional conditions applicable to Wick polynomials and
time-ordered products containing derivatives, and to prove
correspondingly stronger existence and uniqueness theorems.

We will shortly review the axioms that we previously gave in
\cite{hw1, hw2}. However, before doing so, we shall explain a subtle
but important shift in our viewpoint on the nature of Wick polynomials and
time-ordered products. 

A Wick polynomial is a distribution, valued in the quantum field
algebra $\W$ that corresponds to a polynomial expression in the
classical field $\varphi$ and its derivatives. It is therefore natural
to consider the classical algebra, ${\mathcal C}_{\rm class}$, of real
polynomial expressions in the (unsmeared) classical field $\varphi(x)$
and its derivatives, where we impose all of the normal rules of
algebra (such as the associative, commutative, and distributive laws)
and tensor calculus (such as the Leibniz rule) to the expressions in
${\mathcal C}_{\rm class}$, and, in addition, we impose the wave
equation on $\varphi$, i.e., we set $(\nabla^a\nabla_a - m^2 - \xi
R)\varphi(x) = 0$.  It would then be natural to view Wick polynomials
as maps from ${\mathcal C}_{\rm class}$ into distributions with values
in $\W$. However, this viewpoint on Wick polynomials is, in general,
inconsistent because of the existence of anomalies. Indeed, we already
mentioned in the introduction that---as we will explicitly show in section
3.2 below---under our other assumptions, it
will not be consistent to set to zero all Wick monomials containing a
factor of $(\nabla^a\nabla_a - m^2 - \xi R)\varphi(x)$, even though
elements of ${\mathcal C}_{\rm class}$ that contain such a factor
vanish.

This difficulty has a simple remedy: We can instead define a classical
field algebra of polynomial expressions
in the unsmeared field
$\varphi(x)$ and its derivatives where we no longer impose the wave
equation. More precisely, let $\V$ denote the real vector space of all
classical polynomial tensor expressions\footnote{The coefficients 
of these polynomial expressions
may have arbitrary polynomial dependence on the dimensionful parameter
$m^2$ and may have arbitrary
analytic dependence on the dimensionless parameter $\xi$. However, we
will not normally explicitly 
write these possible dependences on the parameters 
appearing in the theory.} involving $\varphi$, its symmetrized
covariant derivatives\footnote{The notation $(\nabla)^k t_{bc \dots d}$
is a shorthand for the symmetrized $k$-th derivative of a tensor, 
$\nabla_{(a_1} \cdots \nabla_{a_k)} t_{bc\dots d}$. We may write any expression
containing $k$ derivatives of a tensor field $t_{bc \dots d}$ in terms of 
symmetrized derivatives of $t_{bc \dots d}$ of $k$th and lower order 
and curvature.} $(\nabla)^k \varphi$, 
the metric, and arbitrary curvature tensors $C$,
\ben
\V = {\rm span}_\mr \{ \Phi = C \cdot 
(\nabla)^{r_1} \varphi \cdots (\nabla)^{r_k} \varphi; \quad k, r_i \in \mn \}. 
\label{Vclassical}
\een
where, as in the case of ${\mathcal C}_{\rm class}$, we impose all of
the normal rules of algebra and tensor calculus to the expressions in
$\V$ but now we do {\em not} impose the field equation associated with
$L_0$.  We denote a generic monomial element in $\V$ by the capital
greek letter $\Phi$. 

We also introduce the space $\F$ of all {\em classical} $D$-form
functionals of the metric $\g$, the field $\varphi$ and its
derivatives, depending in addition on compactly supported (complex) tensor
fields $f$, 
\begin{multline}
\F = {\rm span} \{ A(x) = \eps(x) \nabla^{c_1} \cdots \nabla^{c_m}
f^{a_1 \dots a_r}(x)\Phi_{a_1 \dots a_r c_1 \dots c_m}(x)
\mid \\
\text{$f$ smooth, comp. supported tensor field on $M$}; \Phi \text{ a monomial in} \V\}. 
\label{Fclass} 
\end{multline}
Again, we do not assume in the definition of $\F$ that the classical equations of motion for $\varphi$ hold. In particular, 
we do {\em not} assume that expressions such as $f(\nabla^a \nabla_a - \xi R - m^2) \varphi$ are set to 
zero. We will often suppress the tensor indices and write a classical $D$-form functional $A \in \F$ simply as
\ben
A = f\Phi \in \F,
\een 
or $A = [(\nabla)^k f] \Phi$, if we want to emphasize that the
functional depends on derivatives of $f$.  We then view the Wick
polynomials as linear maps from $\F$ into $\W$. 

Following \cite{boas} and \cite{fd}, we previously explicitly 
adopted the above viewpoint on Wick polynomials in
\cite{hw3}.  This viewpoint does not constitute a significant 
departure from standard
viewpoints, but merely provides a clearer framework for discussing
anomalies. However, as we shall now explain, our viewpoint on
time-ordered products---which corresponds to the viewpoint 
taken in \cite{mwi}---does constitute a significant departure from
viewpoints that are commonly taken. 

As indicated above (see eq.~(\ref{top})), it
would appear natural to view the time-ordered product, $\T$, in
$n$-factors as a multilinear map taking Wick polynomials into
$\W$. Indeed, our previous papers \cite{hw1}-\cite{hw3} contain the
phrase ``Wick powers and their time-ordered products'' in many places.
However, the untenability of this view can be seen from the
following simple example. 
Consider the quantum field theory defined
by the classical Lagrangian density $L = L_0 + L_1$, with 
\ben
\label{c1}
L_1 = f P \varphi \eps 
\een 
for some smooth function, $f$, of compact support, where $P$ 
stands for the Klein-Gordon operator associated with $L_0$, eq.~(\ref{kgP}).
The classical
equations of motion arising from the Lagrangian $L$ are simply
\ben
P\varphi = Pf
\een
i.e., $\varphi$ satisfies the inhomogeneous wave equation with smooth
source $J=Pf$. Clearly, the
interacting quantum field, $\varphi_{L_1}$, also should satisfy the
inhomogeneous wave equation with source $J= Pf \myid$. By inspection, it follows that $\varphi_{L_1}$ should 
be given in terms of the free quantum field $\varphi$ by 
\ben
\varphi_{L_1} = \varphi + f \myid
\label{solsou}
\een
Note that this interacting quantum field theory has a trivial
$S$-matrix (since $\varphi_{L_1} = \varphi$ outside of the support of
$f$), but the local field $\varphi_{L_1}$ is, of course, affected by $f$ in the region where $f \neq 0$.

Now compare eq.~(\ref{solsou}) with what is obtained from perturbation
theory. As already noted above, in perturbation theory,
$\varphi_{L_1}$ is equal to the free quantum field $\varphi(x)$, plus
a sum of corrections terms, where the $n$-th order correction term involves the
quantity
\ben
\label{c2}
\int \T\left(\varphi(x) \prod_{i=1}^n  P\varphi(y_i) \right) f(y_1) \dots f(y_n) 
\eps(y_1) \dots \eps(y_n). 
\een
Since $P\varphi = 0$, it would appear
that perturbation theory yields $\varphi_{L_1} = \varphi$ rather than
eq.~(\ref{solsou}).  Consequently, we are put in the position of having to
choose (at least) one of the following three possibilities: (1) The
exact solution (\ref{solsou}) for the interacting field is wrong. (2)
The Bogoliubov formula for the interacting quantum field is wrong, at
least in the case of interactions involving derivatives of the
field. (3) The time-ordered product eq.~(\ref{c2}) can be nonvanishing
even though the Wick monomial $P\varphi$
vanishes. In our view, choices (1) and (2) are far more unacceptable
than (3), and we therefore choose option (3). The results of this
paper (specifically, the existence theorem of section 6), will establish
that it is mathematically consistent to make this choice.

Thus, we do {\em not} view the time-ordered products (with $n$ factors) as 
an $n$-times multilinear map on Wick polynomials but 
rather as an $n$-times multilinear map
\bena
&&\T_\g: \underbrace{\F \times \cdots \times \F}_{\text{$n$ factors}} \to \W(M, \g),\\
&&(f_1 \Phi_1, \dots, f_n \Phi_n) \to \T_\g \left(\prod_{i=1}^n f_i \Phi_i \right).   
\eena
We note that, for a fixed 
choice of monomials $\Phi_i \in \V$, we get a multilinear 
functional $(f_1, \dots, f_n) \to \T(\prod^n f_j \Phi_j)$ 
mapping test functions on $M$ to the algebra $\W$. In the following, 
we will sometimes use the more suggestive 
informal integral notation\footnote{Note that we do not need to 
specify an integration element in the 
formula below since the quantities $f_i \Phi_i$ already have the character of a density.}
\ben
\label{infint}
\T \left(\prod^n_{i=1} f_i \Phi_i \right) = \int \T(\Phi_1(x_1) \cdots \Phi_n(x_n)) f_1(x_1) \cdots f_n(x_n) 
\een
for this multi-linear map. Note that this notation is exactly
analogous to the usual informal integral notation for distributions
$u(f) = \int u(x) f(x)$ acting on test densities $f$. The Wick monomials are
simply time-ordered products with a single factor, and we will use the notation
\ben
\label{t1f}
\T(f\Phi) = \Phi(f) = \int \Phi(x) f(x).
\een
for these objects. Note, however, we will {\em not} use the much more
standard notation $\T(\prod^n \Phi_i(f_i))$ for time-ordered products,
since this would suggest that the time-ordered products are functions
of the Wick monomials $\Phi_j(f_j)$ rather than of the classical functionals
$f_j \Phi_j$ of the field $\varphi$.

We turn now to a review of the properties satisfied by time-ordered products.

\subsection{Properties of time ordered products: Axioms T1-T9}

In \cite{hw2}, we imposed a list of requirements on time-ordered
products. Since a time-ordered product in a single factor is just a
Wick polynomial, these requirements on time-ordered products also
apply to Wick polynomials. In addition, Wick polynomials are further
restricted by the requirement that if $A \in \F$ is independent of 
$\varphi$, i.e., if $A$ is of the form
\ben
A = (\nabla)^m f C \eps
\een
for some test tensor field $f$ and some monomial $C$ in the Riemann 
tensor and its derivatives, then the corresponding 
Wick polynomial is given by
\ben
\T(A) = \int_M (\nabla)^m f C \eps \cdot \myid, 
\een
where $\myid$ is the identity element in $\W$. Similarly, if 
$A = f\varphi$, then we require that
\ben
\T(f\varphi) = \varphi(f), 
\een 
where $\varphi(f)$ is the free quantum field, i.e., 
the algebra element in $\W$ obeying the relations (i)--(iv) above.

For the convenience of the reader, we now provide the list of axioms
given in \cite{hw2}.
We refer the reader to \cite{hw1} and \cite{hw2} for further
discussion of the motivation for these conditions as well as further
discussion of their meaning and implications.

\smallskip

\paragraph{T1 Locality/Covariance.}
The time ordered products are local, covariant fields, in the following sense. Consider 
an isometric embedding $\chi$ of a spacetime $(M', \g')$ into a spacetime $(M, \g)$ (i.e., $\g' = \chi^* \g$)
preserving the causality structure, and let 
$\alpha_\chi: \W(M', \g') \to \W(M, \g)$ be the corresponding 
algebra homomorphism. Then time ordered products are
required to satisfy
\ben
\alpha_\chi \left[ \T_{\g'} \left( \prod f_i \Phi_i \right) \right] = \T_\g\left( \prod (\chi_* f_i) \Phi_i \right). 
\een 
Here, $\chi_* f$ denotes the compactly supported tensor field on $M$ 
obtained by pushing forward the compactly supported tensor $f$ field on $N$ via the map $\chi$. 
[For example, $\chi_* f(x) = f(\chi^{-1}(x))$
if $f$ is scalar and $x$ in the image of $\chi$.] In particular, for the (scalar) Wick products, the requirement
reads
\ben
\alpha_\chi \left[ \Phi_{\g'}(x) \right] = \Phi_\g(\chi(x)) \quad \text{for all $x \in M'$.}
\een

\paragraph{T2 Scaling.}
The time ordered products scale ``almost homogeneously'' under rescalings
$\g \to \lambda^{-2} \g$ of the spacetime metric in the following
sense. Let $\T_\g$ be a local, covariant time ordered product with $n$ factors, 
and let ${\mathcal S}_\lambda \T_\g$
be the rescaled local, covariant field given by 
${\mathcal S}_\lambda \T_\g \equiv \lambda^{-Dn} \sigma_\lambda 
T_{\lambda^{-2} \g}$, where $\sigma_\lambda: \mathcal{W}(M, \lambda^{-2}\g)
\to \mathcal{W}(M, \g)$ is the canonical isomorphism defined in lemma 4.2 of
\cite{hw1}. The scaling requirement on the time ordered product is then 
that there is some $N$ such that
\begin{equation}
\frac{\partial^N}{\partial^N \ln \lambda}
\lambda^{-d_T} {\mathcal S}_\lambda \T_\g = 0.
\end{equation}
Here, $d_T$ is the engineering dimension of the time-ordered product, 
defined as\footnote{The rule for assigning
an engineering dimension to a field is obtained by requiring that the classical
action be invariant under scaling. Formula (\ref{engdim})
holds only for scalar field theory, i.e., in other theories, 
the dimension of the basic field(s) may be different from $(D-2)/2$.} $d_T = \sum d_{\Phi_i}$, with
\begin{eqnarray}
d_\Phi &=& \frac{(D-2)}{2} \times \#(\text{factors of $\varphi$}) + \#(\text{derivatives}) + 2 \times \#(\text{factors of curvature}) \nonumber \\
&&+ \#(\text{``up'' indices}) - \#(\text{``down'' indices}), 
\label{engdim}
\end{eqnarray}
where $D$ is the dimension of the spacetime $M$.

\paragraph{T3 Microlocal Spectrum condition.}
Let $\omega$ be any continuous state on $\mathcal{W}(M, \g)$,
so that, as shown in \cite{hr}, $\omega$ has 
smooth truncated $n$-point functions 
for $n \neq 2$ and a two-point function $\omega_2(f_1, f_2) 
= \omega(\varphi(f_1) \varphi(f_2))$ of Hadamard from, i.e., $\WF(\omega_2)
\subset {\mathcal C}_+(M, \g)$, where 
\begin{equation}
\label{hadadef}
{\mathcal C}_+(M, \g) = \{(x_1, k_1; x_2, -k_2) \in T^* M^2 \setminus
\{0\} \mid (x_1, k_1) \sim (x_2, k_2); k_1 \in (V^+)_{x_1}\}. 
\end{equation}
Here the notation $(x_1, k_1) \sim (x_2, k_2)$ means that $x_1$ and
$x_2$ can be joined by a null-geodesic and that $k_1$ and $k_2$ are
cotangent and coparallel to that null-geodesic. 
$(V^+)_x$ is the future lightcone at $x$. Furthermore, let 
\begin{equation}
\label{gamphid}
\omega_\T(x_1, \dots, x_n) = 
\omega \left[\T\left(\prod_{i=1}^n \Phi_i(x_i) \right) \right].
\end{equation}
Then we require that 
\begin{equation}
\label{microcond}
\WF(\omega_\T) \subset \cC_\T(M, \g), 
\end{equation}
where the set $\cC_\T(M, \g) \subset T^*M^n \setminus \{0\}$ 
is described as follows (we 
use the graphological notation introduced in \cite{bfk,bf}):  
Let $G(p)$ be a ``decorated embedded graph''
in $(M, \g)$. By this we mean an embedded graph $\subset M$ whose 
vertices are points $x_1, \dots, x_n \in M$
and whose edges, $e$, are oriented null-geodesic curves. Each such null 
geodesic is equipped with a coparallel, cotangent covectorfield $p_e$.  
If $e$ is an edge in $G(p)$ connecting the points $x_i$ and $x_j$ 
with $i < j$, then $s(e) = i$ is its source 
and $t(e) = j$ its target. It is required that
$p_e$ is future/past directed if $x_{s(e)} \notin J^\pm(x_{t(e)})$.
With this notation, we define
\begin{eqnarray}
\label{gamtdef}
\cC_\T(M, \g) &=& 
\Big\{(x_1, k_1; \dots; x_n, k_n) \in T^*M^n \setminus \{0\} \mid 
\exists \,\, \text{decorated graph $G(p)$ with vertices} \nonumber\\
&& \text{$x_1, \dots, x_n$ such that
$k_i = \sum_{e: s(e) = i} p_e - \sum_{e: t(e) = i} p_e 
\quad \forall i$} \Big\}. 
\end{eqnarray}

\paragraph{T4 Smoothness.}
The functional dependence of the time ordered products
on the spacetime metric, $\g$, is such that 
if the metric is varied smoothly, then the time ordered
products vary smoothly, in the following sense. 
Consider a family of metrics $\g^{(s)}$ depending smoothly upon 
a set of parameters $s$ in a parameter space $\P$.
Furthermore, let $\omega^{(s)}$ be a
family of Hadamard states with smooth truncated $n$-point 
functions ($n \neq 2$) depending smoothly on $s$ and with two-point
functions $\omega^{(s)}_2$ depending smoothly on $s$ in the sense 
that, when viewed as a distribution jointly in 
$(s, x_1, x_2)$, we have
\begin{equation}
\label{os}
\WF(\omega_2^{(s)}) \subset \{(s, \rho; x_1, k_1; x_2, k_2) 
\in T^*(\P \times M^2) \setminus \{0\} \mid (x_1, k_1; x_2, k_2) 
\in 
\cC_+(M, \g^{(s)})
\}, 
\end{equation}
where the family of cones $\cC_+(M, \g^{(s)})$ is defined by 
eq.~\eqref{hadadef} in terms of the family $\g^{(s)}$.
Then we require that the family of distributions given by 
\begin{equation}
\label{family}
\omega_\T^{(s)}(x_1, \dots, x_n) = \omega^{(s)}
\left[ \T_{\g^{(s)}} \left(\prod_{i=1}^n \Phi_i(x_i) \right)
\right] 
\end{equation}
(viewed as distributions in the variables $(s, x_1, \dots, x_n)$)
depends smoothly on $s$ with respect to the sets 
$\cC_\T(M, \g^{(s)})$ defined in eq.~\eqref{gamtdef}, in the sense that 
\begin{multline}
\label{smocond}
\WF(\omega_\T^{(s)}) \subset \{(s, \rho; x_1, k_1; \dots; x_n, k_n) 
\in T^*(\P \times M^n) 
\setminus \{0\} \mid \\
(x_1, k_1; \dots; x_n, k_n) \in \cC_\T(M, \g^{(s)}) \}.
\end{multline}
We similarly demand that the time ordered products also have a smooth 
dependence upon the parameters $m^2, \xi$ in the free theory.

\paragraph{T5 Analyticity.}
Similarly, we require that, for an analytic family of analytic
metrics (depending analytically upon a set of parameters), 
the expectation value of the time-ordered products in an
analytic family of states\footnote{As explained in remark (2) on P. 311
of \cite{hw1}, it suffices to consider a suitable analytic family of
linear functionals on $\W$ that do not necessarily satisfy the positivity
condition required for states.} varies analytically in the same sense as in
T4, but with the smooth wave front set replaced by the analytic wave
front set. We similarly demand an analytic dependence upon the 
the parameters $m^2, \xi$.

\paragraph{T6 Symmetry.}
The time ordered products are symmetric under a permutation of 
the factors. 

\paragraph{T7 Unitarity.} 
Let $\bar \T(\prod f_i \Phi_i) = [\T(\prod \bar f_i \Phi_i)]^*$, $\Phi_i \in \V$, be
the ``anti-time-ordered'' product. Then we require
\begin{equation}
\bar \T(f_1 \Phi_1 \dots f_n \Phi_n) = 
\sum_{I_1 \sqcup \dots \sqcup I_j = \{1, \dots, n\}}
(-1)^{n + j} \T\left( 
\prod_{i \in I_1} f_i\Phi_i \right) \dots
\T\left(\prod_{j \in I_j} f_j \Phi_j \right), 
\label{atoprod}
\end{equation}
where the sum runs over all partitions of the set $\{1, \dots, n\}$ into
pairwise disjoint subsets $I_1, \dots, I_j$.

\paragraph{T8 Causal Factorization.}
For time ordered products with more than one factor, we require
the following causal factorization
rule, which reflects the time-ordering of the factors. Consider a
set of test functions $(f_1, \dots, f_n)$ and a partition of $\{1,
\dots, n\}$ into two non-empty disjoint subsets $I$ and $I^c$, with
the property that no point $x_i \in \supp f_i$ with $i \in I$ is in the past of any
of the points $x_j \in \supp f_j$ with $j \in I^c$, that is, $x_i \notin J^-(x_j)$
for all $i \in I$ and $j \in I^c$. Then the corresponding time ordered product
factorizes in the following way:
\ben
\label{cf}
\T\left(\prod_{k=1}^n f_k \Phi_k \right) = \T\left(\prod_{i\in I}  
f_i \Phi_i \right)\,\T\left(\prod_{j \in I^c} 
f_j \Phi_j\right).
\een
In the case of 2 factors,  this requirement reads
(in the informal notation introduced above)
\ben
\T(\Phi(x) \Psi(y)) = 
\begin{cases}
\Phi(x) \Psi(y) & \text{when $x \notin J^+(y)$;}\\
\Psi(y) \Phi(x) & \text{when $y \notin J^-(x)$.}
\end{cases}
\een

\paragraph{T9 Commutator.}
The commutator of a time-ordered product with a free field is given by
lower order time-ordered products times suitable commutator functions,
namely 
\ben
\left[\T\left( \prod^n f_i \Phi_i \right), \varphi(F) \right] =
\i\sum_{i=1}^n \T\left( f_1 \Phi_1 \dots 
(\Delta  F) \frac{\delta(f_i\Phi_i)}{\delta \varphi} \dots f_n\Phi_n \right),
\label{T9}
\een
where $\Delta = \Delta^{\rm adv} - \Delta^{\rm ret}$ is the 
causal propagator (commutator function), and where we are using the 
notation $(\Delta F)(x) = \int_M \Delta(x, y) F(y)$ for the action 
of the causal propagator on a smooth {\em density} $F$ of compact
support\footnote{As previously noted at the beginning of this
section, when writing expressions like $\Delta F$ 
or likewise $\Delta^{\adv/ \ret} F$, we take the point of view 
that the Green's functions are 
linear maps from smooth compactly supported {\em densities} on $M$ 
to smooth {\em scalar} functions on $M$.}.
Here, the functional derivative, $\delta A/\delta \varphi \in \F$, 
of an arbitrary element of $A \in \F$ is given by
\ben
\label{30}
\frac{\delta A}{\delta \varphi} = 
\sum_r (-1)^r \nabla_{(a_1} \cdots \nabla_{a_r)} 
\frac{\partial A}{\partial (\nabla_{(a_1} \cdots \nabla_{a_r)} \varphi)} \, .   
\een
This formula corresponds to the 
usual ``Euler-Lagrange''-type expression familiar from the 
calculus of variations; see appendix B for further discussion.

\medskip

\paragraph{Remark:} In \cite{hw2}, condition T9 was explicitly stated only for
the case where each $\Phi_i$ has no dependence on derivatives of
$\varphi$. Equation~(\ref{T9}) is the appropriate generalization to
arbitrary $\Phi_i$. For the case of a Wick power (i.e., a time-ordered
product in one argument), eq.~(\ref{T9}) can be motivated by the
requirement of maintaining the desired relationship between
Poisson-brackets and commutators.

\medskip

The main results of \cite{hw1} and \cite{hw2} are that there exists a
definition of time-ordered products that satisfies conditions T1-T9
and that, furthermore, this definition is unique up to the expected
renormalization ambiguites. Our goal now is to impose additional
conditions appropriate to time-ordered products whose factors $\Phi_i$
depend upon derivatives of $\varphi$, and to then prove the
corresponding existence and uniqueness theorems. These additional
conditions will arise from the following basic principle already stated in
the introduction: 

\paragraph{Principle of Perturbative Agreement:} {\em If the interaction 
Lagrangian $L_1 = \sum_i
f_i \Phi_i$ (where each $f_i$ is smooth and of compact
support and each $\Phi_i \in \V$) is such that the quantum field
theory defined by the full Lagrangian $L_0 + L_1$ can be solved
exactly, then the perturbative construction of the quantum field
theory as defined by the Bogoliubov formula 
must agree with the exact construction.}

\section{The Leibniz rule}

\subsection[Formulation of the Leibniz rule, T10, and proof of consistency with \\
axioms T1-T9]{Formulation of the Leibniz rule, T10, and proof of consistency with
axioms T1-T9}

Our first new requirement arises from 
considering a classical functional $A \in \F$ of the form
\ben
A = \d B, 
\label{A}
\een
where $B$ is in the analog of the space $\F$ (see eq.~(\ref{Fclass}))
but with ``$D$-form'' replaced by ``$(D-1)$-form'', and where $\d$ is
the exterior differential (mapping $(D-1)$-forms to $D$-forms). An
example of such a $B$ is
\ben
B_{a_1 \dots a_{D-1}}=f^b \epsilon_{ba_1 \dots a_{D-1}} \Phi
\label{Bexample}
\een
where $f^c$ is a test vector field and $\Phi$ is a scalar
element of $\V$. The general such $B$ would be of a similar form,
except that $\Phi$ and $f$ could have additional tensor indices,
derivatives could act on $f$, and $\eps$ could be contracted with an
index of $\Phi$ rather than an index of $f$. For $B$ of the form
eq.(\ref{Bexample}), $A$ would take the explicit form.
\ben
\label{example}
A = (\nabla^c f_c) \Phi \eps + f_c \nabla^c \Phi \eps, 
\een

Classically, the Lagrangian $L = L_0 + A$ defines the same theory as
the Lagrangian $L_0$. Consequently, the ``interacting'' quantum field
theory defined by the interaction Lagrangian $L_1 = A$ should coincide
with the free quantum field theory, i.e., all of the perturbative
corrections should vanish for an interaction Lagrangian of this
form. To ensure this, we shall now add the following condition to our
list of axioms of the previous section:

\paragraph{T10 Leibniz Rule.} 
Let $A \in \F$ be any classical functional of the form eq.~(\ref{A}).
Then for all $f_i \Phi_i \in \F$, we require that 
\ben
\T(A f_1 \Phi_1 \cdots f_n \Phi_n) = 0,
\een
i.e., any time-ordered product containing a factor of $A=\d B$ must vanish.

\paragraph{Remark:} Condition T10 has previously been proposed (in the
context of quantum field theory in flat spacetime)
in \cite{stora} and \cite{mwi,df2} and is referred to as the ``action 
Ward identity'' in these references.

\medskip

Condition T10 for time-ordered products with two or more factors is
clearly necessary and sufficient for the vanishing of {\it all}
perturbative corrections to the interacting fields (including the
interacting time-ordered products). However, causal factorization (T8)
then implies that condition T10 must hold for Wick powers as
well. Thus, condition T10 is necessary and sufficient to guarantee
that the theory defined perturbatively by the interaction Lagrangian
$L_1 = \d B$ yields exactly the free theory.

However, it may not be obvious what, if anything, condition T10 
has to do with the ``Leibniz
rule'', so we shall now explain the relationship of this condition to
more usual formulations of the Leibniz rule. By doing so, we will also
clarify our notation and further elucidate the viewpoint on
time-ordered products introduced in the previous section.

Consider, first, the case of time-ordered products in one factor,
i.e., Wick powers, in which case condition T10 simply states that
for all $B$,
\ben
\T(\d B) = 0.
\een
Therefore, for the case
in which $B$ is given by eq.~(\ref{Bexample})---and hence $A$ is given
by eq.~(\ref{example})---we obtain
\ben
\T((\nabla_a f^a)\eps \Phi + f^a \eps \nabla_a \Phi) = 0.  
\label{wleib}
\een
for all scalar $\Phi \in \V$ and all test vector fields $f^a$. It
should be understood here that $\nabla_a \Phi$ represents the
classical expression corresponding to taking the derivative of
$\Phi$. For example, if $\Phi = \varphi^\alpha$ for some natural
number $\alpha$, then $\nabla_a \Phi = \alpha \varphi^{\alpha-1}
\nabla_a \varphi$. But $\T((\nabla_a f^a)\eps \Phi)$ is the same
thing as the distributional derivative of $-\T(\Phi)$ smeared with
$f^a \eps$. Hence, using our notation $\T(f\Phi) = \Phi(f)$ for Wick
powers, we may re-write eq.~(\ref{wleib}) as 
\ben 
\nabla_a \Phi(f^a \eps) = (\nabla_a \Phi)(f^a \eps).
\label{wleib2}
\een
Here, the quantity $\nabla_a \Phi$, appearing on the left side of this
equation represents the distributional derivative of the algebra valued 
distribution $\Phi$, whereas
the quantity $(\nabla_a \Phi)$ appearing on the right side of this
equation represents the Wick polynomial associated with the classical
quantity $\nabla_a \Phi$. (Note that since these logically distinct
quantities look the same except for the parentheses, our notation
$\Phi(f)$ for Wick powers would be unacceptable if eq.~(\ref{wleib})
was not imposed!) Thus, in the above example where $\Phi =
\varphi^\alpha$, eq.~(\ref{wleib2}) takes the form
\ben 
\nabla_a \varphi^\alpha(f_a \eps)
= \alpha (\varphi^{\alpha-1} \nabla_a \varphi)(\eps f^a),  
\een 
or, in the more common, informal notation 
\ben
\nabla_a[\varphi^\alpha(x)] = \alpha (\varphi^{\alpha-1} \nabla_a
\varphi)(x).
\een
Again, the left side of this equation denotes the distributional
derivative of $\varphi^\alpha$, so this equation does indeed
correspond to the usual notion of the Leibniz rule.  Analogous results
hold for the relations for Wick powers arising from T10 for general forms
of $B$.

The meaning of requirement T10 for time ordered products with more
than one factor can be seen as follows. Again, for simplicity, let $B$
be of the form eq.~(\ref{Bexample}). Condition T10 states that for
all $h_i \Psi_i \in \F$, we have
\ben
\label{award}
- \T( \eps (\nabla_c f^c) \Phi \prod \eps h_j \Psi_j) = \T( \eps f^c (\nabla_c \Phi) \prod \eps h_j \Psi_j).
\een
In the more common, informal notation this equation can be re-written as
\ben
\label{pullout}
\nabla^a_y[\T(\Phi(y)\Psi_1(x_1) \cdots \Psi_n(x_n))] = 
\T((\nabla^a \Phi)(y)\Psi_1(x_1) \cdots \Psi_n(x_n)) \, . 
\een
Here, the left side denotes the distributional derivative of
$\T(\Phi(y)\Psi_1(x_1) \cdots \Psi_n(x_n))$ with respect to the
variable $y$, whereas the factor $(\nabla_a \Phi)$ appearing on the
right side denotes the classical field expression obtained by taking
the derivative of $\Phi$. In other words, for time-ordered products
with more than one factor, the operational meaning of T10 is simply
that derivatives can be ``freely commuted'' through $\T$. Since the arguments
of time-ordered products are classical field expressions, the
Leibniz rule, of course, holds for the expressions hit by the
derivative inside of $\T$.

It is useful to further illustrate the meaning of condition T10---and
the extent to which it differs from conventional viewpoints on
time-ordered products---with a simple example. Let us attempt
to calculate $\T(\varphi(x)\varphi(y))$ according to our axiom
scheme. By causal factorization (T8), $\T(\varphi(x)\varphi(y))$ must
satisfy
\ben 
\label{topdef2}
\T(\varphi(x)\varphi(y)) = 
\begin{cases}
\varphi(x)\varphi(y) & \text{if $x \notin J^+(y)$,} \\
\varphi(y)\varphi(x) & \text{if $y \notin J^+(x)$,} 
\end{cases}
\een
which determines $\T(\varphi(x)\varphi(y))$ except on the ``diagonal''
$x=y$. However, since there do not exist any local and covariant
distributions (T1) with support on the diagonal that have the correct
scaling behavior (T2) as well as the desired smooth and analytic
dependence upon the spacetime metric T4 and T5, 
it follows that $\T(\varphi(x)\varphi(y))$ is unique.
This unique extension of the distribution defined by eq.~\eqref{topdef2} to the
diagonal is
\ben
\T(\varphi(x)\varphi(y)) = \vartheta(x^0 - y^0) \varphi(x)\varphi(y) + 
\vartheta(y^0 - x^0) \varphi(y)\varphi(x)
\een
where $\vartheta$ denotes the step function. (The right side of this
equation is mathematically well defined on account of the wavefront
set properties of $\varphi(x)\varphi(y)$; as already noted in
section 1, the corresponding expression for general Wick monomials (see
eq.~(\ref{top})) is not well defined.) If we apply the Klein-Gordon
operator $P$ to the variable $x$ of this distribution, we obtain
\ben
(P \otimes 1)\T(\varphi(x)\varphi(y)) = i \delta(x,y) \myid
\label{boxT}
\een

Consider, now, the time-ordered product
$\T((P\varphi)(x)\varphi(y))$. By causal factorization (T8), this
distribution satisfies 
\ben 
\T((P\varphi)(x)\varphi(y)) = 0 \,\,\,\, {\rm if} \,\, x \neq y 
\een 
The most obvious extension of this distribution to
the diagonal is, of course, to put $\T((P\varphi)(x)\varphi(y)) =
0$ for all $x, y$, including the diagonal. 
This is the conventional assumption. However, since
$\T((P\varphi)(x)\varphi(y))$ has dimension ${\rm length}^{-D}$,
there {\it does} exist a distribution with support on the diagonal
that satisfies the above required properties, namely $\delta(x,y)
\myid$.  Consequently, within the scheme of axioms T1-T9, we have the
freedom to add a ``contact term'' and define 
$\T((P\varphi)(x)\varphi(y))$ to be an arbitrary
multiple of $\delta(x,y) \myid$. Axiom T10 together with
eq.(\ref{boxT}) requires, in fact, that we make use of this freedom to define
$\T((P\varphi)(x)\varphi(y))$ to be given by 
\ben 
\T((P\varphi)(x)\varphi(y)) = \i \delta(x,y) \myid. 
\label{tpphi}
\een 
Since the Wick power $P\varphi$ vanishes identically, this explicitly
shows that it is inconsistent with axioms T1-T10 to view a time-ordered
product as a multilinear map on Wick polynomials rather than as a multilinear
map on elements of $\F$.

We now prove that, for arbitrary Wick powers and time-ordered
products, it is consistent to impose the Leibniz rule T10 in addition
to our previous axioms T1-T9. In essence, the following
proposition provides a generalization to curved spacetime of the proof of
the ``action Ward identity'' given in \cite{df2}.

\begin{prop}
There exists a prescription for defining time ordered products satisfying our
requirements T1--T10.
\end{prop}

\begin{proof} 
As in \cite{hw2}, we will proceed by an inductive argument 
on the number of factors, $N_T$, appearing in the
time ordered product. Consider, first, the case of Wick monomials, i.e.,
$N_T = 1$.
We previously showed \cite{hw1} that the following prescription of 
``local Hadamard normal ordering'' (i.e., ``covariant point-splitting 
regularization'') satisfies conditions T1-T9: 
Let $H(x,y)$ be a symmetric, locally constructed
Hadamard parametrix.\footnote{See e.g.~eqs.~(7) and~(8) and Appendix~A of 
\cite{mo} for the explicit form of $H$ in $D$ dimensions. Note that
reference~\cite{mo} uses a parametrix, $Z_n$,
that is ``truncated'' at $n$th order, which will give an acceptable 
prescription only when the total number of derivatives, $N_\nabla$, appearing
in the Wick power is sufficiently small. In order to give 
a prescription that is valid for arbitrary $N_\nabla$, one must define
$H$ by the procedure explained below eq.~(69) of \cite{hw1}.}
We define~\cite{hw1}
\begin{equation}
\label{hndef}
:\varphi(x_1) \dots \varphi(x_k):_{H}  \,\, = 
\frac{\delta^k}{\i^k \delta f(x_1) \dots \delta f(x_k)} {\rm exp}\left[
\frac{1}{2}H(f \otimes f) + \i\varphi(f)\right] \Bigg|_{f = 0}
\end{equation}
For an arbitrary 
$\Phi = C (\nabla)^{r_1} \varphi \cdots (\nabla)^{r_k} \varphi \in \V$ 
(where $C$ denotes a curvature term and all tensor indices 
have been suppressed),
we define the corresponding Wick monomial by~\cite{hw1}
\begin{eqnarray}
\label{phih}
\T(f\Phi) = \Phi(f) &\; = \;& \int C(y)
: \varphi(x_1) \dots \varphi(x_k) :_{H}  F(y; x_1, \dots, x_k) \nonumber \\
&\; = \;& \int C(y)
: \prod_{i=1}^k (\nabla)^{r_i} \varphi(y) :_H  \, f(y) \eps(y)
\end{eqnarray} 
where
\ben
F(y; x_1, \dots, x_k) = f(y) (\nabla_{x_1})^{r_1} \cdots (\nabla_{x_k})^{r_k}
\delta(y, x_1, \dots, x_k) \prod_i^k \eps (x_i).
\een
(Note that for the definition of general Wick powers with derivatives,
it is essential for this prescription to be well defined that
$H(x,y)$ be symmetric in $x$ and $y$, so that it does not matter which of
the variables $(x_1, \dots, x_k)$ we select to apply derivatives
to.) The arguments of \cite{hw1} can now be straightforwardly generalized
to show that this prescription satisfies not only conditions T1-T9 but
also satisfies T10. Thus, there is no difficulty in adding condition
T10 to the list of properties that we require for Wick
powers. 

However, since our previous existence proof for time-ordered
products \cite{hw2} does not provide a correspondingly explicit
prescription for their definition, we cannot give a similar, direct
proof that condition T10 can be imposed on time-ordered
products. Instead, we must proceed by re-proving the existence theorem 
of \cite{hw2}, where we now explicitly allow the factors 
appearing in the time-ordered products
to contain derivatives of $\varphi$ and where we now add 
condition T10 to the list of requirements.

We inductively assume that the construction of the time-ordered
products satisfying T1--T10 has been performed up to $<N_T$ factors. 
Our task is to construct the time-ordered products with $N_T$
factors. However, as in~\cite{hw2}, given the time-ordered 
products with $<N_T$ factors, the time ordered
products, $\T$, with $N_T$ factors are determined by causal factorization
as distributions, $\T^0$, on $M^{N_T} \setminus \Delta_{N_T}$, where 
\ben
\Delta_n \equiv \{(x,x,\dots,x) \in M^n \mid x \in M\}
\een
denotes the total diagonal. Furthermore, it is easily verified that 
$\T^0$ satisfies T1--T10 when acting on test functions
supported away from the total diagonal. Consequently, our task is to 
extend $\T^0$ to a distribution in $\cD'(M^{N_T})$, i.e., 
to a distribution defined everywhere, in such way that T1--T7 and T9--T10
are preserved in the extension process. (T8 has already been satisfied by
the requirement that $\T$ be an extension of $\T^0$.) In fact, we need
only show that an extension can be chosen so as to preserve T1--T5 and
T9--T10, since the symmetry property, T6, and unitarity property, T7, can
always be satisfied by a simple re-definition of $\T$ if all of the 
other properties have been satisfied \cite{hw2}.

As in \cite{hw2}, we can reduce this task to a much more manageable one
by the use of a ``local Wick expansion'' for $\T^0$. The appropriate form of 
this local Wick expansion in the case where the arguments of $\T^0$
are general elements\footnote{Since the Hadamard parametrix $H(x,y)$ is only 
defined when $x, y$ are in a convex normal neighborhood of each other, it follows 
that the local Wick expansion is only defined if $F = \otimes_i f_i$ is supported
in a sufficiently small neighborhood of the total diagonal. Note that this does not cause 
any problems in the present context since we are only interested in 
an arbitrarily small neighborhood of the total diagonal for the extension problem.} of $\F$ is
\begin{multline}
\label{lwe}
\T^0\left( \prod_{i=1}^{N_T} f_i \Phi_i \right) 
= \sum_{\alpha_1, \alpha_2, \dots} \frac{1}{\alpha_1! \dots \alpha_{N_T}!} 
\int \prod_j \eps(y_j)\\
t^0\left[\delta^{\alpha_1} \Phi_1 \otimes \dots \otimes \delta^{\alpha_{N_T}} \Phi_{N_T}
\right]
(y_1, \dots, y_{N_T}) \\
f_1(y_1) \dots f_{N_T}(y_{N_T}) 
: \prod_{i=1}^{N_T} \prod_{j} [(\nabla)^{j}\varphi(y_j)]^{\alpha_{ij}} :_H \,\,.
\end{multline}
Here, we 
are using the following notation: The $t^0$ are multilinear mappings
\bena
t^0: \bigotimes^{N_T} \V &\to&  \cD'(M^{N_T} \setminus \Delta_{N_T}) \\
     \otimes_i \Phi_i    &\to&  t^0[\otimes_i \Phi_i](y_1, \dots, y_{N_T}) 
\eena
from classical field expressions to c-number distributions in the product manifold minus
its total diagonal\footnote{More precisely, the $t^0$ are distributions that 
are defined on a suitable
neighborhood of the total diagonal, minus the total diagonal itself; 
see the previous footnote.}. Each $\alpha_i$ is a multi-index $(\alpha_{i1}, \alpha_{i2}, \dots)$, 
and we are using the shorthand $\alpha_i! = \prod_j \alpha_{ij}!$ for such a multi-index.
If $\Phi \in \V$ and $\alpha$ is a multi-index, we are using the notation
\ben
\delta^\alpha \Phi = 
\left\{ \prod_{i} \left( \frac{\partial}{\partial (\nabla)^{i}\varphi} \right)^{\alpha_i}
\right\}
\Phi.  
\een
As in~\cite{hw2}, eq.~(\ref{lwe}) can be proved by induction in $N_\varphi$,
using the commutator property, T9. By use of 
the local Wick expansion, we reduce the problem
of extending $\T^0$ to the problem of extending the expansion coefficients
$t^0$. The time-ordered product defined via eq.~(\ref{lwe}) 
from the extension
of $t^0$ will automatically satisfy the commutator requirement
T9. Thus, we need only show that the expansion coefficients $t^0$ can be 
extended so as to satisfy T1-T5 and T10.

It follows directly from the assumed 
properties T1--T10 of the time ordered products
for $<N_T$ factors that the $t^0$ are distributions that are locally and 
covariantly constructed out of the metric, 
that they satisfy a microlocal spectrum
condition, that they depend smoothly and analytically on the metric, that they
have an ``almost homogeneous scaling behavior'', and that they satisfy the 
Leibniz rule in the sense that
\ben
\label{leib0}
(1 \otimes \cdots \underbrace{\nabla}_{\text{$i$-th slot}} \otimes \dots 1) t^0[ \otimes_i \Phi_i] = 
t^0[\Phi_1 \otimes \cdots \nabla \Phi_i \otimes \cdots \Phi_{N_T}]. 
\een
In parallel with the arguments of \cite{hw2}, properties T1-T5 and T10 
will hold for all time-ordered products with $N_T$ factors
if and only if each $t^0$ can be extended to a distribution $t$ 
defined on all of $M^{N_T}$ in such 
a way that the above properties are preserved in the extension process.

The methods of \cite{hw2} already provide an extension $t$ of $t^0$ that 
satisfies all of the required properties except that
it is not guaranteed to satisfy the Leibniz relation 
\ben
\label{leib1}
(1 \otimes \cdots \underbrace{\nabla}_{\text{$i$-th slot}} \otimes \dots 1) t[ \otimes_i \Phi_i] = 
t[\Phi_1 \otimes \cdots \nabla \Phi_i \otimes \cdots \Phi_{N_T}]. 
\een
We will now
complete the proof by showing how this relation can be satisfied.

Let $\V$ denote the space of classical field expressions,
eq.~(\ref{Vclassical}). Let $\V(N_\varphi, N_\nabla)$ denote the
subspace of $\V$ spanned by the monomial expressions in the curvature, 
in $\varphi$, and in symmetrized derivatives of $\varphi$ 
that have a total
number of precisely $N_\varphi$ powers of $\varphi$ and a total
number of precisely $N_\nabla$ derivatives acting on the factors of
$\varphi$.
Let 
\ben
Q_\nabla: \V(N_\varphi, N_\nabla -1) \rightarrow
\V(N_\varphi, N_\nabla)
\een
denote the map whose action\footnote{
Note that the definition of $Q_\nabla$ appears more natural in 
a framework in which one views $\V$ not as a vector space
over the reals, but instead as a module over the ring of 
polynomial curvature expressions, ${\mathcal R}_{\rm class}$. 
In this context, $Q_{\nabla}$ is simply defined to act as 
the derivative followed by symmetrization on monomials in 
$\V$ without curvature coefficients, and then extended to 
all of $\V$ by ${\mathcal R}_{\rm class}$-linearity.}
on an element
$\Phi \in \V(N_\varphi, N_\nabla - 1)$ is defined by first applying
$\nabla_a$ to $\Phi$, then dropping all terms where $\nabla_a$ acts on
factors of curvature rather than factors of $\varphi$ and, finally,
symmetrizing over all derivatives acting on any given factor of
$\varphi$. 

We note, first, that for $N_\varphi > 0$, the map $Q_{\nabla}$ has
vanishing kernel, i.e., in essence, the derivative of any nonvanishing
expression with a nontrivial dependence on $\varphi$ cannot vanish. To see
this explicitly,
choose an arbitrary but fixed point $x \in M$, 
a coordinate basis $x^\mu = (x^0, x^1, \dots, x^{D-1})$, and 
consider the coordinate components
\ben
\label{coordexp}
\Phi_{\underline \nu}(x) = 
\sum_{\underline \mu_1 \dots \underline \mu_n}
C^{\underline \mu_1 \dots \underline \mu_n}{}_{\underline \nu}(x) \, 
\prod_{i=1}^n \nabla_{\underline \mu_i} \varphi(x)
\een
of a $\Phi \in \V(N_\varphi = n, N_\nabla)$, where $\underline \nu$ etc. is 
a shorthand for a symmetrized combination $(\nu_1 \dots \nu_k)$ of components. 
With each such coordinate component~\eqref{coordexp}, we assign 
a unique element $p_\Phi \in \mc[P_1, \dots, P_n]$, the ring of polynomials in the indeterminates
$P_i^\mu, i= 1, \dots, n, \mu= 0, \dots, D-1$ which are symmetric under exchange of 
$P_i$ and $P_j$, by the following rule: With the $i$-th factor of $\varphi$ in expression~\eqref{coordexp}, we associate the monomial in $P_i = (P_i^0, \dots, P_i^{D-1})$
obtained by replacing each derivative operator by the corresponding component of $P_i$. We
then multiply the 
resulting monomials in $P_i$ for all $i = 1, \dots, n$, we multiply by the corresponding
real constants $C^{\dots}{}_{\dots}(x)$ and we symmetrize in $i$. It is then clear that the components
of $Q_\nabla \Phi$ correspond to the polynomials $(\sum_i P_i^\mu) p_\Phi$, and it is clear 
that $Q_\nabla \Phi$ for $\Phi \in \V(N_\varphi = n, N_\nabla)$ will be zero if and only if all these polynomials are zero. It is easy to see (e.g., by the arguments given on p.~22 of~\cite{df2}) that this can only happen if in fact $p_\Phi = 0$, and hence that
$\Phi = 0$, as we desired to show.

Let $\V^0(N_\varphi, N_\nabla)$ denote the subspace of
$\V(N_\varphi, N_\nabla)$ spanned by expressions in the image of
$\V(N_\varphi, N_\nabla -1)$ under the map $Q_\nabla$, multiplied by
arbitrary curvature tensors.  Let
$\V^1(N_\varphi, N_\nabla)$ be a complementary subspace, so that
\ben
\label{dcmp0}
\V(N_\varphi, N_\nabla) = \V^0(N_\varphi, N_\nabla) \oplus 
\V^1(N_\varphi, N_\nabla). 
\een
Note that $\V^0(N_\varphi, N_\nabla)$
is uniquely defined by our construction, but there are, of course, 
many possible choices\footnote{\label{Nphi2}A particular choice of 
$\V^1(N_\varphi, N_\nabla)$ in the context of flat spacetime theories
was made in~\cite{df2}. 
It is worth noting that when $N_\varphi = 2$, we have
$\V(2, N_\nabla) = \V^0(2, N_\nabla)$ when $N_\nabla$ is odd, i.e., there
are no ``Leibniz independent'' expressions when $N_\nabla$ is 
odd; when $N_\nabla$ is even, a
convenient choice of $\V^1(2, N_\nabla)$ are the expressions of the form
of local curvature terms times 
$\varphi \nabla_{(a_1} \dots \nabla_{a_{N_\nabla})} \varphi$.}
of complementary subspace $\V^1(N_\varphi, N_\nabla)$.
The key point is that if we fix all of the arguments of $t$ except for the
$i$-th and if we know the action of $t$ on all $\Phi_i \in
\V(N_\varphi, N_\nabla -1)$, then --- because $Q_\nabla$ has vanishing kernel ---
the Leibniz rule eq.~(\ref{leib1})
uniquely determines the action of $t$ for 
$\Phi_i \in \V^0(N_\varphi, N_\nabla)$.
On the other hand, the Leibniz rule imposes no constraints whatsoever
on the action of $t$ for $\Phi_i \in \V^1(N_\varphi, N_\nabla)$.
We will therefore refer to $\V^0(N_\varphi, N_\nabla)$
as the ``Leibniz dependent'' subspace of $\V(N_\varphi, N_\nabla)$, and
will refer to $\V^1(N_\varphi, N_\nabla)$ as the subspace of
``Leibniz independent'' expressions. 

We now fix $N_T$ and fix the number, $N_\varphi^i$, of
powers of $\varphi$ in each factor of the argument of $t$, and proceed
by induction on $\{N_\nabla^1, \dots, N_\nabla^{N_T}\}$. The proof
of \cite{hw2} already directly establishes existence of the desired
extension of $t$ when
$N_\nabla^1 = \dots = N_\nabla^{N_T} = 0$, since the Leibniz rule clearly
imposes no additional restrictions in this case. We now
inductively assume existence has been proven for all $N_\nabla^i < n_i$, 
where $i = 1,...,N_T$. The inductive proof will be completed if we can show
that for any $j$, existence continues to hold whenever 
$N_\nabla^j = n_j$
and $N_\nabla^i < n_i$ for $i \neq j$.
To prove existence for
$N_\nabla^j = n_j$, we decompose $\Phi_j \in \V(N_\varphi^j, N_\nabla^j= n_j)$
into its ``Leibniz dependent'' and ``Leibniz independent'' pieces,
eq.~(\ref{dcmp0}). On $\V^1(N_\varphi^j, N_\nabla^j = n_j)$, we define the 
extension of $t$ as in \cite{hw2}, whereas on 
$\V^0(N_\varphi^j, N_\nabla^j = n_j)$
we simply {\em define} the 
extension of $t$ so as to satisfy 
eq.~\eqref{leib1}, i.e., we use the left side of that equation
to define the right side. It is clear that defining 
$t$ in this way 
yields a local and covariant distribution
that depends smoothly and analytically
on the metric, that has an almost 
homogeneous scaling behavior, and that satisfies
the desired microlocal properties, 
since taking covariant derivatives preserves these properties. Consequently,
an extension of $t$ satisfying all of the desired properties (including the
Leibniz rule) exists.

\end{proof}

\subsection{Anomalies with respect to the equations of motion}

We conclude this section by elucidating the difficulties 
(i.e., ``anomalies'') that arise when one
attempts to impose further ``reasonable'' conditions 
concerning the equations of motion in addition to
T1-T10 on Wick powers containing derivatives. Specifically, we shall
show that in two spacetime dimensions, it is impossible to require the
vanishing of $(\nabla_a \varphi) P \varphi$, where $P = \nabla^a
\nabla_a - m^2 - \xi R$ is the Klein-Gordon operator. We also shall show that
in four spacetime dimensions, it is impossible to require the
vanishing of both $\varphi P \varphi$ and $(\nabla_a \varphi) P
\varphi$.  These difficulties are closely related to the well-known
trace anomaly property for a conformally invariant field in these dimensions.
However, in contrast to previous discussions, in our framework 
the relation $T^a{}_a = 0$ holds as a classical algebraic
identity (not requiring the field equations) 
for the conformally invariant field in $D=2$ 
spacetime dimensions. Consequently,
in our framework, there cannot be a trace anomaly for the conformally 
invariant scalar field in two spacetime dimensions; rather, in two dimensions
there is necessarily an anomaly in the conservation of $T_{ab}$.

Consider a prescription (such as the local Hadamard
normal ordering defined above in eq.~\eqref{phih})
for defining Wick polynomials 
in $D$ spacetime dimensions that satisfies T1-T10.  
Using the Leibniz rule, T10, the stress-tensor $T_{ab}$,
eq.~(\ref{tabdef}), may be re-written entirely in terms of
the Wick monomials
\ben
\Psi \equiv \varphi^2, \,\,\,\,\, \Psi_{ab} \equiv \varphi \nabla_a 
\nabla_b \varphi
\een
as follows:
\begin{multline}
\label{tabdef2}
T_{ab} = \frac{1}{2} \nabla_a \nabla_b \Psi - \Psi_{ab} - \frac{1}{2}
g_{ab} (\frac{1}{2} \nabla^c \nabla_c \Psi + m^2 \Psi - {\Psi^c}_c) \\
+ \xi[G_{ab} \Psi -  \nabla_a \nabla_b\Psi + g_{ab} \nabla^c \nabla_c \Psi]. 
\end{multline}
The divergence of $T_{ab}$ is straightforwardly calculated (again using
the Leibniz rule) to be given by
\ben
\nabla^a T_{ab} = (\nabla_b \varphi) P \varphi.
\label{t1}
\een
On the other hand, we also have the obvious relation
\ben
{\Psi^a}_a - (m^2 + \xi R) \Psi 
= \varphi P \varphi.
\label{t2}
\een

The above equations hold for any prescription satisfying T1-T10.  Now let
us calculate $\varphi P\varphi$ and $(\nabla_a \varphi) P \varphi$ by
the local Hadamard normal ordering prescription. In odd dimensions,
these quantities vanish, but in even dimensions the computations
reported in Lemma 2.1 of \cite{mo} yield\footnote{The difference 
between the behavior occurring in even and odd dimensions can be understood
as arising from the following fact: In odd dimensions,
one can construct a local and covariant Hadamard parametrix
$H(x,y)$ that satisfies the wave equation in each variable
up to arbitrarily high order in the geodesic distance between $x$ and
$y$. As a result, the ``local Hadamard normal ordering'' prescription
for Wick powers satisfies T1-T10 {\em and} also satisfies the property
that any Wick power containing a factor of the wave operator must
vanish. By contrast, in even dimensions, it is impossible to construct
a local and covariant Hadamard parametrix $H(x,y)$, that is
symmetric in $x$ and $y$ and satisfies the wave equation to
arbitrarily high order in the geodesic distance between $x$ and $y$. 
Consequently, if one requires $H(x,y)$ to be
symmetric (as is necessary for the prescription for defining general
Wick monomials involving derivatives to be well defined),
then it will fail to satisfy the wave equation, and ``anomalies'' will
occur for the regularized quantities.}
\ben 
\varphi P\varphi = Q \myid 
\een 
\ben 
(\nabla_a \varphi) P \varphi = \frac{D}{2(D+2)} \nabla_a Q \myid 
\een 
where $Q$ is a nonvanishing local
curvature scalar of dimension $({\rm length})^{-D}$ that can be computed
explicitly from the Hadamard recursion relations.

If we wish to require the vanishing of $\varphi P \varphi$ and 
$(\nabla_a \varphi) P \varphi$, our task is to
modify---in a manner consistent with axioms T1-T10---the definitions
of the Wick powers, $\Psi = \varphi^2$ and $\Psi_{ab} =
\varphi \nabla_a \nabla_b \varphi$, 
so that the left sides of eqs.~(\ref{t1}) and (\ref{t2}) vanish. However, 
since $\Psi$ and $\Psi_{ab}$ are each quadratic in $\varphi$, our
previous uniqueness theorem \cite{hw1} establishes that the allowed
freedom in the definition of these quantities consists of local curvature
terms of the correct dimension times the identity, $\myid$. More precisely, 
the allowed freedom to modify the definition of these quantities 
is\footnote{Condition T10, of course, was not imposed in \cite{hw1}. 
However, it is worth
noting that the subspace of classical field expressions spanned by
$\varphi^2$ and $\varphi \nabla_a \nabla_b \varphi$
does not intersect the ``Leibniz dependent'' subspace $\V^0$ (see footnote
\ref{Nphi2}), so
condition T10 actually imposes no extra
conditions on these quantities, i.e.,
the full ambiguity given by eqs.~(\ref{u1}) and (\ref{u2}) is present
even when we impose T10.} 
\ben 
\Psi \rightarrow \Psi + C \myid
\label{u1} 
\een 
\ben 
\Psi_{ab} \rightarrow \Psi_{ab} + C_{ab} \myid
\label{u2} 
\een 
where $C$ is any scalar constructed out of the metric, curvature, 
derivatives of the curvature, 
$m^2$ and $\xi$, with dimension $({\rm length})^{-(D-2)}$
and $C_{ab}$ is any tensor (symmetric in $a$ and $b$)
that is constructed out of the metric, curvature, 
derivatives of the curvature, and  $m^2$
and has dimension $({\rm length})^{-D}$. 
Therefore, in order to modify the local Hadamard normal ordering 
prescription so as to preserve T1-T10 and also make the right sides
of eqs.~(\ref{t1}) and (\ref{t2}) vanish, we must solve the following
equations
\ben
- (\nabla^a C_{ab} - \frac{1}{2}\nabla_b {C^a}_a)
+ \frac{1}{4} \nabla_b \nabla^a \nabla_a C +\frac{1}{2} {R_b}^c \nabla_c C
+ \frac{1}{2}(m^2 + \xi R)\nabla_b C = - \nabla_b Q
\label{v1}
\een
\ben
{C^a}_a - m^2 C - \xi R C = - \frac{D}{2(D+2)} Q
\label{v2}
\een

Let us specialize, now, to the case of two spacetime dimensions,
$D=2$. Since $C$ has dimension zero, the most general choice of $C$ is
simply $C=\alpha$, where $\alpha$ is a constant that is independent
of $m^2$ but may have arbitrary analytic dependence on $\xi$. However,
since $C$ appears in eq.~(\ref{v1}) only in the form $\nabla_b C$, it
cannot contribute to that equation. Similarly, since $C_{ab}$ has
dimension $({\rm length})^{-2}$, it must take the form $C_{ab} =
\beta_1 R_{ab} + \beta_2 R g_{ab} + \beta_3 m^2 g_{ab}$, where
$\beta_1, \beta_2, \beta_3$ are constants that are independent of
$m^2$ but may have arbitrary analytic dependence on $\xi$. However, in
two spacetime dimensions, we have $R_{ab} = \frac{1}{2} R g_{ab}$, and
since $C_{ab}$ appears in eq.~(\ref{v1}) only in the combination $C_{ab} -
\frac{1}{2} {C^c}_c g_{ab}$, it follows immediately that $C_{ab}$ also cannot
contribute to eq.~(\ref{v1}). Since $\nabla_a Q$ is nonvanishing, it is therefore
impossible to solve eq.~(\ref{v1}). Consequently, in two spacetime
dimensions, there does not exist a prescription for defining Wick
powers that satisfies axioms T1-T10 and also satisfies $(\nabla_a
\varphi) P \varphi =0$. Since $\nabla^a T_{ab} = (\nabla_b \varphi) P
\varphi$ by eq.~(\ref{t1}) above, this means, equivalently, that it 
is impossible to
satisfy conservation of stress-energy in two spacetime dimensions
within our axiomatic framework.

By contrast, in spacetime dimension $D>2$, no difficulty arises in
satisfying eq.~(\ref{v1}) alone (in addition to axioms T1-T10), 
since we may always solve this
equation by choosing $C=0$ and taking
\ben
C_{ab} = - \frac{2}{D-2} Q g_{ab}
\een
Thus, in all spacetime dimensions except $D=2$, there is no obstacle
to imposing $(\nabla_a \varphi) P \varphi =0$---or, equivalently,
conservation of stress-energy---within our axiom scheme. However,
difficulties do arise if, in addition, we attempt to impose $\varphi P
\varphi =0$ as well. For example, when $D=4$, the general form of $C$
is 
\ben\label{Cdefn}
C = \alpha_0 R + \alpha_1 m^2, 
\een
where $\alpha_0, \alpha_1$
are constants. If one substitutes this general form of $C$ into
eq.~(\ref{v1}), it turns out that it is still always possible to solve
eq.~(\ref{v1}) for $C_{ab}$. The general solution of eq.~\eqref{v1}
is most conveniently expressed in terms of the quantity $\bar{C}_{ab} \equiv C_{ab} 
- \frac{1}{2} {C^c}_c g_{ab}$, and takes the explicit form
\bena
\label{barCab}
\bar C_{ab} &=& \left( Q + \frac{\alpha_0}{4} \nabla^c \nabla_c R 
+ \frac{1}{2} \alpha_0 m^2 R + \frac{1}{8}\alpha_0(2 \xi + 1) R^2 \right) g_{ab}
+ \frac{1}{2} \alpha_0 R G_{ab}  \nonumber \\
&& + \bar{\beta}_1I_{ab} +
\bar{\beta}_2J_{ab} + \bar{\beta}_3 m^2 G_{ab} + \bar{\beta}_4 m^4
g_{ab}
\eena
where $\bar{\beta}_1,\dots,\bar{\beta}_4$ are arbitrary constants, and $I_{ab}$
and $J_{ab}$ are the two independent conserved local curvature tensors
of dimension $({\rm length})^{-4}$. We now substitute the general form, eq.~\eqref{Cdefn},
of $C$ and this general solution, eq.~\eqref{barCab}, for $C_{ab}$ into
eq.~(\ref{v2}). Since both $I_{ab}$ and $J_{ab}$ have trace
proportional to $\nabla^a \nabla_a R$, 
we obtain an equation of the form
\ben
Q = \gamma_1 \nabla^a \nabla_a R + \gamma_2 R^2 + \gamma_3 m^2 R 
+ \gamma_4 m^4.
\een
However, by explicit calculation, $Q$ contains terms of the form
$C^{abcd} C_{abcd}$ and $R^{ab} R_{ab}$, which cannot be expressed as a sum 
of the curvature terms on the right side. Consequently, there are no
solutions to eq.~(\ref{v2}) when $D=4$. Similar results presumably
hold in all higher even dimensions, but a proof of this would require
both a calculation of $Q$ and an analysis of the conserved local
curvature terms of dimension $({\rm length})^{-D}$.

Finally, we note that our viewpoint with respect to the definition of
the stress-energy tensor differs in two
significant aspects with that of Moretti \cite{mo} and others.  First,
we take the free field quantum stress-energy tensor, $T_{ab}$, to be
defined in terms of Wick monomials by eq.~(\ref{tabdef}) and we do not
allow any modifications of this formula. It is natural that we take
this viewpoint, because it is precisely the quantity defined by
eq.~(\ref{tabdef}) that will directly enter condition T11b of the next
section. Furthermore, if condition T11b is imposed---as is possible
except in 2 spacetime dimensions, as will be proven in section
6---then, as we shall show in section 5, one obtains not only the
conservation of the stress energy tensor $T_{ab}$ in the free theory,
but one also obtains conservation of the interacting stress-energy
tensor $\Theta_{ab}$ in an arbitrary interacting theory. By contrast,
Moretti \cite{mo} allows modifications to the formula for the
stress-energy tensor that are proportional to $\varphi P \varphi$ and
thus vanish classically. If one were only interested in considering
the free field theory and were seeking a definition of its
stress-energy tensor that is conserved and that corresponds to the
classical expression in the classical limit, we see no argument
against allowing such a re-definition\footnote{Indeed, this was the
philosophy taken in the prior work of one of us \cite{wald} on the
stress-energy tensor, before the general theory of Wick products in
curved spacetime had been developed.}. However, it seems unlikely that
this approach could naturally lead to a conserved stress-energy tensor
in interacting theories.

Second, Moretti \cite{mo} takes the Wick monomials appearing in his
(modified) formula for $T_{ab}$ to be defined by a particular, fixed
prescription, namely local Hadamard normal ordering (see
eq.~(\ref{phih}) above). By contrast, we allow an arbitrary
prescription for defining Wick products satisfying T1-T10. However, it
turns out that---with the exception of one case---the freedom
(\ref{u1}) and (\ref{u2}) allowed by T1-T10 in the definition of the
relevant Wick products is sufficient to encompass the modifications to
$T_{ab}$ obtained by Moretti by adding terms proportional to $\varphi
P \varphi$ but keeping the prescription for defining Wick products fixed. 
In other words---with one exception---we achieve the same
final result for $T_{ab}$ as an element of $\W$ by modifying the
prescription for defining Wick products rather than by modifying the
formula for $T_{ab}$ in terms of Wick products. The exception is the
case of 2 spacetime dimensions. Indeed, it can be seen directly from
eq.~(\ref{tabdef2}) that when $D=2$ and when $m=0$, the allowed
freedom (\ref{u1}) and (\ref{u2}) does not permit any modification
whatsoever to $T_{ab}$. Thus, while our results agree with those of
Moretti when $D \neq 2$, they differ when $D=2$.

\section{Quadratic interaction Lagrangians and retarded response}

\subsection{Formulation of the general condition T11}

In this section, we formulate the new requirement, T11, that arises from 
our Principle of Perturbative Agreement when the
perturbation, $L_1$, of the free Lagrangian, $L_0$, is at most quadratic in 
the field $\varphi$ and contains at most 2 derivatives of
$\varphi$. The most general such 
nontrivial interaction Lagrangian for a real scalar field is
\ben
L_1 = \frac{1}{2} [j(x) \varphi + w^{ab}(x)  \nabla_a \varphi \nabla_b \varphi
+ v(x) \varphi^2] \eps
\label{intL}
\een
where $j$, $w^{ab}$, and $v$ are smooth. Without loss of
generality, we will assume that $j$, $w^{ab}$, and $v$ are of compact 
support on $M$, since the conditions that we derive are local conditions
that will not depend on their support properties, 
and it is simpler to consider the compact support
case. The term $j \varphi \eps$ in eq.~(\ref{intL}) corresponds to the
addition of an external current; the term $v \varphi^2 \eps$ corresponds
to the presence of an external potential (or, equivalently, a spacetime
variable mass); finally, the change in $L$ produced by a 
changing the spacetime metric from $g_{ab}$ to ${g'}_{ab}$ corresponds to 
taking $w^{ab} \eps = {g'}^{ab} \eps' - g^{ab} \eps$ 
and $v \eps = m^2(\eps' - \eps) + \xi (R'\eps' - R\eps)$. Note that we have
not included a term of the form $A^a \varphi \nabla_a \varphi \eps$ in $L_1$, 
since such a term is Leibniz equivalent to $v \varphi^2 \eps$ with 
$v = - \nabla_a A^a$. However, if we were considering a complex scalar field,
then we would have an additional term in $L_1$ of the form 
$\i A^a (\bar{\varphi} \nabla_a \varphi - \varphi \nabla_a \bar{\varphi}) \eps$.

The quantum field theory of a scalar field $\varphi$ with Lagrangian 
$L = L_0 + L_1$ can be constructed in the following two independent ways.
First, it can be constructed perturbatively about $L_0$ by means of the
Bogoliubov formula. This formula is most conveniently expressed in terms of 
retarded products, so we first recall the definition of retarded and advanced
products, $\R$ and $\A$, in terms of time-ordered products.
If $A,B \in \F$, we define
\ben
\label{ARdef}
\R(e^{\i A}; e^{\i B}) = \S(B)^{-1} \S(A+B), \quad \A(e^{\i A}; e^{\i B}) = \S(A+B) \S(B)^{-1},  
\een
where $\S$ and $\S^{-1}$ are given by the formal series expressions
\ben
\S(A) = \T(e^{\i A}), \quad \S(A)^{-1} = \bar \T(e^{\i A}), 
\label{S}
\een
and where $\bar \T$ denotes the ``anti-time-ordered'' product given 
by eq.~(\ref{atoprod}).
Equation~(\ref{ARdef}) is to be interpreted as an infinite sequence 
of well defined
equations obtained by formally expanding the exponentials on the left
side, formally substituting eq.~(\ref{S}) for $\S$ and $\S^{-1}$ on the
right side, and then equating the terms with the same number of powers
of $A$ and $B$.
For example, if we set $A=\sum f_i \Phi_i$
and $B = h \Psi$ and expand 
to first order in $B$, we get the formula
\ben
\label{1}
\R\left(\prod f_i \Phi_i; h\Psi\right) = -\Psi(h) \T\left(\prod f_i \Phi_i \right) + \T\left(h\Psi \prod f_i \Phi_i\right)
\een
as well as a similar formula for the advanced products. The retarded and 
advanced products have the important support property
\begin{multline}
\label{rsupp}
\supp \A/\R(\prod \Phi_i(x_i); \prod \Psi_j(y_j)) \subset \\
\{(x_1, \dots, x_n; y_1, \dots, y_m) \mid 
\text{some $x_i \in J^{-/+}(y_j)$ for at least one $j$.} \}. 
\end{multline}
which follows from the causal factorization property, 
T8, of the time ordered products. 

Now let $L_1$ be any compactly supported 
interaction Lagrangian density that is polynomial
in $\varphi$ and its derivatives. Let $\Phi_i \in \V$. Then
the Bogoliubov formula defines the
{\em interacting} time-ordered product $\T_{L_1} (\prod f_i \Phi_i)$
as an element of the non-interacting algebra $\W(M,\g)$ by 
the power series expression
\ben
\label{pertseries}
\T_{L_1} \left( \prod_{i=1}^m f_i \Phi_i \right) 
= \R\left(\prod_{i=1}^m f_i \Phi_i; e^{\i L_1} \right) = 
\sum_n \frac{\i^n}{n!} \R\left(\prod_{i=1}^m f_i \Phi_i; 
\underbrace{L_1 \cdots L_1}_{\text{$n$ factors}} 
\right). 
\een
This definition of time-ordered products for the interacting quantum field
corresponds to the boundary condition that the 
interacting field
be equal to the corresponding non-interacting field outside the causal future
of the support of the interaction Lagrangian $L_1$. The 
interacting field corresponding
to the opposite boundary condition (with the future of the support 
of $L_1$ replaced
by the past) is given by the analogous formula involving advanced products. 
As is well known, the perturbation series (\ref{pertseries}) is expected
not to converge\footnote{However, see the remarks at the end of 
section 7.} for general $L_1$, so this relation is, in general, only
a formal one.

The Bogoliubov formula (\ref{pertseries}) holds for an arbitrary
interaction Lagrangian. However, if $L_1$ takes the simple form
(\ref{intL}), then there is a second means of constructing
``interacting'' time-ordered products: We simply construct them
directly for the theory given by the Lagrangian $L'_0 = L_0 + L_1$. As
already indicated, the Lagrangian $L'_0$ corresponds to a free scalar
field in a curved spacetime with metric $\g'$ 
in the presence of an external potential
$V'$ and an external source $J'$. We have already constructed the quantum
field theory of $\varphi$ in an arbitrary, globally hyperbolic spacetime
$(M,\g)$.
The generalization of this construction to include an external potential,
$V$, is accomplished in an entirely straightforward manner as follows: In
the construction of the algebras ${\mathcal A} (M, \g, V)$ and $\W(M, \g, V)$, 
we simply replace the original
Klein-Gordon operator $P = \nabla^a \nabla_a - \xi R - m^2$ by 
$\nabla^a \nabla_a - \xi R - m^2 -V$. In the definition of the classical
field algebra, $\V$, we must now also allow arbitrary factors of $V$ and its 
symmetrized covariant derivatives.
In the axioms for time-ordered products, we simply make the obvious 
modifications to T1, T2, T4, and T5 to allow for the presence of $V$. The
proof of existence of a prescription for defining time-ordered products
satisfying T1-T10 then goes through without any substantive changes.

Given the theory of the real scalar field $\varphi$ in the absence
of an external current, $J$, the corresponding 
theory in the presence of an external
current may be uniquely constructed as follows (where, for simplicity, we
assume $V=0$ in the following discussion): First, the CCR algebra, 
${\mathcal A}(M, \g, J)$, is constructed by starting with
the free algebra generated by symbols 
$\varphi_J(f)$ and $\varphi_J(h)^*$, and factoring
by the relations (i), (ii) and (iv) of section 2, 
together with the relation (iii')
\ben
\varphi_J(Pf) = \int fJ \eps \cdot \myid, 
\label{phiJ}
\een
where $P$ is the Klein-Gordon operator, eq.~(\ref{kgP}). 
Thus, the only 
modified relation compared to the case of vanishing source is (iii'),
which ensures that 
the field $\varphi_J$ satisfies the Klein-Gordon equation with source $J$ 
in the distributional sense. Note that the commutation relations (iv) 
of the field with 
source are identical to the commutation relations of the field without source.
To construct $\W(M, \g, J)$, we define the generators $W_n(u)$ 
by eq.~(\ref{a2}) with $\varphi_J$ replacing $\varphi$ but with
$\omega_2$ taken to be a Hadamard state of the theory with vanishing source,
so that $\omega_2$
still satisfies the homogeneous equation in each entry.

The Wick monomials may now be defined for the theory with external source
by exactly the same formula, eq.~\eqref{phih}, as used in the 
theory without source, with the only
difference being that the normal ordered expressions of the field without 
source that appear on the right side of \eqref{phih} must be 
replaced by the local Hadamard normal ordered 
expressions $: \varphi_J(x_1) \cdots \varphi_J(x_k) :_H$ of the 
field $\varphi_J$ with source.
The latter are defined exactly as for 
the case of a vanishing source by eq.~\eqref{hndef}, where $H$ is the same
local Hadamard parametrix for the Klein-Gordon operator $P$ as used
in the theory without a source. 
It then follows from causal factorization, T8, and the commutator property,
T9, that a local Wick expansion of the form 
eq.~(\ref{lwe}) holds for time-ordered products, 
$\T^0_J(\Phi_1(x_1) \cdots \Phi_n(x_n))$ when $x_i \neq x_j$ for all $i,j$,
where the expansion coeficients, $t^0[\otimes \Phi_i]$, are identical to those
of the theory without source. We may therefore {\em define} 
time-ordered products
in the theory with source by choosing the extension,
$t[\otimes \Phi_i]$, 
to be {\em independent} of the source $J$. The resulting definition of
time-ordered products satisfies axioms T1-T10, with the obvious modifications
to T1, T2, T4, and T5 to allow for the presence of $J$. This construction 
of the theory of a real scalar field in the presence of an external source,
$J$, in terms of the theory defined when $J=0$ corresponds to 
demanding that the 
renormalization prescription be independent\footnote{$J$-dependent 
renormalization prescriptions that satisfy the appropriate versions of
T1-T10 could be defined by choosing a local Hadamard parametrix 
for the Klein-Gordon operator $P$ that depends nontrivially on $J$
in a local and covariant manner (e.g., choosing 
$H'(x,y) = H(x,y) + J(x) J(y)$) and/or by choosing the extension, $t$, of
$t^0$ to depend nontrivially on $J$ in a local and covariant manner.}
of $J$.

Our Principle of Perturbative Agreement demands that for $L_1$ given by
eq.~(\ref{intL}), the perturbative theory defined by the Bogoliubov
formula (\ref{pertseries}) must agree with the exact theory for the
Lagrangian $L'_0 = L_0 + L_1$.  However, we cannot yet easily compare
these constructions, since the Bogoliubov formula expresses the
interacting time-ordered products as elements of the algebra
$\W(M,\g)$, whereas the exact construction yields time-ordered
products as elements of a different algebra, $\W(M,\g', V', J')$.
Therefore, in order to compare these two expressions for the
time-ordered products, we must map $\W(M,\g', V', J')$ into $\W(M,\g)$
in such a way that, in the past (i.e., outside the future of the
supports of $\g - \g'$, $V'$, and $J'$), the interacting time-ordered
products in $\W(M,\g', V', J')$ are mapped into the corresponding
time-ordered products in $\W(M,\g)$. The desired map, denoted
$\tau^\ret$, was constructed in lemma 4.1 of \cite{hw1} in the case
where $V'=J'=0$. We now briefly review its construction and generalize
it to the case of nonvanishing external potential and current.

Let $(M, \g)$ be a globally hyperbolic spacetime and $(M,\g', V', J')$
be such that $(M, \g')$ also is globally hyperbolic and outside some
compact set $K \subset M$, we have $\g' = \g$, $V=0$, and $J=0$. Let
$\Sigma_t$ be a foliation of $(M, \g)$ by Cauchy surfaces.
Choose $t_1$ such that $\Sigma_{t_1}$ does not intersect
the causal future of $K$ and choose $t_0 < t_1$, so that $\Sigma_{t_0}
\subset I^-(\Sigma_{t_1})$.
(It follows automatically that
$\Sigma_{t_0}$ and $\Sigma_{t_1}$ are also Cauchy surfaces for $(M, \g')$.)
Let $\psi$ be a smooth function on $M$ such that $\psi(x) = 0$
for $x$ in the future of $\Sigma_{t_1}$  and $\psi(x) = 1$ for 
$x$ in the past of $\Sigma_{t_0}$.
The action of $\tau^\ret$ on
$\varphi_{(\g', V', J')}$ may now be defined as follows: Let $f$ be a test
function on $M$. We define 
\ben
F = \Delta_{(\g', V')} f
\een
and we define 
\ben
f' = P'(\psi F)
\een
with $P'={\nabla'}^a {\nabla'}_a - \xi R' - m^2 -V'$ and with 
$\Delta_{(\g', V')}$ being the advanced minus retarded Green's function
associated with $P'$. Then $f'$ is a test
function with support lying between $\Sigma_{t_0}$ and $\Sigma_{t_1}$. Furthermore,
it is easily checked that $\Delta_{(\g', V')} (f - f') = 0$, which implies
that $f-f' = P'h$ where $h \equiv \Delta^\adv_{(\g', V')}(f-f')$ is
of compact support \cite{wald}, and hence is a test function. Consequently, by 
eq.~(\ref{phiJ}), we have
\ben
\varphi_{(\g', V', J')} (f) = \varphi_{(\g', V', J')} (f') 
+ \int hJ' \eps' \cdot \myid
\een
Since the supports of $\Delta^\adv_{(\g', V')} f'$ and $J'$ do not
overlap, and since $\Delta^\adv_{(\g', V')}(x,y) = \Delta^\ret_{(\g', V')}(y, x)$ (up to factors of $\eps'$) 
we may re-write this equation as
\ben
\varphi_{(\g', V', J')} (f) = \varphi_{(\g', V', J')} (f') 
+ \int f\Delta^\ret_{(\g', V')}J' \eps' \cdot \myid
\label{tauretphi}
\een
All of the above equations are relations in the algebra $\W(M,\g', V', J')$.
We want to define the map $\tau^\ret:\W(M,\g', V', J') \rightarrow \W(M,\g)$
to be such that it maps $\varphi_{(\g', V', J')} (f')$ to 
$\varphi_\g (f')$ when the support of $f'$ lies outside the future of $K$
and to be such that $\tau^\ret(\myid) = \myid$.
Therefore, for arbitrary $f$, we define
\ben
\tau^\ret[\varphi_{(\g', V', J')} (f)] = \varphi_\g (f') 
+ \int f\Delta^\ret_{(\g', V')}J' \eps' \cdot \myid
\label{tauretphi2}
\een

The above formula for the action of $\tau^\ret$ on $\varphi_{(\g', V', J')}$
can be rewritten in a more useful form as follows.
Let $S: \cD(M) \to \cD'(M)$ be the map defined by its distributional 
kernel, 
\ben
\label{a8}
S(f_1, f_2) = \int_{\Sigma_-} (F_1 \nabla_a F_2 - F_2 \nabla_a F_1) \, \d \sigma^a, 
\een
where $F_1 = \Delta_\g f_1$, $F_2 = \Delta_{(\g',V')} f_2$, and $\Sigma_-$
is any Cauchy surface that does not intersect the future of $K$.  
Define $A^\ret: \cD(M) \to \cD'(M)$ by
\ben
\label{a7}
A^\ret f = -P'(\psi S f). 
\een
Then we have
\ben
\tau^\ret[\varphi_{(\g', V', J')} (f)] = \varphi_\g (A^\ret f) 
+ \int f\Delta^\ret_{(\g', V')}J' \eps' \cdot \myid
\een
The map $\tau^\ret$ can be uniquely extended to all of 
${\mathcal A}(M, \g', V', J')$ as a *-isomorphism and, thereby, to $W_n(u)$ when $u$ is smooth.
The further extension of $\tau^\ret$ to $W_n(u)$ for $u$ a distribution in
the space eq.~(\ref{a3}) then can be accomplished as in 
lemma 4.1 of \cite{hw1}, with the trivial modifications 
resulting from the presence of $V$ and the straightforward modifications
resulting from the presence of $J$.

Our Principle of Perturbative Agreement now leads to the following requirement:

\paragraph{T11 Quadratic interaction Lagrangians.} Let $(M,\g)$ and $(M,\g')$
be globally hyperbolic spacetimes such that $\g'= \g$ outside of a compact
set $K$. Let $V'$ and $J'$ have support in $K$. Then for all $\Phi_i \in \V$
we have
\ben
\tau^\ret \left[\T_{(\g',V',J')} \left( \prod_{i=1}^m f_i \Phi_i \right) \right] 
= \sum_n \frac{\i^n}{n!} \R_\g \left(\prod_{i=1}^m f_i \Phi_i; 
\underbrace{L_1 \cdots L_1}_{\text{$n$ factors}} 
\right). 
\label{T11full}
\een
where $L_1$ is the interaction Lagrangian of the form
eq.~(\ref{intL}) given by $L_1 = L_0 - L'_0$.

\medskip

In order to express this condition in a more explicit and useful form,
we consider separately the subcases of (a) external source variation,
(b) metric variation, and (c) external potential variation. These will 
lead to sub-requirements T11a, T11b, and T11c. 

\subsection{External source variation: Axiom T11a}

We now apply condition T11 to the case of an interaction 
Lagrangian of the form $L_1 = J\varphi \eps$, where
$J$ is a compactly supported smooth function (``external current''). 
In this case, condition T11 reduces to
\ben
\label{nonpertj}
\tau^{\ret}\left[ \T_{J}\left(\prod_{i=1}^m f_i \Phi_i \right) \right]
= \sum_n \frac{\i^n}{n!} \R_{J=0} \left(\prod_{i=1}^m f_i \Phi_i; 
\underbrace{J\varphi \eps \cdots J\varphi \eps}_{\text{$n$ factors}} 
\right), 
\een
However, if only an 
external current is present, eq.~(\ref{tauretphi2}) becomes simply
\ben
\tau^{\ret}[\varphi_{J}(f)] = \varphi_{J=0} (f) - \int f  \Delta^{\ret} J \eps \cdot \myid.
\een 
More generally, for $A_i \in \F$, we have
\ben
\tau^\ret \left[ 
\T_{J}\left(
\prod_i A_i(\varphi)
\right)
\right]
= \T_{J=0}
\left(
\prod_i A_i(\varphi - \Delta^\ret J \eps)
\right), \quad A_i \in \F, 
\label{tauretphi3}
\een
which can be proved by induction the number of factors of 
the time ordered product, making 
use of the fact that the c-number coefficients in the Wick 
expansion of $\T_J$ are independent of $J$. Expanding eq.~(\ref{nonpertj})
to first order in $J$ and using eq.~(\ref{tauretphi3}), we obtain

\paragraph{T11a Free field factor.}
We have 
\ben
\label{5}
\R(\prod f_j \Phi_j; J \varphi \eps) = 
\i \sum_j \T\left(f_1\Phi_1\dots 
(\Delta^\ret J\eps)
\frac{\delta(f_j\Phi_j)}{\delta \varphi} \dots f_n \Phi_n\right). 
\een

\medskip

\paragraph{Remarks:} {\bf (1)} Condition T11a corresponds to condition N4 
of~\cite{df} in the context of QED in flat spacetime.

\smallskip
\noindent
{\bf (2)} We can use eq.~\eqref{1} to re-write eq.~(\ref{5})
purely in terms of time-ordered products as
\ben
\label{5'}
\T(J \varphi \eps \prod f_j \Phi_j) = \varphi(J\eps)\T(\prod f_j \Phi_j) +
\i \sum_j \T\left(f_1\Phi_1\dots 
(\Delta^\ret J \eps)
\frac{\delta(f_j\Phi_j)}{\delta \varphi} \dots f_n \Phi_n\right). 
\een
Hence, condition T11a implies that
a time-ordered product with $n+1$ factors such that
at least one of the factors is $\varphi$
can be expressed in terms of 
time-ordered products with fewer factors. Using this fact, one may show
that condition T11a implies all of the relations obtained by expanding
eq.~(\ref{nonpertj}) to any order in $J$. Thus, condition T11a is equivalent
to eq.~(\ref{nonpertj}) and contains the full content of condition T11 in
the case of an external current.

\smallskip
\noindent
{\bf (3)} The requirement T11a could also have be formulated in terms of 
advanced products by replacing ``$\R$'' by ``$\A$'' on the left side of 
eq.~\eqref{5} and by replacing the retarded propagator $\Delta^\ret$ by 
the advanced propagator $\Delta^\adv$ on the right side. 
It is not difficult to show (using the commutator property T9) 
that this would yield an equivalent requirement. 

\smallskip
\noindent
{\bf (4)} If we choose $J = (\nabla^a \nabla_a - m^2 - \xi R)h$ 
and use the Leibniz rule T10 
as well as $(\nabla^a \nabla_a - m^2 - \xi R)\Delta^\ret = \delta$, 
eq.~\eqref{5'} yields
\ben
\label{6}
\T\left(\eps h(\nabla^a \nabla_a - m^2 - \xi R)\varphi \prod f_i \Phi_i\right) = 
\i\sum_i \T\left(f_1\Phi_1\cdots 
h \frac{\delta (f_i\Phi_i)}{\delta \varphi} 
\dots f_n\Phi_n\right), 
\een
In theorem \ref{mainthm} in the next section, we will see that eq.~\eqref{6} 
is necessary and sufficient for the interacting field to satisfy the 
interacting equations of motion.

\smallskip
\noindent
{\bf (5)} In the simple case of two free-field factors, eq.~(\ref{5'})
reduces (in unsmeared form) to simply
\ben
\T(\varphi(x) \varphi(y)) = \varphi(x) \varphi(y) + \i \Delta^\ret(x,y)\myid.  
\een
This agrees with eq.~(\ref{top}), which was deduced from
axioms T1-T9. Similarly, condition T11a directly yields
\ben
\T(P\varphi(x) \varphi(y)) = i \delta(x,y) \myid,  
\een
which agrees with eq.~(\ref{tpphi}), which was deduced from the Leibniz rule.
These agreements are comforting, but they illustrate that
many nontrivial consistency checks will arise when we attempt to impose
condition T11a along with T1-T10. A proof that T1-T10, T11a, and T11b (see
below) can all be 
consistently imposed will be given in section 6.

\subsection{Metric variation: Axiom T11b}

We now apply condition T11 to the case where the interaction Lagrangian
corresponds to a variation in the spacetime metric, i.e., 
$L_1 = L_0(\g) - L_0(\g')$, where $\g$ and $\g'$ are both globally 
hyperbolic and differ only in a compact
subset $K$. In this case, condition T11 becomes
\ben
\label{nonpert}
\tau^{\ret}\left[ \T_{\g'}\left(\prod_{i=1}^m f_i \Phi_i \right)
\right] = \sum_n \frac{\i^n}{n!} \R_\g\left(\prod_{i=1}^m f_i \Phi_i;
\underbrace{L_1 \cdots L_1}_{\text{$n$ factors}} \right).  
\een 

As in the case of an external current considered in the previous
subsection, it is useful to pass to an infinitesimal version of 
this equation.
To accomplish 
this, we introduce a smooth 1-parameter family of metrics
$\g^{(s)}$ differing from $\g = \g^{(0)}$ only within $K$. 
To first order in $s$, the interaction Lagrangian density is then given by 
$L_1 = (s/2)  \eps h_{ab} T^{ab}$, where 
$h_{ab} = \frac{\partial}{\partial s} g^{(s)}_{ab}$, and 
where $T_{ab}$ is the stress energy tensor~\eqref{tabdef}.
For all $f_i \Phi_i \in \F$
we define
\ben
\delta_\g^{\ret}\left[\T_\g\left( \prod f_i \Phi_i \right) \right] 
= \frac{\partial}{\partial s} \tau^{\ret}_{\g^{(s)}} 
\left[ \T_{\g^{(s)}} \left(\prod f_i \Phi_i \right) \right] \bigg|_{s=0}. 
\een
In appendix A, we show that the right side of this equation exists
as a well defined element of $\W(M,\g)$.
By differentiating eq.~(\ref{nonpert}) with respect to $s$ and setting $s=0$,
we obtain the following
infinitesimal version of condition T11 in the case of metric variations:

\paragraph{T11b Stress-energy factor:}
Let $\T_{\g^{(s)}}$ be the 1-parameter family of time-ordered 
products associated with 
a smooth 1-parameter family of globally hyperbolic
metrics $\g^{(s)}$ on $M$ that vary only within some compact subset $K$, 
and such that $\g \equiv \g^{(0)}$. Then we require that
for all $f_i \Phi_i \in \F$, 
\bena
\label{Rdef}
\delta_\g^{\ret}\left[\T_\g\left( \prod f_i \Phi_i \right) \right]
&=& \Ihaves \R_\g\left(\prod f_i \Phi_i; \eps h_{ab} T^{ab}\right) \nonumber\\
&+& \sum_i \T_\g\left( f_1 \Phi_1 \cdots h_{ab} \frac{\delta (f_i \Phi_i)}{\delta g_{ab}} \cdots f_n \Phi_n \right), 
\eena
where $h_{ab}$ is the compactly supported tensor field given by 
\ben
h_{ab}=
\frac{\partial}{\partial s}
g^{(s)}_{ab} \bigg|_{s=0},
\een
and the functional derivative, $\delta A/\delta g_{ab}$, 
of a classical functional 
$A \in \F$ with respect to the metric, $\g$, is defined in appendix B and is
explicitly given by the formula
\ben
\frac{\delta A}{\delta g_{ab}} = \sum_r (-1)^r 
\nablao_{(c_1} \cdots \nablao_{c_r)} 
\frac{\partial A}{\partial (\nablao_{(c_1} \cdots \nablao_{c_r)} g_{ab})} \, . 
\label{delAdelg}
\een
In this formula, $\nablao_a$ is an arbitrary fixed, background 
derivative operator, and it is understood that we have re-written 
the dependence of $A$ on 
$\nabla_a$ and the curvature in terms of 
$\nablao_a$ and $\nablao_a$-derivatives of $\g$.
Note that 
when none of the functionals $f_i\Phi_i$ explicitly depend upon the metric 
(including dependence on $\nabla_a$ or curvature terms), then the term in the 
second line of eq.~\eqref{Rdef} is absent.

\medskip

\paragraph{Remarks:} {\bf (1)} We are not aware of condition
T11b having been proposed previously.

\smallskip
\noindent
{\bf (2)} Condition T11b represents only the ``first order'' part of 
the identity~\eqref{nonpert}, so one might wonder if one would 
get any new requirements by expanding eq.~\eqref{nonpert} to 
higher orders. However, it can be checked by an explicit 
calculation that this is not the case, i.e., 
that all of the higher order relations implicit in eq.~(\ref{nonpert}) 
already follow from the first order condition stated
as T11b. This is not surprising since 
condition T11b is required to hold for metric variations about 
{\em all} (globally hyperbolic) spacetimes. 

\smallskip
\noindent
{\bf (3)} Using eq.~(\ref{1}) we can re-write condition T11b purely in terms
of time-ordered products as 
\bena
\label{Rdeftop}
\delta_\g^{\ret}\left[\T_\g\left( \prod f_i \Phi_i \right) \right]
&=& \Ihaves \T_\g\left(\eps h_{ab} T^{ab} \prod f_i \Phi_i \right) 
-  \Ihaves T^{ab}(\eps h_{ab}) \T_\g\left(\prod f_i \Phi_i \right) \nonumber\\
&+& \sum_i \T_\g\left( f_1 \Phi_1 \cdots h_{ab} \frac{\delta (f_i \Phi_i)}{\delta g_{ab}} \cdots f_n \Phi_n \right). 
\eena

\smallskip
\noindent
{\bf (4)} 
In Euclidean field theory for the action (\ref{s0}) on a complete
Riemannian manifold there will, in general, be a unique Green's function for
the (now elliptic) operator $P$. Hence, there would be no distinction between
retarded and advanced variations, nor between retarded and advanced products.
There also would be a unique, preferred vacuum state, 
$\langle \, \, \, \rangle_0$. The Euclidean version
of condition T11b would be:
\bena
\label{Rdefeuc}
\delta_\g 
\langle \Phi_1(f_1) \cdots \Phi_n(f_n) \rangle_0
&=& \frac{1}{2} 
\langle \Phi_1(f_1) \cdots \Phi_n(f_n) T^{ab}( \eps h_{ab}) 
\rangle_0  \nonumber\\
&+& \sum_i \left\langle \Phi_1(f_1)  \cdots \frac{\delta (f_i \Phi_i)}{\delta g_{ab}}( h_{ab}) \cdots \Phi_n(f_n) \right\rangle_0 . 
\eena
This corresponds to the formula that one would formally obtain by assuming
that the correlation functions $\langle \Phi_1(f_1) \cdots \Phi_n(f_n) 
\rangle_0$
can be defined by a
path integral
\bena
\langle \Phi_1(f_1) \cdots \Phi_n(f_n) \rangle_0 = 
\int [D\varphi] \, \Phi_1(f_1) \cdots \Phi_n(f_n) e^{-S(\varphi, \g)} \, .
\eena
Of course, in curved spacetime, there is no direct relationship between 
the formulations of the Euclidean and Lorentzian versions of
quantum field theory, since a (non-static) Lorentzian spacetime will
not, in general, be a real section of a complex analytic 
manifold with complex analytic metric that also admits a real Riemannian
section. Nevertheless, we may view condition T11b as a mathematically precise
formulation---applicable in the Lorentzian case---of a relation that can be
formally derived from the Euclidean path integral.

\smallskip
\noindent
{\bf (5)} The requirement T11b could also have be formulated in terms of 
advanced variations and advanced products by replacing 
$\delta_\g^{\ret}$ by $\delta_\g^{\adv}$ on the left side of eq.~(\ref{Rdef})
and replacing ``$\R$'' by ``$\A$'' on the right side of that equation, i.e.,
\bena
\label{Adef}
\delta_\g^{\adv}\left[\T_\g\left( \prod f_i \Phi_i \right) \right]
&=& \Ihaves \A_\g\left(\prod f_i \Phi_i; \eps h_{ab} T^{ab}\right) \nonumber\\
&+& \sum_i \T_\g\left( f_1 \Phi_1 \cdots h_{ab} \frac{\delta (f_i \Phi_i)}{\delta g_{ab}} \cdots f_n \Phi_n \right), 
\eena
However, this formulation of condition T11b can be seen to be equivalent to 
eq.~(\ref{Rdef}) as follows. If $\g^{(s)}$ is the one-parameter family of
metrics appearing in eq.~(\ref{Rdef}), then 
\ben
\label{bdef}
\beta_s \equiv \tau^\adv_{\g^{(s)}} \circ (\tau^\ret_{\g^{(s)}})^{-1}, 
\een
is an automorphism of $\W(M, \g)$ 
for all $s$ with the property that $\beta_0 = id$.
It was proven in~\cite{bfv}, that, for all $a \in \W$,
we have
\ben
\label{bdef'}
\frac{\partial}{\partial s}
\beta_s(a) \bigg|_{s=0} = \Ihaves [T^{ab}(\eps h_{ab}), a] \, ,
\een
where $T^{ab}$ is the stress-energy tensor\footnote{
Equation~\eqref{bdef'} holds for any valid prescription for defining 
Wick powers satisfying T1--T10, since the ambiguity in $T_{ab}$ is 
proportional to $\myid$.}.
In particular, for any time-ordered product, we have\footnote{Note that 
the advanced and retarded variations are only defined on 
local covariant field quantities such as the time-ordered products. By 
contrast, eq.~(\ref{bdef'}) holds for an arbitrary element $a \in \W$.}
\ben
\delta^\adv(\T) - \delta^\ret(\T) = \Ihaves [T^{ab}(\eps h_{ab}), \T]. 
\een
Equivalence of eqs.~(\ref{Rdef}) and (\ref{Adef}) then follows 
from this equation and the definitions of advanced and retarded products.

\subsection{External potential variation}

Finally, for completeness, we state the infinitesimal version of condition
T11 for the case of a variation of the external potential:

\paragraph{T11c $\varphi^2$ factor.}
Let $(M,\g)$ be globally hyperbolic and let $V^{(s)}$ be a smooth
one-parameter
family of smooth functions which vary only in a fixed compact set $K$.
Write $V = V^{(0)}$ and write $U = (\partial V^{(s)}/\partial s)|_{s=0}$.
Then we require that for all $f_i \Phi_i \in \F$, 
\bena
\label{VRdef}
\delta_V^{\ret}\left[\T_{(\g,V)} \left( \prod f_i \Phi_i \right) \right]
&=& \frac{\i}{2} 
\R_{(\g,V)} \left(\prod f_i \Phi_i; \eps U \varphi^2\right) \nonumber\\
&+& \sum_i \T_{(\g,V)} \left( f_1 \Phi_1 \cdots U \frac{\delta (f_i \Phi_i)}{\delta V} \cdots f_n \Phi_n \right) \, . 
\eena

\medskip

\paragraph{Remarks:} {\bf (1)} In writing condition T11c, we have 
generalized the definition of $\tau^\ret$ in the obvious way so that
it now maps $\W(M,\g', V', J')$ to $\W(M,\g, V)$. The second term 
on the right side of
eq.~(\ref{VRdef}) is present in this formula
because, as mentioned above, in the construction 
of the theory with an external potential,
elements of $\V$ are allowed to depend explicitly upon $V$.

\smallskip
\noindent
{\bf (2)} For the most part, condition T11c imposes
restrictions on the definition of time-ordered products only if one has
defined the exact theory in an arbitrary external potential. However, even 
if one considers only the theory defined by the action (\ref{s0}) (which 
does not include an external potential), condition T11c 
does impose a restriction on the
definition of time-ordered products in the case where $U$ is constant on
the union of the supports of the $f_i$. 
For simplicity, we shall not consider this or any other consequences
of condition T11c in the remainder of this paper. However, it should be
straightforward to generalize the proof of section 6 to show that condition
T11c can be consistently imposed for quantum field theory in curved
spacetime with an arbitrary external potential.

\smallskip
\noindent
{\bf (3)} As mentioned at the beginning of this section, if we considered a
complex scalar field, then we could also have
a term in $L_1$ of the form 
$\i A^a (\bar{\varphi} \nabla_a \varphi - \varphi \nabla_a \bar{\varphi}) \eps$,
corresponding to the presence of an external electromagnetic field. Our
Principle of Perturbative Agreement would then lead to a corresponding 
additional condition (``T11d'') on time-ordered products. 
We expect that this additional condition can also be
consistently imposed (in addition to the analogs of all of our other 
conditions) for a complex scalar field. We also expect that the
imposition of this condition will imply current conservation for the free
and interacting fields in analogy with Theorems 5.1 and 5.3
below.

\section{Some key consequences of our new requirements} 

In this section, we will derive some important consequences of our new 
requirements for both the free field theory and the interacting field theory.
The demonstration that our requirements can, in fact, be imposed (except in 
two spacetime dimensions) will be given in section 6. 

\subsection{Consequences for the free field}

In this subsection, we will derive some key consequences of condition T11b for
the stress-energy tensor $T_{ab}$ associated with 
the free field Lagrangian $L_0$. 

\begin{thm}
Suppose that the prescription for defining Wick products 
and time ordered products
satisfies conditions T1--T10 together with condition T11b. 
Then the stress tensor $T_{ab}$ defined via that prescription 
is automatically conserved
\ben
\nabla_a T^{ab}(x) = 0. 
\label{freestc}
\een
More generally, we have the following free field 
``Ward identity'' for $T_{ab}$:
\bena
\label{13}
0 &=& \T
\left((\eps \nabla^a \xi^b ) T_{ab} \prod^n f_i \Phi_i \right)\nonumber \\
&+& \i \sum_i \T\left( f_1 \Phi_1 \cdots 
\frac{\delta (f_i \Phi_i)}{\delta \varphi}\pounds_\xi \varphi
\cdots f_n \Phi_n \right).
\eena
\end{thm}

\begin{proof}
We first show that the divergence of the stress tensor is a 
c-number, i.e., proportional 
to the identity operator. Let $F$ be a density of compact support, 
and let $\xi^a$ 
be a compactly supported vector field. Then 
$(\nabla_a T^{ab})(\eps \xi_b) = -(1/2) T_{ab}(\eps \pounds_\xi g^{ab})$, 
and the commutator property, T9, yields
\ben
\label{tabdiv0}
[T^{ab}(\eps \pounds_\xi g_{ab}), \varphi(F)] = \i 
\T\left( (\Delta F)\frac{\delta}{\delta \varphi} \eps (\pounds_\xi g_{ab}) T^{ab} \right). 
\een
We want to show that the right side of this equation is, in fact, 
equal to 0. To see this, 
we write $\eps (\pounds_\xi g_{ab}) T^{ab} = 2 (\pounds_\xi g_{ab}) \,  
\delta L_0/\delta g_{ab}$ and use the fact, proven in appendix B, that
functional derivatives of $L_0$ with respect to $\varphi$ and $\g$ commute
modulo exact forms in the sense of eq.~(\ref{comder}).
We therefore obtain
\bena 
(\Delta F)\frac{\delta}{\delta \varphi} \eps (\pounds_\xi g_{ab}) T^{ab} 
&=& 2 (\pounds_\xi g_{ab}) \frac{\delta}{\delta g_{ab}} \bigg( (\Delta F) 
\frac{\delta L_0}{\delta \varphi} \bigg) + \d B_0 \nonumber \\
&=& 2 (\pounds_\xi g_{ab}) \frac{\delta}{\delta g_{ab}} \bigg( (\Delta F) 
\eps P\varphi \bigg) + \d B_0 \, ,
\eena
where in this equation $\Delta F$ is viewed as being evaluated at the
metric $\g$ about which the variations are being taken (i.e., the
$\delta/\delta g_{ab}$ does not act on $\Delta$). However, we have
\bena
(\pounds_\xi g_{ab}) \frac{\delta}{\delta g_{ab}} \bigg( (\Delta F) 
\eps P\varphi \bigg) &=& 
\pounds_\xi \bigg( (\Delta F) \eps P\varphi \bigg) - 
\bigg( \pounds_\xi (\Delta F) \bigg) \eps P \varphi - (\Delta F) \eps 
P(\pounds_\xi \varphi) \nonumber \\
&=& - \bigg( \pounds_\xi (\Delta F) \bigg) \eps P \varphi - P(\Delta F) \pounds_\xi \varphi
+ \d B_1 \nonumber \\
&=& - \bigg( \pounds_\xi (\Delta F) \bigg) \eps P \varphi + \d B_1 \, ,
\eena
where in the second line we used the facts that $\pounds_\xi$ applied to any
$D$-form is exact and that $P$ is self-adjoint, whereas in the last line
we used the fact that $P(\Delta F) = 0$. 
Using the Leibniz rule, T10, we 
see that the right side of eq.~\eqref{tabdiv0} is equal to 
$-2 \i\varphi[\eps P(\pounds_\xi \Delta F)]$, 
which indeed vanishes
since the quantum field $\varphi$ satisfies the Klein-Gordon
equation. 
Thus, $T^{ab}(\eps \pounds_\xi g_{ab})$ commutes 
with $\varphi(F)$ for all compactly supported $F$. 
By Proposition 2.1 of \cite{hw1}, every 
element of $\W$ with this property has to be proportional to $\myid$. 
Since $\nabla^a T_{ab}$
is locally and covariantly
constructed out of the metric by T1, we must therefore have that
\ben
\label{tabdiv}
T^{ab}(\eps \pounds_\xi g_{ab}) = \int_M C_a \xi^a \eps \cdot \myid
\een 
for some local curvature term $C_a$. Furthermore, by our scaling axiom 
T2, $C_a$ must be  
a polynomial in the Riemann tensor and its covariant derivatives 
of dimension 
${\rm length}^{-D-1}$. We will now show that if condition T11b holds, then $C_a$, 
in fact, has to vanish. 

To prove this, we consider the retarded variation of the local covariant field 
$T^{ab}(\eps \pounds_\xi g_{ab})$ with 
the metric variation taken to be of the ``pure gauge'' form
$h_{ab} = \pounds_\eta g_{ab}$, where $\eta^a$ 
is a compactly supported vector field. Condition T11b in the simple case
of only one factor $f_1 \Phi_1 = \eps (\pounds_\xi g_{ab})T^{ab}$ yields
\ben
\delta^\ret(T^{ab}(\eps \pounds_\xi g_{ab})) = \Ihaves \R(\eps (\pounds_\xi g_{ab}) T^{ab}; \eps (\pounds_\eta g_{cd}) T^{cd})
+ \T \left( (\pounds_\eta g_{cd}) \frac{\delta }{\delta g_{cd}} (\eps (\pounds_\xi g_{ab}) T^{ab}) \right). 
\label{T11bgauge}
\een
By eq.~\eqref{tabdiv}, the left side of this equation is just
\ben
\delta^\ret(T^{ab}(\eps \pounds_\xi g_{ab}) = 
\int_M \xi^a \pounds_\eta (C_a \eps) \cdot \myid = 
- \int_M [\eta, \xi]^a C_a \eps \cdot \myid \, , 
\een
where a partial integration was done in the last step. 
Now subtract from eq.~(\ref{T11bgauge}) the same equation with  
$\eta^a$ and $\xi^a$ interchanged. We obtain
\begin{multline}
-2 \int_M [\eta, \xi]^a C_a \eps \cdot \myid = 
  \Ihaves \R \left( \eps (\pounds_\xi g_{ab})  T^{ab}; \eps (\pounds_\eta g_{cd}) T^{cd} \right)
- \Ihaves \R \left( \eps (\pounds_\eta g_{cd}) T^{cd}; \eps (\pounds_\xi g_{ab})  T^{ab} \right)\\
+ 2\T \left( (\pounds_\eta g_{cd}) \frac{\delta }{\delta g_{cd}} \left\{ (\pounds_\xi g_{ab})  \frac{\delta}{\delta g_{ab}} L_0
\right\}
-           (\pounds_\xi g_{cd})  \frac{\delta }{\delta g_{cd}}  \left\{ (\pounds_\eta g_{ab}) \frac{\delta}{\delta g_{ab}} L_0
\right\}
\right) \, ,
\label{antisym}
\end{multline} 
where we again substituted $\eps T^{ab} = 2 \delta L_0/\delta g_{ab}$.
The terms on the right side can be simplified as follows: 
For the retarded products, we use the identity
$\R(f\Phi, h\Psi) - \R(h\Psi, f\Phi) = [\Psi(h), \Phi(f)]$, 
which holds for any Wick products $\Phi, \Psi$. 
In the case at hand, $\Phi$ and $\Psi$ are equal to the 
divergence of the stress tensor and therefore  
have a vanishing commutator by eq.~\eqref{tabdiv}. Hence, there 
is no contribution from the terms in eq.~(\ref{antisym}) involving
retarded products. The last term on the right side can be 
simplified using the identity
\ben
(D_\xi D_\eta - D_\eta D_\xi)A = D_{[\eta, \xi]}A + \d C
\label{DD}
\een
holding for the ``derivative operator'' 
\ben
D_\xi A = (\pounds_\xi g_{ab}) \frac{\delta A}{\delta g_{ab}}
\een
on classical functionals $A$ of the metric such as $L_0$. 
A proof of this identity is given in appendix B.
Inserting this relation (with $A = L_0$)
and using again eq.~\eqref{tabdiv}, we find that the last term in 
eq.~\eqref{antisym} is just $-\int_M [\eta, \xi]^a C_a \eps \cdot \myid$.
Thus, we obtain
\ben
\int_M C_a [\xi, \eta]^a \eps = 0. 
\een
This equation holds for all smooth compactly supported vector fields 
$\xi^a, \eta^a$. Variation with 
respect to $\xi^a$ yields
\ben
(\nabla_a \eta^b) C_b + (\nabla_b \eta^b) C_a + \eta^b \nabla_b C_a = 0
\een
for all $\eta^a$, at every point in $M$. Now focus on an arbitrary, but
fixed point $x \in M$. At $x$, the quantities $\eta^a$ and 
$K_a{}^b = \nabla_a \eta^b$ can be chosen independently 
to be arbitrary tensors. 
Choosing first $K_a{}^b = 0$ and $\eta^a$ arbitrary, we conclude that
$\nabla_b C_a = 0$ at $x$. Thus, at any $x\in M$ we must have 
\ben
(K_a{}^b + K_c{}^c \delta_a{}^b) C_b = 0
\een
for all $K_a{}^b$, which is possible only when $C_b = 0$. This completes
the proof of stress-tensor conservation, eq.~(\ref{freestc}).

To prove the more general Ward identity, eq.~(\ref{13}), we again consider
condition T11b for the case of a ``pure gauge'' metric variation
$h_{ab} = \pounds_\xi g_{ab} = 2\nabla_{(a} \xi_{b)}$, 
but we now consider arbitrary factors of
$f_i \Phi_i$. We obtain
\bena
\label{7}
\delta^\ret
\left[ \T\left(\prod f_i \Phi_i\right) \right] &=& \i \R
\left(\prod^n f_i \Phi_i; (\eps \nabla^a \xi^b ) T_{ab} \right) \nonumber \\
&+& \sum_i \T\left( f_1\Phi_1 \cdots (\pounds_\xi g_{ab}) \frac{\delta (f_i \Phi_i)}{\delta g_{ab}} 
\cdots f_n\Phi_n \right).
\eena 
But, for our ``pure gauge'' metric variation, we 
have\footnote{Note that the Lie derivative of a tensor field 
$f = f_{a \dots c}{}^{b\dots d}$ is defined by 
$\pounds_\xi f = \frac{\partial}{\partial s} \chi_s^* f$, 
which in turn is equal to  
$-\frac{\partial}{\partial s} \chi_{s \, *} f$, 
since $\chi_s^* = (\chi_s^{-1})_* = \chi_{-s \, *}$.}
\bena
\label{8}
\delta^\ret_\g \left( \T_{\g}\left(\prod f_i \Phi_i\right) \right)
&=& \frac{\partial}{\partial s}
\T_\g ( 
(\chi_{s \, *} f_1) \Phi_1 \cdots (\chi_{s \, *} f_n) \Phi_n ) \bigg|_{s=0} \nonumber\\
&=&-\sum_i \T_\g(f_1\Phi_1 \cdots(\pounds_\xi f_i) \Phi_i 
\cdots f_n\Phi_n), 
\eena
since the time ordered products are local, covariant fields.
Moreover, writing the retarded product on the 
right side of eq.~\eqref{7} in terms of 
time ordered products and using the relation
$T_{ab}(\eps \nabla^a \xi^b) = 0$ which we just proved above, 
we find
\bena\label{9}
\text{right side of eq.~\eqref{7}} &=& \i \T
\left((\eps \nabla^a \xi^b ) T_{ab} \prod^n f_i \Phi_i \right) \nonumber \\
&+& \sum_i \T\left( f_1 \Phi_1 \cdots (\pounds_\xi g_{ab}) \frac{\delta(f_i\Phi_i)}{\delta g_{ab}} 
\cdots f_n \Phi_n \right).
\eena
We now use the fact---proven in appendix B---that for any 
classical field $D$-form $f_i \Phi_i \in \F$,
we have
\ben
(\pounds_\xi f_i) \Phi_i + \frac{\delta (f_i\Phi_i)}{\delta g_{ab}} \pounds_\xi g_{ab}
+ \frac{\delta (f_i \Phi_i)}{\delta \varphi}\pounds_\xi \varphi = \d H
\label{LiefPhi}
\een
for some $(D-1)$-form $H$ that is locally constructed out of $g_{ab}$, 
$\varphi$, $f_i$, and $\xi^a$.
Inserting this relation into eq.~\eqref{8} and~\eqref{9}, and using T10, 
we get the desired relation eq.~(\ref{13}).
\end{proof}

\paragraph{Remarks:} {\bf (1)}
There exist completely reasonable classical field theories 
for which $T_{ab}$ in the quantum field theory cannot
be made divergence free within our axiom scheme.
As we have seen in 
subsection 3.2 above, one example of such a theory is the free
scalar field in $D=2$ spacetime dimensions. 
Hence Theorem 5.1 implies that T11b cannot be 
satisfied in addition to conditions T1-T10 for 
scalar field theory for $D=2$. 

\smallskip
\noindent
{\bf (2)} There appear to be two independent possible obstructions
to the imposition of the analog of condition T11b in a general 
free quantum field theory. 
First, there may exist algebraic identities (i.e., relations that do
not involve the field equations)
satisfied by the classical stress-energy tensor. Within our axiom scheme,
these identities must be respected by the quantum stress-energy tensor, and,
consequently,
the freedom to modify the definition of Wick powers by arbitrary local
curvature terms of the correct dimension does not translate into a similar
ability to modify the definition of $T_{ab}$ so as to make
it conserved, as is necessary for T11b to hold.
As we have seen in 
subsection 3.2 above, this occurs for the free
scalar field in $D=2$ spacetime dimensions.
However,
we will see in the course of our analysis in section
6.2.6 below that, in principle, 
there also can exist a ``cohomological obstruction''
to the imposition of condition T11b. Although such a cohomological 
obstruction does not occur in scalar field theory, it can occur in parity 
violating theories, and it appears to be the cause of the failure of
conservation of $T_{ab}$ for the theories described in \cite{aw}  
in $D=4k+2$ spacetime dimensions. (In these theories, one finds that
$\nabla^a T_{ab} = A_b \myid$, where $A_b$ (the ``gravitational anomaly'')
is a curvature polynomial that does not arise as the divergence of a 
symmetric curvature tensor, i.e., $A_b \neq \nabla^a A_{ab}$ for any 
symmetric $A_{ab}$.) Thus, the analog of condition T11b also cannot 
be satisfied in 
theories analyzed in \cite{aw}, but the root cause of the failure of condition
T11b for these theories appears to be different in nature from the root cause
of the failure of condition T11b for a scalar field in $D=2$ dimensions.

\smallskip
\noindent
{\bf (3)} Note that our Ward identity, eq.~(\ref{13}), is a
relation between elements in 
the free field algebra $\W(M, \g)$, rather than an relation 
between correlation 
functions, which is the more conventional way to express Ward identities.

\medskip

A similar type of argument to that used in the first part of
the above proof can be 
applied in the context of 
a conformally coupled massless field to yield the following nontrivial 
consistency (or ``cocycle'') relation for the trace of the 
stress-energy tensor:
\begin{thm}
Suppose that conditions T1--T10 and T11b are satisfied.
Then, for the case of a massless, conformally coupled scalar field 
[i.e., $m=0, \xi = (D-2)/4(D-1)$], 
we have $T^a{}_a(x) = C(x)\myid$, where $C(x)$ is 
a local curvature term of mass dimension $D$ that satisfies the ``cocycle 
condition''
\ben
\label{cocycle}
\int_M [k(\delta_f C) - f(\delta_k C)] \eps = 0, 
\een
for any smooth compactly supported functions $f, k$ on $M$, where 
\ben
\delta_f C(\g) = \frac{\partial}{\partial s} C(e^{sf} \g) \bigg|_{s=0}
\een
denotes the infinitesimal variation of a curvature term under a change in the conformal factor.
\end{thm}
\begin{proof}
That the trace of the quantum stress tensor in the
massless, conformally coupled case is proportional 
to the identity, $T^a{}_a = C\myid$,
follows,
as in the proof of the previous theorem, from the fact that
$[T^a{}_a(\eps f), \varphi(F)] = 0$, which is an immediate consequence
of the commutator property, T9, and the field equation for $\varphi$. 
By the scaling property T2, $C$ must be a curvature polynomial 
of dimension ${\rm length}^{-D}$. To show that T11b implies that
this curvature polynomial satisfies 
the cocycle condition eq.~\eqref{cocycle}, we consider the retarded 
variation of the field $T^a{}_a(\eps f)$ with 
respect to the metric perturbation 
$h_{ab} = k g_{ab}$, where $k$ is some smooth, 
compactly supported function (i.e., we 
consider an infinitesimal change $k$ in the conformal factor 
of the metric $g_{ab}$). Condition T11b in the simple case
where only the single factor $f_1 \Phi_1 = \eps f T^a{}_a$ 
is present now yields
\ben
\delta^\ret(T^a{}_a(\eps f)) = \Ihaves \R(\eps fT^a{}_a; \eps kT^b{}_b)
+ \T \left( (kg_{cd}) \frac{\delta }{\delta g_{cd}} (\eps f T^a{}_a) \right). 
\een
For the left side we use the fact
that $T^a{}_a(\eps f) = \int fC \eps \cdot \myid$, so
the retarded variation with respect to $h_{ab} = kg_{ab}$ 
yields $\int f \delta_k(C \eps) \cdot \myid$.
Now antisymmetrize the above equation 
in $k$ and $f$. We obtain
\begin{multline}
\int_M [k(\delta_f C) - f(\delta_k C)] \eps \cdot \myid = 
  \Ihaves \R \left( \eps f T^a{}_a; \eps k T^b{}_b \right)
- \Ihaves \R \left( \eps k T^b{}_b; \eps f T^a{}_a \right)\\
+ 2\T \left( (f g_{cd}) \frac{\delta }{\delta g_{cd}} \left\{ (kg_{ab})  \frac{\delta}{\delta g_{ab}} L_0 \right\}
-           (k g_{ab})  \frac{\delta }{\delta g_{ab}} \left\{ (fg_{cd})  \frac{\delta}{\delta g_{cd}} L_0 \right\}
\right). 
\end{multline}
As in the proof of Theorem 5.1, the two retarded products on the right
side combine to yield
the commutator $(\i/2) [T^a{}_a(f\eps), T^b{}_b(k\eps)]$, which 
in turn vanishes because the trace of the 
stress tensor is proportional to $\myid$ (note that the retarded products 
individually might be non-vanishing). 
The last term on the right side also vanishes since taking 
the variation of $L_0$ with respect to the conformal factors 
$f$ and $k$ clearly does not depend on the order in which 
they are taken. Consequently, the right side vanishes, and we obtain 
the desired cocycle property~\eqref{cocycle} for the 
trace of the stress tensor, $C$.
\end{proof}

\paragraph{Remarks:} {\bf (1)} The cocycle condition on the conformal
anomaly that we have derived
here from axioms T1-T10 and T11b is the same condition as would be 
formally derived by assuming that the (expectation value of the) quantum
stress-energy tensor can be calculated by taking the
variation of some ``effective action'' with respect to the metric.
Conditions of this nature are known in the literature under the name
``Wess-Zumino consistency conditions'' \cite{wz}.

\smallskip
\noindent
{\bf (2)} The method of proof of the above theorem 
only relies upon 
properties T1--T10 and T11b, and therefore can 
be generalized to arbitrary field theories 
that satisfy suitable
analogs of these conditions, and whose stress energy tensor has 
a c-number trace. We shall consider such ``non-perturbative'' 
results elsewhere. 

\smallskip
\noindent
{\bf (3)}  In odd spacetime dimensions, there simply are no scalar 
polynomials in the curvature of dimension ${\rm length}^{-D}$, 
so there is no trace anomaly.
By contrast, in even dimensions, there always exist
curvature scalars $C$ of dimension 
${\rm length}^{-D}$ satisfying the cocycle condition.
For example, any term scaling as $C \to \Omega^D C$ under a conformal 
transformation
$g_{ab} \to \Omega^2 g_{ab}$ is a solution to the cocycle 
condition~\eqref{cocycle}, 
so when $D=2k$ with $k>1$, any monomial expression 
in the Weyl tensor which contains $k$ factors of the Weyl tensor
is a solution.
In $D=4$ spacetime dimensions, there are 3 linearly independent local, covariant
scalars with dimension $({\rm length})^{-4}$ which solve the cocycle condition\footnote{
If our axioms were weakened so as
to require the locality and covariance condition T1 only for orientation preserving isometries,
then the ``parity violating'' curvature term 
$C_4 = \epsilon_{abcd} R^{ab}{}_{pq} R^{pqcd}$ would be allowed and
would also satisfy the cocycle condition.},
namely, $C_1 = C_{abcd} C^{abcd}$,
$C_2 = R_{abcd} R^{abcd} - 4 R_{ab} R^{ab} + R^2$ (the Euler density), 
and $C_3 = \nabla^a \nabla_a R$. Since there are 4
linearly independent local covariant curvature terms with 
that dimension (namely, $R^2$ in addition to the above 3 terms), 
this shows explicitly 
that the cocycle condition is non-trivial, and thus is potentially useful
for restricting the form of the trace anomaly.
Of course, the existence of solutions to the cocycle condition does not 
automatically imply that there is actually a trace anomaly, as the 
coefficients of these terms might be zero.
However, for the massless, 
conformally coupled scalar field in $D=4$ dimensions, the arguments given
in subsection 3.2 can be used to show that the coeficients of $C_1$ and
$C_2$ must be nonvanishing for any prescription satisfying
axioms T1-T10 and T11b. (The coeficient of $C_3$ can be set to zero using
the renormalization freedom allowed by these axioms.)
By the same type of arguments, it can presumably be 
established that $C$ cannot be zero in any even dimension $D>2$, 
although the calculations
that must be carried out to show this rapidly become very 
complicated as the number of dimensions
$D$ of the spacetime increases. 
All solutions to the cocycle condition in $D=6$ have 
been found in ref.~\cite{bpb}. 
We are not aware of an efficient algorithm to determine the 
general solution to the 
cocycle condition in arbitrary dimensions, and this appears to be an 
interesting mathematical problem.

\subsection{Consequences for interacting fields}

In this subsection, we will derive some important consequences of our 
requirements with regard to 
perturbatively defined interacting field theories. 
(Other consequences such as the existence of the renormalization 
group are derived in~\cite{hw3}.) We will consider interacting theories 
described by a classical Lagrangian of the form
\ben
L = L_0 + L_1
\een
where $L_0$ is given by eq.~(\ref{s0}) and where the interaction Lagrangian
is of the form
\ben
L_1 = \frac{1}{2} \sum \kappa_i \Phi_i.
\label{interL}
\een
The $\kappa_i$ denote
coupling parameters and each
$\Phi_i \in \V$ is
any polynomial in the field $\varphi$ and its derivatives as well
as the Riemann tensor and its derivatives.
In particular, we do not require that $L_1$ be renormalizable. 
Associated with the Lagrangian $L$ is the 
classical stress-energy tensor $\Theta_{ab}$ given by
\ben
\label{Thetadef}
\Theta^{ab} = 2\eps^{-1} \frac{\delta L}{\delta g_{ab}} = T^{ab} +  2\eps^{-1}
\frac{\delta L_1}{\delta g_{ab}}, 
\een
where $T_{ab}$ is the stress tensor~\eqref{tabdef} associated 
with the free Lagrangian $L_0$.

As reviewed in subsection 4.1 above, if $\theta$ is a smooth function of
compact support,
interacting fields in the quantum field theory associated with the
interaction Lagrangian $\theta L_1$ 
can be defined in perturbation theory
in terms of time-ordered products of the free theory by the Bogoliubov 
formula, eq.~(\ref{pertseries}). As shown in~\cite{bf} and in section 3.1
of~\cite{hw3}, one may always then take the limit as $\theta \rightarrow 1$
in a suitable way so as to (perturbatively) define the interacting theory with 
interaction Lagrangian $L_1$. (No restrictions on the asymptotic properties
of the spacetime $(M,\g)$ are needed in order to take this limit.) 
As explained in \cite{hw3}, the 
resulting interacting fields---denoted $\Phi_{L_1}(x)$---and interacting
time-ordered products---denoted $\T_{L_1} (\prod f_i \Phi_i)$---live (after 
smearing with a smooth compactly supported test functions) 
in an suitable abstract algebra $\B_{L_1}(M, \g)$ of formal power series
of elements of $\W(M, \g)$. 

The classical stress tensor $\Theta^{ab}$ is conserved when the 
classical field equations associated
with the Lagrangian, $L$, hold for $\varphi$. 
However, it is a priori far from clear that the quantized
interacting field operator $\varphi_{L_1}$ satisfies the 
interacting field equations, and, even if it does, it is far from
clear that the interacting
stress-energy operator $\Theta_{L_1}^{ab}$ is conserved. 
The following theorem---which constitutes one of the main results of this
paper---establishes that if axioms T1-T10 hold, then condition T11a 
guarantees that $\varphi_{L_1}$ satisfies the 
interacting field equations, and condition T11b guarantees that the interacting
stress-energy operator $\Theta_{L_1}^{ab}$ is conserved.

\begin{thm}
\label{mainthm} Suppose that the prescription for defining time-ordered
products in the free theory with Lagrangian $L_0$ satisfies axioms T1-T10.
Let $L_1$ be any interaction Lagrangian of the form eq.~(\ref{interL}).
Then the following properties hold for the interacting theory:

\begin{enumerate}

\item Let $B$ be a $(D-1)$-form on $M$ depending 
polynomially on the classical field $\varphi$ and its 
derivatives and the Riemann tensor and its derivatives. Then the map 
$\Phi_{L_1 + \d B}(f) \to \Phi_{L_1}(f)$ 
defines an isomorphism
\ben
\B_{L_1 + \d B}(M, \g) \cong \B_{L_1}(M, \g),
\een
i.e., the theory is unchanged if a total divergence is 
added to the Lagrangian. 

\item
The Leibniz rule holds for the interacting fields in the sense that
\ben
\nabla_a [\Phi_{L_1}(x)] = (\nabla_a \Phi)_{L_1}(x), 
\een
where the expression $(\nabla_a \Phi)$ on the right denotes the 
field expression obtained by applying the Leibniz rule. More generally, the Leibniz rule 
also holds for the interacting time-ordered products.

\item
If, in addition to T1-T10, axiom T11a also holds, then 
the equations of motion are satisfied in the interacting theory, i.e.,
\ben
\label{eom'}
(\nabla^a \nabla_a - m^2 - \xi R)\varphi_{L_1}(x)  = -(\delta L_1/\delta \varphi)_{L_1}(x) 
\een
in the sense of distributions valued in $\B_{L_1}(M, \g)$. 
Here, the variation on the right
is the usual Euler-Lagrange type variation defined in eq.~\eqref{30} above.

\item
If, in addition to T1-T10, axiom T11b also holds,
then the interacting stress-energy tensor is conserved
\ben
\label{cons}
\nabla_a \Theta_{L_1}^{ab}(x) = 0. 
\een
More generally, for all $J_i \Psi_i \in \F$, and all vector 
fields $\xi^a$ of compact support, the 
following interacting field Ward identity holds:
\bena
\label{interward}
0 &=& \T_{L_1}
\left((\eps \nabla^a \xi^b ) \Theta_{ab} \prod^n J_i \Psi_i \right)\nonumber \\
&+& \i \sum_i \T_{L_1}\left( J_1 \Psi_1 \cdots 
\frac{\delta (J_i \Psi_i)}{\delta \varphi}\pounds_\xi \varphi
\cdots J_n \Psi_n \right) \, .
\eena
\end{enumerate}
\end{thm}

\begin{proof}
{}From the construction of the interacting fields given in section 3.1
of \cite{hw3}, it is clear that it suffices to prove
the statements in the theorem for a cutoff interaction 
$\theta L_1$, where $\theta$ is a smooth function of compact support which 
is equal to 1 in the spacetime region under consideration. 
Statements (1) and (2) of the theorem
are seen to be an immediate consequence of the Leibniz rule 
T10 applied to the individual 
terms in the formula
\ben
\Phi_{\theta L_1}(f) \equiv \sum_{n \ge 0} \frac{\i^n}{n!} \R\bigg(f\Phi; (\theta L_1)^n \bigg), 
\een
where $\theta = 1$ on the support of $f$.

In order to prove statement (3), we must show that for any smooth function
$f$ of compact support, we have
\ben
\varphi_{\theta L_1}(\eps (\nabla^a \nabla_a - m^2 - \xi R)f)  = -(\delta L_1/\delta \varphi)_{\theta L_1}(f),  
\een
where, again, $\theta = 1$ on the support of $f$.
In terms of retarded products, we need to show that 
\ben
\label{eom}
0=
\R\left(\eps f(\nabla^a \nabla_a - m^2 - \xi R)\varphi; e^{\i \theta L_1} \right) + \R\left(f 
\frac{\delta L_1}{\delta \varphi}; e^{\i \theta L_1} \right), 
\een
However, using the definition of the retarded products in terms of 
time-ordered products 
[see eq.~\eqref{ARdef}], and using the fact that $\theta \equiv 1$ 
on the support of $f$,
we see that eq.~\eqref{eom} is equivalent to 
\ben
\T\left(\eps f(\nabla^a \nabla_a - m^2 - \xi R)\varphi \prod^n \theta L_1 \right) 
= \i n \T\left(f \frac{\delta (\theta L_1)}{\delta \varphi} \prod^{n-1} \theta L_1 \right) 
\een
for all natural numbers $n$.
But this equation is equivalent to \eqref{6} above, which was previously 
shown to hold as a direct consequence of condition T11a. Thus, we have 
succeeded in showing that the equations of motion~\eqref{eom'} 
hold in the interacting theory.

To prove eq.~(\ref{cons}) of statement (4), we must show that for any smooth
vector field $\xi^a$ of compact support, we have
\ben
\left(\Theta_{ab}\right)_{\theta L_1}(\eps \nabla^a \xi^b) = 0. 
\een
where $\theta = 1$ on the support of $\xi^a$.
In terms of retarded products, we need to show that 
\ben\label{12}
0 = \R\left((\eps \nabla^a \xi^b) T_{ab}; e^{\i \theta L_1}\right) + 
2\R\left((\nabla^a \xi^b) \frac{\delta L_1}{\delta g^{ab}};e^{\i \theta L_1} 
\right).
\een
Equation~\eqref{12} is seen to be equivalent to 
\bena
\label{10}
\T\left((\eps \nabla^a \xi^b) T_{ab} \prod^n \theta L_1 \right) &=& 
\i n\T\left((\pounds_\xi g_{ab}) 
\frac{\delta (\theta L_1)}{\delta g^{ab}} \prod^{n-1} \theta L_1 \right)   
\eena
for all natural numbers $n$.
Now, if we apply eq.~(\ref{LiefPhi}) to the case $f_i = \theta$,
$\Phi_i =  L_1$
and use the fact that $\theta = 1$ on the support of $\xi^a$, we obtain
\ben
\label{lele}
\frac{\delta (\theta L_1)}{\delta g_{ab}} \pounds_\xi g_{ab}+ 
\frac{\delta (\theta L_1)}{\delta \varphi} \pounds_\xi \varphi = \d B. 
\een
Therefore, by the Leibniz rule T10, we can rewrite eq.~\eqref{10} 
in the equivalent form
\bena
\label{11}
\T\left( (\eps \nabla^a \xi^b)  T_{ab} \prod^n \theta L_1 \right) &=& 
-\i n\T\left(
\frac{\delta (\theta L_1)}{\delta \varphi} \pounds_\xi \varphi
\prod^{n-1} \theta L_1 \right) .  
\eena
But this equation holds as a consequence of the free field Ward identity
eq.~(\ref{13}), which was proven to hold when condition T11b is satisfied.
Thus, we have shown that eq.~(\ref{cons}) holds, i.e., the interacting
stress-energy is conserved in the interacting theory.

To prove the interacting Ward identity, eq.~(\ref{interward}), we will
need the generalization of eq.~(\ref{13}) for the case where $L_1$ may
also depend upon an external source $J_{ab \dots c}$, 
\ben
L_1 = L_1(g_{ab}, (\nabla)^k \varphi, J_{ab \dots c}). 
\een
We assume the source to be a smooth compactly supported tensor field on 
$M$ although certain distributional 
sources would also be admissible\footnote{It can be shown as a consequence
of the microlocal spectrum condition that any distributional
source with spacelike $\WF(J)$ would be admissible, i.e., lead to 
well defined interacting field expressions. For example $J$ given by 
the delta distribution supported on a timelike smooth submanifold $S$, 
$\delta_S(f) = \int_S f n \cdot \eps$, (with $n^a$ the normal to $S$) is 
acceptable.}. 
The generalization of the conservation law eq.~(\ref{cons}) appropriate to 
this case is\footnote{The factor of 2 in front of the second term arises
because $\Theta_{ab}$ is twice the metric variation.} 
\ben
\label{cons'}
\Theta_{L_1}^{ab}(\pounds_\xi g_{ab}) + 
2 \left(\frac{\delta L_1}{\delta J_{ab \dots c}} \right)_{L_1}(\pounds_\xi J_{ab \dots c}) = 0 
\een
for any compactly supported test vector field $\xi^a$. In unsmeared form, this
equation can be written as
\begin{multline}
\nabla_q \Bigg[ 
\Theta^{qr}_{L_1}(x) + \left( J^r{}_{ab \dots c} \frac{\delta L_1}{\delta J_{qab \dots c}} \right)_{L_1}(x) \\
+ \left(J_a{}^r{}_{b \dots c} \frac{\delta L_1}{\delta J_{aqb \dots c}}\right)_{L_1} (x)
+ \dots 
+ \left( J_{ab \dots c}{}^r \frac{\delta L_1}{\delta J_{ab \dots cq}}\right)_{L_1} (x) \Bigg] \\
= \left( (\nabla^r J_{ab \dots c}) \frac{\delta L_1}{\delta J_{ab \dots c}}\right)_{L_1}(x) .
\end{multline}
A similar formula holds when there is any 
finite number of external sources.

Now consider, for a given $\Phi$, the $m$-parameter family of interaction
Lagrangians $K_1 = L_1 + \sum \lambda_j J_j \Psi_j$. 
We differentiate the interacting field $\Phi_{K_1}$
with respect to the parameters $\lambda_i$---identifying 
at the same time $\Phi_{K_1}$ with an
element in $\B_{L_1}(M, \g)$ via a suitably defined isomorphism $\tau^{\ret}:
\B_{K_1}(M, \g) \to \B_{L_1}(M, \g)$ associated with the respective 
algebras of interacting fields. We thereby obtain the retarded 
products\footnote{In a similar way, we could write the advanced 
products in the interacting field theory considering instead
the corresponding *-isomorphism 
$\tau^{\adv}:\B_{K_1}(M, \g) \to \B_{L_1}(M, \g)$, but this would make 
no difference in the argument.} in the 
interacting field theory associated with $L_1$, 
\ben
\frac{\partial^m}{\partial \lambda_1 \dots \partial \lambda_m}
\tau^{\ret} \left[ 
\Phi_{L_1 + \sum \lambda_j J_j \Psi_j}(F) 
\right] \bigg|_{\lambda_i = 0}= 
\R_{L_1} \left(F\Phi; \prod_j J_j \Psi_j \right). 
\label{intretprod}
\een
Now consider the special 
case in which $\Phi$ is the stress energy tensor associated with 
the Lagrangian density $L_0 + K_1$, 
i.e., we choose
\ben
\Phi^{ab} = \Theta^{ab} + 
2\eps^{-1}  \sum \lambda_j \frac{\delta(J_j \Psi_j)}{\delta g_{ab}} ,
\een
where $\Theta_{ab}$ is the stress-energy tensor eq.~\eqref{Thetadef} 
associated with $L_0 + L_1$. We also choose the ``$F$'' in 
eq.~(\ref{intretprod}) to be
$F_{ab} = \eps \nabla_a \xi_b$, where $\xi^a$ is a smooth,
compactly supported vector field. 
We use formula~\eqref{cons'} (with $L_1$ replaced by $K_1$) to 
calculate $\Phi^{ab}_{K_1}(F_{ab})$, and differentiate the resulting identity
with respect to the parameters $\lambda_i$. This gives
\begin{multline}
\label{robodef}
\i \R_{L_1} \left(\eps (\nabla_a \xi_b)\Theta^{ab}; \prod_j J_j \Psi_j \right)
= \\
\sum_j \R_{L_1} \left(\frac{\delta(J_j \Psi_j)}{\delta g_{ab}} \pounds_\xi g_{ab}; \prod_{i \neq j} J_i \Psi_i \right)
+ \sum_j \R_{L_1} \left( \Psi_j \pounds_\xi J_j; \prod_{i \neq j} J_i \Psi_i \right). 
\end{multline}
We now again apply eq.~(\ref{LiefPhi}), this time with 
$f_i \Phi_i = J_j \Psi_j$, to obtain
\ben
\frac{\delta(J_j \Psi_j)}{\delta g_{ab}} \pounds_\xi g_{ab} + \Psi_j \pounds_\xi J_j + \frac{\delta(J_j \Psi_j)}{\delta \varphi} 
\pounds_\xi \varphi = \d B, 
\een
and we apply the Leibniz rule to the retarded products in~\eqref{robodef}
(which also holds for the interacting quantities, since these are expressible 
in terms of the time ordered products in the free theory). Finally,
we express the retarded products $\R_{L_1}$ in 
the interacting theory in terms of time ordered products $\T_{L_1}$ 
in the interacting theory (using a formula completely 
analogous to eq.~\eqref{Rdef}). When this is done, we
arrive at the Ward identity, eq.~(\ref{interward}), for the 
interacting field theory associated with the interaction Lagrangian $L_1$.
\end{proof}

\paragraph{Remarks} {\bf (1)} We note explicitly that
Theorem 5.3 does {\em not} say 
that an interacting field $\Phi_{L_1}$ vanishes when the 
classical field expression 
$\Phi$ is of the form 
$\Phi = \Psi \frac{\delta L}{\delta \varphi}$, with $\Psi$ containing
factors of $\varphi$. 
In other words, the theorem does not say that a general 
interacting field $\Phi_{L_1}$ vanishes if it 
would vanish in the classical interacting theory associated with 
$L$ by the classical equations 
of motion. Rather, the theorem asserts only that this is
true in the special cases 
$\Phi = \frac{\delta L}{\delta \varphi}$ 
and $\Phi_b = \nabla^a \Theta_{ab}$. Indeed, we have already seen in
subsection 3.2 above that even in the free theory, field expressions
of the form $\Psi \frac{\delta L_0}{\delta \varphi}$ will, in general
be nonvanishing.

\smallskip
\noindent
{\bf (2)} Note that as in the case of the free theory, the interacting 
Ward identity, eq.~(\ref{interward}), is a relation between elements in 
the algebra $\B_{L_1}(M, \g)$ of interacting fields, 
rather than a relation between correlation 
functions associated with a state. Note also that the interacting
Ward identity has
the same form as the Ward identity~\eqref{13} in the 
free quantum field theory, except that  
the free stress-energy tensor $T_{ab}$ is 
replaced by the interacting stress-energy
tensor $\Theta_{ab}$. Note, however, that the Ward identity in the free theory
is an operator identity between elements in the algebra $\W(M, \g)$, 
whereas the 
interacting Ward identity is an identity in the 
interacting field algebra $\B_{L_1}(M, \g)$.

\smallskip
\noindent

{\bf (3)} In our informal distribution notation~\eqref{infint} for the 
time-ordered products, the Ward
identity~\eqref{interward} takes the form
\begin{multline}
\label{interward'}
\nabla^y_a \T_{L_1}
\left(\Theta^{ab}(y) \prod^n \Psi_i(x_i) \right)  \\
= \i \sum_i \delta(y, x_i) \, \T_{L_1}\left( \Psi_1(x_1) \cdots 
\left( (\nabla^b \varphi) \frac{\delta}{\delta \varphi} \Psi_i \right)(x_i) 
\cdots \Psi_n(x_n) \right).
\end{multline}

\section[Proof that there exists a prescription for time-ordered 
products \\
satisfying T11a and T11b in addition to T1-T10]{Proof that there exists a prescription for time-ordered 
products satisfying T11a and T11b in addition to T1-T10}

Our remaining task is to prove that requirements T11a and---in $D>2$ 
dimensions---T11b can be consistently imposed in addition to
requirements T1-T10. Specifically, we shall prove the following:

\begin{thm} In all spacetime dimensions $D>2$, there exists a 
prescription for defining time-ordered products of the quantum
scalar field with Lagrangian $L_0$, eq.~(\ref{s0}), that satisfies 
conditions T1-T10, T11a, and T11b. When $D=2$, there exists a prescription
satisfying T1-T10 and T11a, but condition T11b cannot be imposed in
addition to T1-T10.
\end{thm}

We have already proven that condition T11b cannot be satisfied in addition
to T1-T10 in $D=2$ spacetime dimensions (see remark (1) following Theorem
5.1 above), so we need only prove the existence statements here.

We have already proved in proposition 3.1 above that T1--T10
always can be satisfied. Our strategy will therefore be to use
the remaining ``renormalization freedom'' to additionally satisfy
T11a and T11b. This remaining renormalization freedom may be 
precisely characterized as follows:
In our previous work~\cite{hw1} (see also~\cite{hw3})
we proved a uniqueness theorem for 
time-ordered products satisfying T1--T9
whose factors do not contain derivatives of the fields. 
This result can be
straightforwardly generalized to the case when derivatives are present
and the prescription also satisfies T10. The generalized result is as 
follows: Let $\T$ and $\T'$ be arbitrary prescriptions for defining 
time-ordered products satisfying T1--T10. 
Then they must be related in the following way:
\begin{multline}
\label{unique1}
\T'\left( \prod_{i=1}^n A_i \right) = 
\T\left( \prod_{i=1}^n A_i \right) + 
\sum_{I_0 \cup I_1 \cup \dots \cup I_k = \{1, \dots, n\}}
\T \left(
\prod_{k>0} \O_{|I_k|}\left(\bigotimes_{j \in I_k} A_j \right)
\prod_{i \in I_0} A_i 
\right). 
\end{multline}
Here the $\O_r$ are linear maps (essentially the ``counterterms'', see eq.~\eqref{countert} below) 
$\O_r: \otimes^r \F \to \F$ that can be written in the following form:
\begin{multline}
\label{unique2}
\O_r(\otimes f_i \Phi_i)(x) = 
\sum_{\alpha_1, \alpha_2, \dots} \frac{1}{\alpha_1! \dots \alpha_{r}!} 
\int \prod_j \eps(y_j)\\
c\left[\delta^{\alpha_1} \Phi_1 \otimes \dots \otimes \delta^{\alpha_{r}} \Phi_{r}
\right]
(x; y_1, \dots, y_r) \\
f_1(y_1) \dots f_{N_T}(y_r) 
\prod_{i=1}^{r} \prod_{j} [(\nabla)^{j}\varphi(y_i)]^{\alpha_{ij}}, 
\end{multline}
where we are using the same notation as in the Wick expansion~\eqref{lwe}. 
The $c$ are linear maps on $\otimes^r \V$ taking values in the distributions
over $M^{r+1}$. These distributions are always writable as 
a sum of derivatives of the delta function $\delta(x; y_1, \dots, y_r)$, times
polynomials in the Riemann tensor and its covariant derivatives and $m^2$. The 
engineering dimension of each such term appearing in $c[\otimes_i \Phi_i]$
(with the dimension of the delta function counted as $rD$) must 
be equal precisely to the sum of the engineering dimensions of the $\Phi_i$, 
defined as in the scaling requirement, T2. The $c$ must satisfy the reality condition
\ben
\label{ucond}
\overline{c\left[
\otimes_{i=1}^n \Phi_i  
\right]}
= (-1)^{n+1}
c\left[
\otimes_{i=1}^n \Phi_i  
\right]
\een
as a consequence of the unitarity property satisfied by $\T$ and $\T'$, and 
they must satisfy the symmetry condition
\ben
\label{scond}
c\left[
\otimes_{i=1}^n \Phi_i  
\right](y; x_1, \dots, x_n) = c[\otimes_{i=1}^n \Phi_{\pi i}](y; x_{\pi 1}, \dots, x_{\pi n}) 
\quad \text{$\forall$ permutations $\pi$}, 
\een
as a consequence of the symmetry of the time ordered products.
Finally, the imposition of the Leibniz
rule, T10, on the time ordered products $\T$ and $\T'$ yields the 
following additional constraint on the $c$:
\ben
\label{leib3}
c\left[
\Phi_1 \otimes \dots \nabla \Phi_i \otimes \dots \Phi_n
\right]
= 
(1 \otimes \cdots \underbrace{\nabla}_{\text{$i$-th slot}} \otimes \dots 1) c\left[
\otimes_i \Phi_i
\right].
\een

Formula eq.~\eqref{unique1} can be restated more compactly using the 
generating functional ${\mathcal S}(A)$ for the time ordered products
defined in eq.~\eqref{S}: 
\ben
\label{countert}
{\mathcal S}'(A) = {\mathcal S}(A + \frac{1}{\i}{\mathcal O} (e^{\i A})), 
\een
where 
\ben
{\mathcal O}(e^{\i A}) = \sum_{n \ge 0} \frac{\i^{n}}{n!} {\mathcal O}_n
\left( \bigotimes^n A \right) 
\een
is a formal power series in $\F$. In other words,  
if $L_1 = A$ is the interaction Lagrangian, then $L_2 \equiv (1/\i) {\mathcal O} (e^{\i L_1})$ 
corresponds precisely to the (finite) counterterms that must be added to $L_1$
in order to compensate for the change in the renormalization prescription
from $\T$ to $\T'$.

Our task is to show that T11a and T11b can be satisfied 
by making changes within the allowed
class of changes that we have just characterized in terms of the $c$.

\subsection{Proof that T11a can be satisfied}

It is not difficult to prove that T11a can always be 
satisfied in any dimension $D$, including $D=2$. In fact, 
T11a automatically holds for the Wick powers 
(i.e., time-ordered products with one factor) when the latter are
defined via the local normal ordering prescription given 
in eq.~\eqref{phih}. To show that T1--T10 together with T11a can be 
satisfied for arbitrary time-ordered products, 
we proceed inductively in the number of powers of $\varphi$ as follows. 
We assume that we are
given a prescription which satisfies T1-T10 for 
arbitrary time-ordered products, 
and we assume, inductively, that T11a also holds for all time-ordered products
$\T(f_1\Phi_1 \dots f_n \Phi_n)$ that
contain a total number $N_\varphi < k$ powers of $\varphi$. 
{}From the identity
$\R(\eps J_1 \varphi; \eps J_2 \varphi) = \i \Delta^\ret(\eps J_1, \eps J_2)$, 
we easily see that T11a 
is satisfied when $N_\varphi = 1$, which case occurs only when $n=1$ and 
$\Phi_1$ is linear in $\varphi$. 

Consider now a set of fields $\Phi_1, \dots, \Phi_n$
with $N_\varphi = k$, and let $G_n(J; f_1, \dots, f_n)$ be the 
difference between the left and right sides of 
T11a (see eq.~(\ref{5})). We wish to show that it is possible to change 
our prescription, if necessary, so that $G_n' = 0$ for 
the new prescription $\T'$, while maintaining T1--T10 on all time-ordered
products and maintaining T11a on the time-ordered products with $N_\varphi < k$. 
It can easily be seen, from the causal 
factorization property and the definition of the retarded products, that $G_n(J; f_1, \dots, f_n)=0$ for test functions
$J, f_1, \dots, f_n$ supported off the total diagonal $\Delta_{n+1}$ in the product manifold $M^{n+1}$. 
Furthermore, using the inductive assumption and T9, one can verify by an explicit calculation that the commutator 
$[G_n(J; f_1, \dots, f_n), \varphi(F)]$ vanishes for any compactly supported $F$. Thus, by prop.~2.1 of~\cite{hw1}, 
$G_n$ must be proportional to the identity operator and can therefore be identified with a multilinear functional taking valuese
in the complex numbers. By conditions T1-T5, 
this functional must actually be a distribution (i.e., 
it must be continuous in the appropriate sense)
which is local and covariantly constructed from the metric, with a 
smooth/analytic dependence upon the metric and $m^2,\xi$, 
and with an almost homogeneous scaling behavior. Therefore, by 
the arguments in~\cite{hw1}, $G_n$ has to be 
a sum of covariant derivatives of the delta-distribution on $M^{n+1}$, 
multiplied by polynomials in $m^2$, covariant 
derivatives of the Riemann tensor, and analytic functions of $\xi$, 
of the appropriate dimension. 
It follows from unitarity T7 that $G_n$ satisfies the reality condition $\bar G_n = (-1)^{n+1} G_n$. 

We now set $c[\varphi \otimes (\otimes_i \Phi_i)] = -\i G_n$, 
and define $c[(\nabla)^k \varphi \otimes 
(\otimes_i \Phi_i)]$ via eq.~\eqref{leib3}. 
We use these $c$
to define a new prescription $\T'$ via eqs.~\eqref{unique1} 
and~\eqref{unique2}. It is clear that the new prescription
satisfies $G_n' = 0$ and hence satisfies T11a for $N_\varphi = k$.
Thus, our inductive proof will be complete if we can show that the
$c$ satisfy all of the properties that are necessary for the new 
prescription to satisfy T1-T10 on all time-ordered products.
However, it is clear from its definition that $c$ satisfies
all of these properties, with the possible exception of the 
symmetry property~\eqref{scond}.
We now complete the proof by showing 
that $c$ also satisfies this symmetry property. The symmetry property
of $c[\varphi \otimes (\otimes_i \Phi_i)]$
holds trivially except in the case 
where we have a 
factor, say $\Phi_1$, of the form $\Phi_1 = \varphi$ and we consider
the interchange of $\Phi_1$ and $\varphi$. Thus, let us consider the difference
between the 
left and right sides of eq.~(\ref{5}) with free field factor 
$\eps J_2 \varphi$, in the case when $f_1 \Phi_1 = \eps J_1 \varphi$.
Antisymmetrizing in $J_1$ and $J_2$, we get 
\bena
&&   G_n(J_1; J_2, f_2, \dots, f_n) - G_n(J_2; J_1, f_2, \dots, f_n) \nonumber\\
&=& \Bigg(\i \Delta^\ret(\eps J_2, \eps J_1) - \i \Delta^\ret(\eps J_1, \eps J_2) - [\varphi(\eps J_1), \varphi( \eps J_2) ] 
\Bigg) \T \left(\prod_{i=2}^n f_i \Phi_i \right)  \nonumber\\
&+& \sum_{i,j = 2}^n \Bigg[ \T \left( \cdots (\Delta^\ret J_1 \eps) \frac{\delta ( f_i \Phi_i ) }{\delta \varphi} 
\cdots (\Delta^\ret J_2 \eps) \frac{\delta ( f_j \Phi_j ) }{\delta \varphi} \cdots 
\right) -  (J_1 \leftrightarrow J_2) \Bigg] \nonumber\\
&+& \text{other terms},
\eena  
where ``other terms'' stand for expressions that vanish under the inductive assumption that T11a is true 
for $N_\varphi < k$. The first expression on the right side vanishes,
because the commutator of $\varphi$ with itself is given by $\i
\Delta$ [see eq.\eqref{commutiv}], and because $\Delta^\ret(\eps J_2, \eps
J_1) = \Delta^\adv(\eps J_1, \eps J_2)$. The second expression on the
right side vanishes because the time ordered products are symmetric.
This shows that $G_n(J_1; J_2, f_2, \dots, f_n)$ 
is symmetric in $J_1, J_2$, implying  
that $c[\varphi \otimes \varphi \otimes \Phi_2 \otimes 
\dots \Phi_n]$ is symmetric in the spacetime arguments associated with 
the factors of $\varphi$, as we desired to show. This completes the proof.

We have therefore obtained a construction of 
time-ordered products satisfying T1--T10 and T11a. We will work with such a prescription in everything that
follows. 
Any other prescription satisfying these properties will differ from the given one by formulas~\eqref{unique1} 
and~\eqref{unique2}, where the distributions $c$ must now satisfy the additional constraint
\ben
\label{11a}
c[\varphi \otimes (\otimes_i \Phi_i)] = 0
\een
due to the imposition of the further requirement T11a.

\subsection{Proof that T11b can be satisfied when $D>2$}

For the remainder of this section, we restrict consideration to 
spacetimes of dimension $D>2$, and we will prove that the remaining 
requirement, T11b, can be satisfied
together with all other requirements T1--T10, T11a. 

Condition T11b is far from obvious even for the Wick products, 
and it is not satisfied by our local normal ordering 
prescription~\eqref{phih} (which satisfies T1--T10, T11a),
as can be seen from the fact that the
stress tensor $T_{ab}$ when defined via the local normal 
ordering prescription fails to be conserved (see subsection 3.2), whereas
any prescription satisfying T11b automatically gives rise to a conserved 
stress tensor by thm.~5.1. 
Thus, in order to construct a prescription satisfying T11b together 
with all other requirements, we have to reconsider even the definition of Wick powers. 
\medskip

For these reasons, it is not surprising that our proof of T11b is technically much more complex
than the proof of T10 or T11a given in the previous sections. 
Nevertheless, the basic logic underlying the proof is actually rather simple and transparent. 
We now outline this basic logic, leaving the details to the following six subsections 6.2.1--6.2.6.
As with many other constructions in this paper, it is convenient not to attempt to construct the time 
ordered products satisfying T11b in one stroke for an arbitrary number $N_\varphi$ of factors of $\varphi$, but to proceed inductively in the number of factors. Starting off with the trivial case,  we therefore assume that a prescription satisfying T11b has been give up to less than $k$ factors. At $N_\varphi = k$ 
factors we consider the algebra valued map $D_n$ which is precisely the failure of 
T11b to be satisfied: For a given collection of $f_i \Phi_i \in \F$ with a total number of 
$k$ factors of $\varphi$, and any smooth, compactly supported variation $h_{ab}$ of the metric, this is given by 
\bena\label{Ddef}
D_n^{ab}(h_{ab}; f_1, \dots, f_n) &\equiv&
\delta^\ret
\left[ \T\left(\prod_{i=1}^n f_i \Phi_i\right) \right] \nonumber\\
&-& \Ihaves \R\left(\prod_{i=1}^n f_i \Phi_i; \eps h_{ab} T^{ab}\right)
\nonumber\\
&-& \sum_i \T\left( f_1 \Phi_1 \cdots h_{ab} \frac{\delta (f_i \Phi_i)}{\delta g_{ab}} \cdots f_n \Phi_n \right),  
\eena
where the retarded variation is taken with respect to the infinitesimal variation $h_{ab}$ of the metric. 
(Note that $D_n$ involves time ordered products with $N_T = n+1$ factors.)
The basic idea of the proof is to show that the given prescription $\T$ for the time ordered products
can be adjusted, if necessary, to a new prescription $\T'$---related 
to the original one by eq.~\eqref{unique1}---in such a way that $D_n'=0$ for this new prescription, and so that the coefficient distributions $c$ implicit in eq.~\eqref{unique1} obey the constraints described 
above. This then would show that T1--T10, T11a, and T11b will hold for the modified 
prescription for time ordered products with up to $k$ factors of $\varphi$.  

The obvious strategy for doing this is, of course, to absorb $D_n$ into a redefinition of the 
appropriate time ordered products involving a stress energy factor by simply subtracting it from the given prescription, and we will indeed follow this basic strategy. However, while it is straightforward to show that subtracting $D_n$ from the corresponding time ordered products $\T$ with a stress energy factor will automatically produce a new prescription $\T'$ satisfying $D_n' = 0$, it is not at all obvious that $\T'$ will continue to satisfy the other requirements T1--T10 and T11a. In order to demonstrate that this is 
indeed the case, we proceed by establishing a number of properties about $D_n$ in the following
subsections. The upshot is that $D_n$ is `sufficiently harmless', in the sense that subtracting it 
from the given prescription $\T$ will produce a $\T'$ which continues to have the desired properties
T1--T10 and T11a. 

In more detail, we proceed as follows:
\begin{enumerate}
\item
In subsection~6.2.1, we first show that $D_n$ is a functional of $h_{ab}, f_1, \dots, f_n$ that 
is supported on the total diagonal. 
\item
In subsection~6.2.2, we then establish that, at the 
induction order considered, $D_n$ is a c-number. 
\item
In subsection~6.2.3, we show that $D_n$ is local and covariant, with an appropriate 
scaling behavior. 
\item
In subsection~6.2.4, we show that $D_n=0$ if 
one of the field factors is equal to $\varphi$. 
\item
In subsection~6.2.5., we  establish that 
$D_n$ is not merely a linear functional, but in fact a distribution (i.e., 
continuous in the appropriate sense) with a smooth dependence upon the 
metric and with an appropriate scaling behavior under scaling of  
the metric. 
\item
In subsection~6.2.6., we show that $D_n$ has the appropriate symmetry property
when one of the factors $\Phi_i$ is equal to a stress energy tensor $T_{cd}$.
\end{enumerate}
These properties imply that $D_n$ is, in fact, a delta 
function, multiplied by appropriate 
curvature polynomials (with appropriate symmetry properties). 
Since the freedom to redefine time-ordered products consists 
precisely in adding such delta function expressions, we can absorb $D_n$ into a 
redefinition of time ordered products (here it is used that $D>2$), while preserving
T1--T10 and T11a. This is described in detail in subsection~6.2.7.
 
\medskip

We now elaborate these arguments. As for the induction start, when there are 
no factors of $\varphi$ in the fields $f_1 \Phi_1, \dots, f_n \Phi_n$ on which $D_n$ depends, we obviously must have $n=0$. In this 
case,  $D_0^{ab}(h_{ab}) = -(\i/2) \R(\eps h_{ab} T^{ab})$, 
since $\delta^\ret(\myid) = 0$. But 
any retarded product with only one factor vanishes
by definition, so there is nothing to show for $N_\varphi = 0$. Let us therefore inductively assume 
that $D_n=0$ for any set of $f_1 \Phi_1, \dots, f_n \Phi_n$, with a 
total number $N_\varphi$ of $\varphi$ less than $k$. 

\subsubsection{Proof that $D_n$ is supported on the total diagonal}
First, we will show that $D_n$ is supported on the total diagonal 
$\Delta_{n+1}$
in the product manifold $M^{n+1}$. For this, choose a test
function $h_{ab} \otimes f_1 \otimes \dots \otimes f_n$ 
whose support does not intersect 
$\Delta_{n+1}$. Then, without loss of generality, we can assume
that one of the following cases occurs:

\begin{enumerate}
\item
There is a Cauchy surface $\Sigma$ in $M$ such that $\supp h_{ab} \subset J^+(\Sigma)$ and $\supp f_i \subset
J^-(\Sigma)$ for all $i= 1, \dots, n$. 
\item
The same as the previous one, but with ``$+$'' and ``$-$'' interchanged.
\item
There is a Cauchy surface $\Sigma$, and a proper, non-empty subset $I \subset \{1, \dots, n\}$ with 
the property that $\supp h_{ab}, \supp f_i \subset J^+(\Sigma)$ for all $i \in I$, and such that 
$\supp f_j \subset J^-(\Sigma)$ for all $j$ in the complement $J$ of $I$. 
\item
The same as the previous one, but with ``$+$'' and ``$-$'' interchanged.
\end{enumerate}

We now analyze these cases one-by-one. To simplify the notation, let us use the shorthand
\ben
A_i = f_i \Phi_i \in \F. 
\een
In case (1), the support  of infinitesimal variation $h_{ab}$ is outside the 
causal past of the support of the $A_i$, and we consequently have that 
$\delta^\ret[\T( \prod A_i )] = 0$.  Thus, the first term in $D_n$ vanishes. 
But the other terms also vanish: The second because of the support properties of the retarded 
products, eq.~\eqref{rsupp}, and the third one because $\supp f_i \cap \supp h_{ab}$ is empty.

\medskip

In case (2), it follows that $\delta^\adv[\T( \prod A_i )] = 0$ by the same argument as above. 
Thus, the first term in $D_n$ is equal to
\bena
\delta^\ret \left[
\T \left( \prod A_i \right)
\right]
&=& 
\delta^\ret \left[
\T \left( \prod A_i \right)
\right] - 
\delta^\adv \left[
\T \left( \prod A_i \right)
\right] \nonumber \\
&=& -\Ihaves \left[ T^{ab}(\eps h_{ab}), \T \left(\prod A_i\right) \right]\nonumber\\
&=& \Ihaves \R\left(\prod A_i; \eps h_{ab} T^{ab}\right) - \Ihaves \A\left(\prod A_i; \eps h_{ab} T^{ab}\right) \nonumber \\
&=& \Ihaves \R\left(\prod A_i; \eps h_{ab} T^{ab}\right)   
\eena 
where in the second line we have used eqs.~\eqref{bdef} and~\eqref{bdef'}, where 
in the third line we have used an identity for retarded and advanced products, 
and where in the fourth line we used that $\supp f_i \subset J^+(\supp h_{ab})$
and the support property of the advanced products. The calculation 
shows that the first term and the second term in $D_n$ cancel. 
But the third term vanishes, because $\supp f_i \cap \supp h_{ab}$ is empty, showing that $D_n = 0$ 
in case (2). 

\medskip

In case (3), we use the causal factorization property T8 of the time 
ordered products and the homomorphism 
property of $\tau^\ret$ (that is, 
$\tau^\ret(ab) = \tau^\ret(a) \tau^\ret(b)$) to write
\bena
\text{first term in eq.~\eqref{Ddef}} &=& 
\delta^\ret
\left[ \T \left(\prod_{i \in I} A_i\right) 
       \T \left(\prod_{j \in J} A_j\right) \right] \\
 &=& 
\delta^\ret 
\left[ \T\left(\prod_{i \in I} A_i\right) \right]
\T\left(\prod_{j \in J} A_j\right) \nonumber 
+ \T\left(\prod_{i \in I} A_i\right) 
\delta^\ret 
\left[ \T \left(\prod_{j \in J} A_j\right) \right].
\eena
Since neither $I$ nor $J$ are empty by assumption, and since $A_1, \dots, A_n$ together
have at most a total of $N_\varphi = k$ factors of $\varphi$, it follows that 
$A_i, i \in I$ as well as $A_j, j \in J$ each have strictly less than $k$ factors of $\varphi$.
Hence we can use our inductive assumption which gives $D_{|I|} = D_{|J|} = 0$. It follows 
that
\begin{multline}
\text{first term in eq.~\eqref{Ddef}} =\\
\Ihaves \R\left(\prod_{i \in I} A_i; \eps h_{ab} T^{ab}\right) \T \left( \prod_{j \in J} A_j \right) +
\Ihaves \T\left(\prod_{i \in I} A_i\right) \R\left( \prod_{j \in J} A_j ; \eps h_{ab} T^{ab}\right)\\
+ \sum_{k \in I}
\T\left(\frac{\delta A_k}{\delta \varphi} \prod_{i \in I, i \neq k} A_i \right) \T \left( \prod_{j \in J} A_j \right) \\
+ \T \left( \prod_{i \in I} A_i \right)
\sum_{l \in J} \T \left(\frac{\delta A_l}{\delta \varphi} \prod_{j \in J, j \neq l} A_j \right) \, .
\end{multline}
For the second and third 
term in eq.~\eqref{Ddef}, we likewise use the causal factorization property and the definition of the retarded 
product. It is then seen that these terms precisely cancel the first term in eq.~\eqref{Ddef}, 
showing that $D_n = 0$ in case~(3). 

\medskip

Case (4) can be treated in the same way as the previous one.

\subsubsection{Proof that $D_n$ is a c-number}

We next want to show that the 
algebra element $D_n \in \W$ is in fact proportional
to the identity operator. By~\cite[prop.~2.1]{hw1}, an element $a \in \W$ is 
proportional to the identity if and only if $[a, \varphi(F)]=0$ for all smooth, 
compactly supported densities $F$. Thus, we will be done if we can show that 
\ben\label{16}
[D_n^{ab}(h_{ab}; f_1, \dots, f_n), \varphi(F)] = 0.
\een
Inductively, we know this is true when $N_\varphi < k$ since $D_n$ itself 
vanishes then. We now prove that it is also true when $N_\varphi = k$.

We begin by calculating the commutator with the first term in $D_n$ [see eq.~\eqref{Ddef}], which, 
using the homomorphism property of $\tau^\ret$ is equal to 
\begin{multline}\label{17}
\left[
\delta^\ret \left\{ 
\T \left(\prod f_i \Phi_i\right) \right\} , \varphi(F) \right] = \\
\delta^\ret \left\{
\left[ \T \left(\prod f_i \Phi_i\right), \varphi (F) \right] \right\}
-
\left[ \T \left(\prod f_i \Phi_i\right), 
\delta^\ret \Big\{ \varphi(F) \Big\} \right].
\end{multline}
We now simplify the first term on the right side of this expression using the commutator property
of the time ordered products, T9, and we simplify the second term on the right side using that
\ben
\label{24}
\delta^\ret[ \varphi(F) ] = -\varphi(\delta(\eps P) \Delta^\adv F), 
\een
which follows from a direct calculation using the definition of 
$\tau^\ret$ (see~\cite{bfv}). Here, $\delta(\eps P)$ 
is the infinitesimal variation of the densitized Klein-Gordon operator under a
change in the metric, 
\ben
\label{Kdef}
\delta(\eps P)_\g f = \frac{\partial}{\partial s} (\eps P)_{\g + s \h} f \Bigg|_{s=0}.
\een
(Note that $\delta(\eps P)$ is a second order differential operator mapping smooth 
scalar functions to densities.) Substituting eq.~\eqref{24} into eq.~\eqref{17} gives
\begin{multline}
\left[
\delta^\ret_{\g}
\left\{ \T_{\g}\left(\prod f_i \Phi_i\right) \right\}, \varphi_\g(F) \right]
= \\
\i\frac{\partial}{\partial s}
\tau^\ret_{\g^{(s)}}
\left( \sum_{i=1}^n \T_{\g^{(s)}}\left(
f_1 \Phi_1 \dots 
(\Delta_{\g^{(s)}}  F) \frac{\delta(f_i \Phi_i)}{\delta \varphi} \dots f_n\Phi_n
\right)\right) \Bigg|_{s=0}\\
+
\left[ \T_\g\left(\prod f_i \Phi_i\right), 
\varphi_\g (\delta(\eps P) \Delta^\adv F) \right].
\end{multline}
The first term on the right side involves only $N_\varphi = k-1$ factors of $\varphi$, 
and therefore can be simplified using the inductive assumption that $D_n = 0$
in that case. The second 
term on the right side can again be simplified using the commutator property. This gives\footnote{
We use the convention that whenever the expression $\Delta F$ appears in an expression to which 
$\delta/\delta g_{ab}$ is applied, we will view $\Delta F$ as independent of $\g$, i.e., $\delta/\delta g_{ab}$
does not act on $\Delta F$ in such an expression.} 
\begin{multline}
\label{20}
\left[
\delta^\ret_{\g}
\left\{ \T_{\g}\left(\prod f_i \Phi_i\right) \right\}, \varphi_\g(F) \right]
= \\
- \frac{1}{2} \sum_{i=1}^n \R_\g\left(
f_1 \Phi_1 \dots 
(\Delta  F) \frac{\delta(f_i \Phi_i)}{\delta \varphi} \dots f_n\Phi_n
; \eps h_{ab} T^{ab}\right)\\
+ \i \sum_{i=1}^n \T_\g\left(
f_1 \Phi_1 \dots 
h_{ab} \frac{\delta}{\delta g_{ab}} \left\{ (\Delta F) \frac{\delta}{\delta \varphi} (f_i \Phi_i) \right\} \dots f_n\Phi_n
\right)\\
+ \i \sum_{i=1}^n \T_\g\left(
f_1 \Phi_1 \dots 
\frac{\partial}{\partial s} \Delta_{\g^{(s)}} F \bigg|_{s=0}  \frac{\delta (f_i \Phi_i)}{\delta \varphi} 
\dots f_n\Phi_n 
\right)\\
+ \i \sum_{i=1}^n \sum_{k \neq i} \T_\g\left(
f_1 \Phi_1 \dots 
h_{ab} \frac{\delta(f_k \Phi_k)}{\delta g_{ab}} \dots 
(\Delta F) \frac{\delta(f_i \Phi_i)}{\delta \varphi} \dots f_n\Phi_n
\right)\\
+ \i \sum_{i=1}^n
\T_\g\left(f_1 \Phi_1 \dots(\Delta \delta(\eps P)  \Delta^\adv F) \frac{\delta(f_i \Phi_i)}{\delta \varphi} \dots f_n\Phi_n 
 \right).
\end{multline}
We next calculate the commutator of the second term in $D_n$ [see eq.~\eqref{Ddef}] with 
$\varphi(F)$, by expanding the retarded product in terms of 
time-ordered products 
and using, for each of the resulting terms the commutator property of the 
time-ordered products, 
\begin{multline}
-\Ihaves \left[ 
\R\left(
\prod f_i \Phi_i; \eps h_{ab} T^{ab}
\right), \varphi(F)
\right] = \frac{1}{2}
\R\left(
\prod f_i \Phi_i; (\Delta F) 
\frac{\delta(\eps h_{ab} T^{ab})}{\delta \varphi}
\right)\\
+ \frac{1}{2} \sum_{i=1}^n 
\R\left(f_1 \Phi_1 \cdots (\Delta F) \frac{\delta(f_i \Phi_i)}{\delta \varphi} \cdots f_n \Phi_n; \eps h_{ab} T^{ab} \right).  
\end{multline}
Using that the variational derivatives $\delta/\delta g_{ab}$ and $\delta/\delta \varphi$ commute up 
to an exact form (see eq.~\eqref{comder}), and using eq.~\eqref{funder} of appendix B, we have
\bena
\frac{1}{2} (\Delta F)
\frac{\delta(\eps h_{ab} T^{ab})}{\delta \varphi}
&=&  
h_{ab} \frac{\delta}{\delta g_{ab}} \left\{ (\Delta F) \frac{\delta L_0}{\delta \varphi} \right\} + \d B_1\nonumber\\
&=& 
\frac{\partial}{\partial s}\left\{ (\Delta_\g F) (\eps P)_{\g + s \h} \varphi \right\} \Bigg|_{s=0}  + \d B_2. 
\eena
where $B_1, B_2$ are local, $(D-1)$ form functional of $\varphi$ and the metric, and 
where $P$ is the Klein-Gordon operator. Since $P$ is hermitian, the right side can 
be rewritten further as 
\ben
\frac{1}{2} (\Delta F)
\frac{\delta(\eps h_{ab} T^{ab})}{\delta \varphi}
=
\frac{\partial}{\partial s}\left\{ (\eps P)_{\g + s \h} (\Delta_\g F) \varphi + \d C \right\} \Bigg|_{s=0} + \d B_2
= \delta(\eps P) (\Delta F) \varphi + \d B_3 
\een 
remembering that $\delta (\eps P)$ is metric variation of the densitized Klein-Gordon operator.
Thus, by the Leibniz rule, T10, we get
\begin{multline}
-\Ihaves \left[ 
\R\left(
\prod f_i \Phi_i; \eps h_{ab} T^{ab}
\right), \varphi(F)
\right] = 
\R\left(
\prod f_i \Phi_i; (\delta(\eps P) \Delta F) \varphi
\right)\\
+ \frac{1}{2} \sum_{i=1}^n 
\R\left(f_1 \Phi_1 \cdots (\Delta F) \frac{\delta(f_i \Phi_i)}{\delta \varphi} \cdots f_n \Phi_n; 
\eps h_{ab} T^{ab}  \right).  
\end{multline}
We apply T11a to the first term on the right side of this equation. This gives
\begin{multline}
\label{22}
-\i \left[ 
\R\left(
\prod f_i \Phi_i; \eps h_{ab} T^{ab}
\right), \varphi(F)
\right] 
= 
\i \sum_{i=1}^n \T
\left(f_1 \Phi_1 \cdots (\Delta^\ret \delta(\eps P) \Delta F) 
\frac{\delta(f_i \Phi_i)}{\delta \varphi} \cdots f_n \Phi_n \right)
\\
+ \sum_{i=1}^n 
\R\left(f_1 \Phi_1 \cdots (\Delta F) \frac{\delta(f_i \Phi_i)}{\delta \varphi} \cdots f_n \Phi_n; 
\eps h_{ab} T^{ab}  \right).  
\end{multline}
We finally take the commutator of the third term in $D_n$ with $\varphi(F)$, and use the 
commutator property to simplify. This gives
\begin{multline}\label{21}
-\left[
\sum_i \T\left( f_1 \Phi_1 \cdots h_{ab} \frac{\delta (f_i \Phi_i)}{\delta g_{ab}} \cdots f_n \Phi_n 
\right)
, \varphi(F)
\right]\\
= -\i \sum_{i=1}^n \T\left(
f_1 \Phi_1 \dots (\Delta  F) \frac{\delta}{\delta \varphi} \left\{
h_{ab} \frac{\delta}{\delta g_{ab}} (f_i \Phi_i) \right\} \dots f_n\Phi_n
\right)\\
- \i \sum_{i=1}^n \sum_{k \neq i} \T\left(
f_1 \Phi_1 \dots 
h_{ab} \frac{\delta(f_k \Phi_k)}{\delta g_{ab}} \dots 
(\Delta F) \frac{\delta(f_i \Phi_i)}{\delta \varphi} \dots f_n\Phi_n
\right).
\end{multline}
We have now calculated the commutator of all three terms in $D_n$ with $\varphi(F)$, given by 
eqs.~\eqref{20},~\eqref{22} and~\eqref{21} respectively. 
If we add these contributions up, then we see that 
the commutator $[D_n, \varphi(F)]$ will vanish if we can show that
\ben\label{23}
\i\Delta^\ret \delta(\eps P) \Delta F + \i\Delta \delta(\eps P) \Delta^\adv F = 
-\i\frac{\partial}{\partial s} \Delta_{\g^{(s)}} F \bigg|_{s=0} 
\een
for all compactly supported densities $F$. 
However, this identity follows immediately from $\Delta = 
\Delta^\adv - \Delta^\ret$ together with the identity
\ben
\label{41}
\frac{\partial}{\partial s} \Delta^\ret_{\g^{(s)}} F \bigg|_{s=0} = -\Delta^\ret \delta(\eps P) \Delta^\ret F
\quad \text{and ``adv'' $\leftrightarrow$ ``ret''}, 
\een
for all compactly supported densities $F$, 
which in turn is seen to be true owing to the relation $(\partial/\partial s) (\eps P \Delta^\ret)_{\g^{(s)}} = 0$
(and the analogous relation for the advanced propagator).

\subsubsection[Proof that $D_n$ is local and covariant and scales almost \\
homogeneously]{Proof that $D_n$ is local and covariant and scales almost homogeneously}

It is `obvious' that $D_n$ is a c-number functional that is constructed entirely from the metric, because
all the terms in the defining equation for $D_n$ have this property. $D_n$ 
depends moreover locally and covariantly on the metric in 
the sense that if $\chi: N \to M$ is any causality and orientation preserving isometric embedding, 
then
\ben
D^{ab}_n[M, \g](\chi_* h_{ab}, \chi_* f_1, \dots \chi_* f_n) = D^{ab}_n[N, \chi^*\g](h_{ab}, 
f_1, \dots, f_n)
\een
for all test (tensor-)fields with compact support on $N$. This property follows because
the second and third term in the definition of $D_n$ are local and covariant quantities by T1, 
and because the map $\tau^\ret$ appearing in the first term also has this property by construction.

Moreover, the functionals $D_n$ also have an almost homogeneous scaling behavior under rescalings 
of the metric in the sense that 
\ben
\frac{\partial^N}{\partial^N \ln \lambda} (\lambda^d \cdot 
D_n[M, \lambda^2 \g]) = 0
\een
for some natural number $N$, where $d$ is the sum of the engineering dimension of the fields
appearing in $D_n$. Again, for the second and third term in the definition of $D_n$, this property 
follows since we are assuming that our time ordered products (and hence retarded products) 
have an almost homogeneous scaling behavior in the sense of T2. For the first term in the 
definition of $D_n$, this follows from the fact that 
if $(M, \g)$ and $(M, \g')$ are spacetimes whose metrics differ only within some compact 
set $K$, and if $\tau^{\adv /\ret}$ are the corresponding algebra isomorphisms from 
$\W(M, \g)$ to $\W(M, \g')$, then 
\ben
\sigma_\lambda' \circ \tau^{\adv/ \ret} = \tau^{\adv/ \ret} \circ \sigma_\lambda, 
\een
where $\sigma_\lambda$ is the natural isomorphism from $\W(M, \g)$ to $\W(M, \lambda^2 \g)$
introduced above in T2, and where $\sigma'_\lambda$ is the corresponding 
isomorphism for 
$\g'$.

\subsubsection{Proof that $D_n = 0$ when one of the $\Phi_i$ is equal to $\varphi$}

Let us now assume that one of the fields $\Phi_i$ is equal to $\varphi$, say 
$\Phi_n = \varphi$, and as before, that the total number $N_\varphi$ of free field factors 
in $\Phi_1, \dots, \Phi_{n-1}, \Phi_n = \varphi$ is equal to $k$. We will show that $D_n$ is 
automatically zero in this case under our inductive assumption that $D_n = 0$ 
when $N_\varphi < k$.

We first look at the second term in $D_n$ in eq.~\eqref{Ddef}, 
setting $A_i = f_i \Phi_i$ for $i < n$ and
$f_n = F$ to facilitate the notation. After some algebra, repeatedly using T10, T11a 
and eq.~\eqref{1}, we get
\bena
\label{29}
-\Ihaves \R\left( F\varphi \prod^{n-1} A_i; \eps h_{ab} T^{ab} \right) &=& 
\frac{1}{2} \sum_j \R\left( A_1 \dots (\Delta^\ret F)
\frac{\delta A_j}{\delta \varphi} \dots A_{n-1}; \eps h_{ab} T^{ab}  \right)\nonumber \\
&-& \Ihaves \varphi(F) \, \R \left( \prod^{n-1} A_i;  \eps h_{ab} T^{ab} \right)\nonumber \\
&-& \varphi(\delta(\eps P) \Delta^\adv F) \, \T\left( \prod^{n-1} A_i \right)\nonumber \\
&+& \i\sum_j \T\left( A_1 \dots (\Delta^\ret \delta(\eps P) \Delta^\ret F)
\frac{\delta A_j}{\delta \varphi} \dots A_{n-1} \right).  
\eena
Here $\delta(\eps P)$ is the first order variation of the Klein-Gordon operator with respect to 
our family of metrics, see eq.~\eqref{Kdef}. 
For the third term in $D_n$, we get, using T10, T11a,
\bena
\label{third}
&& -\sum_j \T\left( F\varphi A_1 \dots h_{ab}
\frac{\delta A_j}{\delta g_{ab}} \dots A_{n-1} \right)
\nonumber \\
&=& -\i  \sum_{j \neq k}\T\left( A_1 \dots h_{ab} 
\frac{\delta A_j}{\delta g_{ab}} \dots (\Delta^\ret F)
\frac{\delta A_k}{\delta \varphi}
\dots A_{n-1} \right)\nonumber\\
&-& \i \sum_{j}\T\left( A_1 \dots (\Delta^\ret F) \frac{\delta}{\delta \varphi} \left\{ h_{ab} 
\frac{\delta}{\delta g_{ab}} A_j \right\} \dots 
 A_{n-1} \right)\nonumber \\
&-& \varphi(F) \sum_{j}\T\left( A_1 \dots h_{ab} 
\frac{\delta A_j}{\delta g_{ab}} \dots 
 A_{n-1} \right). 
\eena
For the first term in $D_n$ we get, using eqs.~\eqref{24} and the definition~\eqref{Kdef} of 
$\delta(\eps P)$, 
\bena
\label{31}
\delta^\ret \left[
\T\left( F\varphi \prod^{n-1} A_i
\right)
\right] 
&=&  \varphi(\delta(\eps P) \Delta^\adv F) \, \T\left( \prod^{n-1} A_i \right)\nonumber \\
&+& \i \sum_j \delta^\ret \left[
\T\left( A_1 \dots (\Delta^\ret F)
\frac{\delta A_j}{\delta \varphi} \dots A_{n-1} \right)
\right] \nonumber\\
&+& \varphi(F) \, \delta^\ret \left[
\T\left( \prod^{n-1} A_i
\right)
\right]. 
\eena
We can simplify the terms on the right side using the inductive assumption. Adding up the 
contributions eqs.~\eqref{29}, \eqref{third} and~\eqref{31} to $D_n$, and using eq.~\eqref{41}, 
we find that all terms cancel. Thus we have shown that $D_n = 0$
when $N_\varphi = k$ and when one of the factors $\Phi_i$ is $\varphi$. 

\subsubsection[Proof that $D_n$ satisfies a wave front set condition and
depends \\
smoothly and analytically on the metric]{Proof that $D_n$ satisfies a wave front set condition and
depends smoothly and analytically on the metric}
 
We now show that $D_n$ is a distribution on $M^{n+1}$---i.e., $D_n$ is a 
multilinear functional that is {\em continuous} in the appropriate 
sense---and that it satisfies the wave front set condition 
\ben
\label{wfsc}
\WF(D_n)\restriction_{\Delta_{n+1}} 
\perp T(\Delta_{n+1}).  
\een
Moreover, we will show that if $\g^{(s)}$ is a smooth (resp. analytic) family of metrics depending smoothly (resp. analytically)
upon a set of parameters $s$ in a parameter space $\P$, and if
$D_n^{(s)}$ are the corresponding
distributions (viewed now as a single distribution 
on $\P \times M^{n+1}$), then
\ben
\label{swfsc}
\WF(D_n^{(s)})\restriction_{\P \times \Delta_{n+1}} 
\perp T(\P \times \Delta_{n+1}),  
\een 
(with the smooth wave front set $\WF$ replaced by the analytic wave front set $\WF_A$ 
in the analytic case).

Since $D_n$ is a c-number, it is equal to the expectation value of eq.~\eqref{Ddef} in a any
state $\omega$ on $\W(M, \g)$. To simplify things we take $\omega$ to be a 
quasifree Hadamard state, and write $D_n$ as
\bena\label{Ddef'}
D_n^{ab}(h_{ab}; f_1, \dots, f_n) &=&
r^{ab}_n(h_{ab}; f_1, \dots, f_n) \nonumber\\
&-& \Ihaves \omega\left( \R\left(\prod_{i=1}^n f_i \Phi_i; \eps h_{ab} T^{ab}\right) \right)
\nonumber\\
&-& \sum_i \omega \left(
\T\left( f_1 \Phi_1 \cdots h_{ab} \frac{\delta (f_i \Phi_i)}{\delta g_{ab}} \cdots f_n \Phi_n \right)
\right),  
\eena
where we have set
\ben
r_n^{ab}(h_{ab}; f_1, \dots, f_n) = 
\omega \left[ \delta^\ret
\left\{ \T\left(\prod_{i=1}^n f_i \Phi_i\right) 
\right\} \right]. 
\een

To prove the desired properties, eqs.~\eqref{wfsc} and~\eqref{swfsc}, 
of $D_n$, we show that each term on the right
side of eq.~\eqref{Ddef'} 
satisfies these properties separately.
This is relatively straightforward for the second and third terms.
It follows from our microlocal spectrum condition, T3, that the 
second and third terms in $D_n$ each satisfy
\begin{multline}
\label{firstsec}
\WF(\text{2nd and 3rd terms in eq.~\eqref{Ddef'}}) \subset \{  (y, p; x_1, k_1; \dots; x_n, x_n) \mid \\
\exists \,\, \text{Feynman graph $G(q)$ with vertices $y,x_1, \dots, x_n$}\\
\text{and edges $e$ such that if $y = s/t(e)$ then $t/s(e) \in J^+(y)$}\\
k_i = \sum_{e: s(e) = x_i} q_e - \sum_{e: t(e) = x_i} q_e, \quad
p =  \sum_{e: s(e) = y} q_e - \sum_{e: t(e) = y} q_e \} \equiv \cC_\R(M, \g).
\end{multline}
Since
\ben
\cC_\R(M, \g) \restriction_{\Delta_{n+1}} \perp T(\Delta_{n+1}),  
\een
on the total diagonal, it immediately follows that the second and third 
terms in $D_n$ 
satisfy the analog of eq.~\eqref{wfsc}. 
Moreover, if we consider a smooth family of metrics $\g^{(s)}$ 
and a corresponding  
family of quasifree Hadmard states $\omega^{(s)}$ depending 
smoothly upon $s$ in the sense of eq.~\eqref{os}, 
then it similarly follows from T4 that 
the second and third terms 
in $D_n^{(s)}$ (with $\omega$ in those expressions replaced by $\omega^{(s)}$)
have a smooth dependence upon $s$. It then follows immediately that 
the second and third terms in 
$D_n^{(s)}$ satisfy the smoothness condition~\eqref{swfsc}. 
The corresponding statement in the 
analytic case similarly follows from condition T5.

Having dealt with the second and third terms 
on the right side of eq.~\eqref{Ddef'}, 
our claims will be established by proving the following proposition:
\begin{prop} The first term on the right side of eq.~\eqref{Ddef'} 
satisfies
\ben
\label{wfsc'}
\WF(r_n)\restriction_{\Delta_{n+1}} 
\perp T(\Delta_{n+1}),  
\een
as well as
\ben
\label{swfsc'}
\WF(r_n^{(s)})\restriction_{\P \times \Delta_{n+1}} 
\perp T(\P \times \Delta_{n+1}),  
\een
where $r_n^{(s)}$ is defined by the same formula as $r_n$ except that 
$\omega$ is replaced by the smooth family $\omega^{(s)}$ in that formula, and
the metric $\g$ is replaced everywhere by $\g^{(s)}$. 
The analogous statement also holds true with regard to the analytic 
wave front set. 
\end{prop}
\noindent
{\it Proof.}
We know that $r_n$ is 
a multilinear functional which is also a distribution 
in $f_1 \otimes  \cdots \otimes  f_n$ for any fixed $\h$ of compact support. Also, since $D_n$ is 
already known to vanish for test functions $\h \otimes f_1 \otimes \cdots \otimes f_n$ 
whose support has no intersection with the total diagonal $\Delta_{n+1}$ in $M^{n+1}$, it follows
that $r_n(\h, f_1, \dots, f_n)$ is equal to minus the second and third term in eq.~\eqref{Ddef'}. 
Therefore, since these terms are individually known to be distributions, we know that $r_n$ is in 
fact a distribution off the total diagonal. However, our constructions so 
far do not tell us that $r_n$ is also a distribution {\rm on} the
total diagonal, let alone whether it satisfies the 
wave front set conditions eqs.~\eqref{wfsc'} and~\eqref{swfsc'} there. 
Thus, in order to prove the above proposition, we must look at the detailed structure
of $r_n$ near the total diagonal. 

For this, we first use our local Wick expansion~\eqref{lwe}
to write the time ordered products in the following form when  $f_1 \otimes  \cdots \otimes  f_n$ is supported in a sufficiently small 
neighborhood $U_n$ total diagonal in $M^n$ (which we assume from now on):
\begin{multline}
\label{wickexp}
\T\left(\prod_{i=1}^n f_i\Phi_i \right) 
= \sum_{\alpha_1, \alpha_2, \dots} \frac{1}{\alpha_1! \dots \alpha_{n}!} 
\int \prod_j \eps(y_j)\\
t\left[\delta^{\alpha_1} \Phi_1 \otimes \dots \otimes \delta^{\alpha_{n}} \Phi_{n}
\right]
(y_1, \dots, y_n) \\
f_1(y_1) \dots f_{n}(y_n) 
: \prod_{i=1}^{n} \prod_{j} [(\nabla)^{j}\varphi(y_i)]^{\alpha_{ij}} :_H \\
 =\sum_r \int w(y_1, \dots, y_n; x_1, \dots, x_r) \prod_{i=1}^n f_i(y_i) \, :  \prod_{j=1}^r \varphi(x_j) :_H.
\end{multline}
Here, the distributions $w \in \cD'(U_n \times M^r)$ are defined by the last equation in terms of sums of 
products of $t[\dots]$ and suitable delta functions and their derivatives. Since these distributions are in 
turn locally and covariantly constructed from the metric, it follows that also the distributions $w$ have 
this property, and we will write $w = w_\g$ 
when we want to emphasize this fact. From the $\delta$-functions
implicit in the definition of $w$, one easily finds the support property
\begin{multline}
\label{tsupp}
\supp w \subset \{  (y_1, \dots, y_n; x_1, \dots,x_r) \mid \\
\text{$\exists$ partition $\{1, \dots, r\} = I_1 \cup \dots \cup I_n$ such that $x_i = y_l \,\, \forall i \in I_l$}\}, 
\end{multline}
and from the wave front set property of the $t$, one finds furthermore the wave front set property
\begin{multline}
\label{twfs}
\WF(w) \subset \{  (y_1, k_1; \dots; y_n, k_n; x_1, p_1; \dots; x_r, p_r) \mid \\
\text{$\exists$ partition $\{1, \dots, r\} = I_1 \cup \dots \cup I_n$ such that $x_i = y_l \,\, \forall i \in I_l$}\\
\text{if $q_l \equiv k_l + \sum_{i \in I_l} p_i$, then 
$(y_1, q_1; \dots; y_n, q_n) \in \cC_\T(M, \g)$}
\} =: {\mathcal I}_\T(M, \g) 
\end{multline}
for the $w$. We also note that the $w$
scale almost homogeneously under a rescaling of the metric,  
and vary smoothly under smooth variations of the metric in the sense that 
if $\g^{(s)}$ is a family of metrics depending smoothly on $s$ in some parameter space $\P$, then the 
distributions $w^{(s)} = w_{\g^{(s)}}$ (viewed as distributions on $\P \times U_n \times M^r$) have wave front set
\begin{multline}
\label{tswfs}
\WF(w^{(s)}) \subset \{  (s, \rho; y_1, k_1; \dots; y_n, k_n; x_1, p_1; \dots; x_r, p_r) \mid \\
(y_1, k_1; \dots; y_n, k_n; x_1, p_1; \dots; x_r, p_r) \in {\mathcal I}_\T(M, \g^{(s)})\}.  
\end{multline}
These properties follow immediately from the corresponding properties satisfied by the $t$ (as a 
consequence of T3 and T4)  
as well as the delta functions. 

We now insert eq.~\eqref{wickexp} into the definition of $r_n$. This gives
\begin{multline}
r_n(\h; f_1, \dots, f_n) = 
\\
\sum_r \int
w^{(0)} (y_1, \dots, y_n; x_1, \dots, x_r) \prod_i f_i(y_i)\, 
\omega\left[ \frac{\partial}{\partial s}
\tau^{\ret}_{\g^{(s)}} \left( : \prod_j \varphi(x_j) :_{H^{(s)}} \right) \Bigg|_{s=0} 
\right]  
\\
+ \sum_r \int \frac{\partial}{\partial s} w^{(s)}(y_1, 
\dots, y_n; x_1, \dots, x_r) \bigg|_{s=0} \prod_i f_i(y_i)\,
\omega\left( :  \prod_j \varphi(x_j) :_{H^{(0)}}  \right) \\
\equiv I_1 + I_2
\end{multline}
where $\frac{\partial}{\partial s} \g^{(s)} = \h$.
Furthermore, when $\g$ is replaced everywhere in the above 
formula by a family $\g^{(s)}$ depending on a parameter $s \in \P$, 
we obtain a corresponding expression for $r^{(s)}_n$.
The proof of the proposition will be complete 
if we can show that the first term, $I_1$, and second term, $I_2$,
on the right side separately satisfy the wave front set 
condition eq.~\eqref{wfsc'}, and the smoothness condition 
eq.~\eqref{swfsc'}, i.e., if we can prove the following lemma:

\begin{lemma}
$I_1$ and $I_2$ are distributions satisfying
\ben
\label{wfsc''}
\WF(I_j) \restriction_{\Delta_{n+1}} 
\perp T(\Delta_{n+1}), 
\een
as well as 
\ben
\label{swfsc''}
\WF(I_j^{(s)}) \restriction_{\P \times \Delta_{n+1}} 
\perp T(\P \times \Delta_{n+1}) \, . 
\een
\end{lemma}

The remainder of this subsection consists of the proof of this lemma.

\paragraph{Proof of lemma~6.1 for $I_1$:}

We begin by showing eq.~\eqref{wfsc''} 
for $I_1$. For this, consider the smooth 1-parameter family of metrics $\g^{(s)}$ with 
$\frac{\partial}{\partial s} \g^{(s)} = \h$, 
and let $\omega^{(s)}$ be the unique quasifree Hadamard state on $\W(M, \g^{(s)})$ with 
the property that $\omega^{(s)}$ coincides with $\omega$ on $M \setminus J^+(K)$, where $K$ is the compact
region where $\h$ is supported. Furthermore, let $H^{(s)}$ be the local Hadamard parametrix 
associated with this 1-parameter family of metrics, and,
in a sufficiently small neighborhood $U_2$ of the diagonal, define
\ben
d^{(s)}(x_1, x_2) = \omega^{(s)}_2(x_1, x_2) - H^{(s)}(x_1, x_2).   
\een
Then one finds from the definition 
of $\tau^\ret$ that
\ben
\omega\left[ 
\tau^{\ret}_{\g^{(s)}} \left( : \prod_j \varphi(x_j) :_{H^{(s)}} \right)  
\right]
= 
\sum_{\text{pairs $ij$}}
d^{(s)}(x_i, x_j),  
\een
and hence that 
\ben
\label{i1}
I_1(\h; f_1, \dots, f_n) = \sum_r \int
w^{(0)}(y_1, \dots, y_n; x_1, \dots, x_r)  
\prod_i f_i(y_i) \,
\frac{\partial}{\partial s} 
\sum_{\text{pairs $ij$}}
d^{(s)}(x_i, x_j)
\bigg|_{s=0}.  
\een
We estimate the wave front set of $I_1$ by analyzing the wave front set of the individual 
terms in eq.~\eqref{i1}. The wave front set of $w$ is already known, 
whereas the 
wave front set associated with the distributions $d^{(s)}$ is given 
by the following lemma.

\begin{lemma}
$d^{(s)}$ is jointly smooth in $s$ and its spacetime 
arguments within a sufficiently small neighborhood $U_2$ of the diagonal in 
$M \times M$. Furthermore, in such a neighborhood, 
if 
\ben
(\delta d)(\h, f_1, f_2) = \frac{\partial}{\partial s} d^{(s)}(f_1, f_2) \Bigg|_{s=0}, 
\een
then 
\bena
\WF(\delta d) 
&\subset& \{ (y, p; x_1, k_1; x_2, k_2) \mid \text{either of the following holds:}\nonumber\\
&& \text{($(y, p) \sim (x_1, -k_1)$, $k_2 = 0$) or ($(y, p) \sim (x_2, -k_2)$, $k_1 = 0$)}\nonumber\\
&& \text{or ($x_1 = x_2 = y$ and $p = -k_1 -k_2$)}\}. 
\eena
\end{lemma}
\noindent
{\it Proof.}
The bidistribution $d^{(s)}$ is symmetric, and is a bisolution 
of the Klein-Gordon equation modulo
a smooth function, because $H^{(s)}$ is a bisolution modulo a smooth function. In fact, 
\bena
\label{fnweq}
( P^{(s)} \otimes 1 ) d^{(s)}(x_1, x_2)  &=& G^{(s)}(x_1, x_2)\nonumber \\ 
( 1 \otimes P^{(s)} ) d^{(s)}(x_1, x_2)  &=& G^{(s)}(x_2, x_1), 
\eena
where $G^{(s)}$ is equal to the action of the Klein-Gordon 
operator on the first variable in $H^{(s)}$ (and can thereby be calculated by 
Hadamard's recursion procedure, at least in analytic spacetimes), and 
where $P^{(s)}$ is the Klein-Gordon operator associated with $\g^{(s)}$. It follows that 
$G^{(s)}$ is jointly smooth (resp. analytic, in analytic spacetimes) in 
$s$ and its spacetime arguments. Furthermore, since $\omega^{(s)}$ is independent of $s$
everywhere in $M \setminus J^+(K)$, and since $H^{(s)}$ is independent of $s$
on any convex normal neighborhood which does not intersect $K$, 
it follows that 
$d^{(s)}$ is independent of $s$ on any convex normal neighborhood which has no
intersection with $J^+(K)$. Using these facts, we will now show that $d^{(s)}(x_1, x_2)$ 
is jointly smooth in $s,x_1,x_2$. 

For this, we consider a globally hyperbolic subset $N$ of $M$ with compact closure, 
which contains $K$, and which has Cauchy surfaces $S_-$ resp. $S_+$ 
not intersecting $J^+(K)$ resp. $J^-(K)$ (for all metrics $\g^{(s)}$ with $s$ sufficiently small).
Without loss of generality, we may assume that $K$ is so small that $N$ can 
be chosen to be convex and normal (again for all metrics $\g^{(s)}$ 
with $s$ sufficiently small). 
By what we have said above, $d^{(s)}$ does not depend upon $s$ in a neighborhood of $S_-$.
Within $N$, we define the bi-distribution
\ben
\alpha_{ab}^{(s)}(x_1, x_2) = (\Delta^{\adv \, (s)} f_1)(x_1) (\Delta^{\adv \, (s)} f_2)(x_2)
\, \stackrel{\leftrightarrow}{\nabla}_a
   \stackrel{\leftrightarrow}{\nabla}_b \, 
d^{(s)}(x_1, x_2), 
\een
where $\nabla_a$ acts on $x_1$ and $\nabla_b$ acts on $x_2$.
We now take the divergence of $\alpha_{ab}^{(s)}(x_1, x_2)$ both in $x_1$ and $x_2$ and integrate
the resulting expression over $U \times U$, where $U \subset N$ is the region enclosed by 
$S_-$ and $S_+$. By Stokes' theorem and the 
support property of $\Delta^\adv$, we have 
\ben
\int_{U \times U} \nabla^a \nabla^b \alpha_{ab}(x_1, x_2) \, \eps(x_1)\eps(x_2)
= \int_{S_- \times S_-} \alpha_{ab}(x_1, x_2) \, {\rm d}\sigma^a(x_1) {\rm d}\sigma^b(x_2), 
\een
for any test (densities) $f_1, f_2$ supported in $U$. 
(Here, $\d \sigma^a$ is the usual integration element induced
by $\eps$, and we are suppressing the dependence 
upon $s$ to lighten the notation.) 
Now perform the differentiation on the left side, 
using $(\nabla^a \nabla_a - \xi R - m^2) \Delta^\adv
= \delta$, using the fact that the advanced 
propagator on the right side can be replaced 
by the causal propagator, and using the symmetry properties of 
$G$ implied by eq.~\eqref{fnweq}. We obtain
\begin{multline}
\label{50} 
d(f_1, f_2) = 
\int_{S_- \times S_-} (\Delta f_1)(x_1) (\Delta f_2)(x_2) \, 
\stackrel{\leftrightarrow}{\nabla}_a
\stackrel{\leftrightarrow}{\nabla}_b
\, d(x_1, x_2) \, {\rm d}\sigma^a(x_1) {\rm d}\sigma^b(x_2)
\\
- G(\Delta^\adv f_1, f_2) - G(\Delta^\adv f_2, f_1) - 
\frac{1}{2} G(P f_1, f_2) - \frac{1}{2}G(P f_2, f_1) ,  
\end{multline}
where it should be
remembered that all quantities depend upon $s$.
This equation expresses $d^{(s)}(f_1, f_2)$ in terms of the advanced and retarded propagators for the 
metric $\g^{(s)}$, $G^{(s)}$, and initial data of $d^{(s)}$ on $S_-$. 
Now the retarded and advanced propagators have a smooth
dependence upon $s$ in the sense that
\ben
\WF(\Delta^{(s) \, \ret/ \adv}) \subset \{(s, \rho; x_1, k_1; x_2, k_2) \mid (x_1, k_1; x_2, k_2)
\in \cC_{\R/\A}(M, \g^{(s)})\}, 
\een
and $G^{(s)}$ is explicitly seen to be jointly smooth in $s$ 
and its spacetime arguments. 
Moreover, near $S_-$, $d^{(s)}$ is a smooth function independent of $s$, since $\omega^{(s)}$ 
is equal to the Hadamard state $\omega$ there. It follows from 
these facts, together with the expression eq.~\eqref{50} for $d^{(s)}$  
and the wave front set calculus, that $d^{(s)}$ 
is jointly smooth in $s$ and its spacetime arguments within $N$. 

We next analyze the $s$-derivative of $d^{(s)}$. We denote the variation
of any functional, $F$, of the metric by 
\ben
\delta F_\g ({\bf h}; f_1, \dots, f_m) = 
\frac{\partial}{\partial s} F_{\g^{(s)}}(f_1, \dots, f_m) \bigg|_{s=0}, \quad \frac{\partial}{\partial s} \g^{(s)} \bigg|_{s=0} = \h.  
\een
Now take the $s$ derivative of both sides of eq.~\eqref{50} at $s=0$. It follows that $\delta d$
can be written as a sum of terms involving $\delta \Delta^\adv$ and $\delta \Delta^\ret$, $\delta G$ and $\delta P$ 
(the variation of the KG-operator) linearly. The
wave front set of $\delta G$ can be computed explicitly and is given by 
\bena
\label{dj}
\WF(\delta G) 
&\subset& \{ (y, p; x_1, k_1; x_2, k_2) \mid \text{either of the following holds:}\nonumber\\
&& \text{($y=x_1$ and $p = -k_1$, $k_2 = 0$) or ($y=x_2$ and $p = -k_2$, $k_1 = 0$)}\nonumber\\
&& \text{or ($x_1 = x_2 = y$ and $p = -k_1 -k_2$)}\}, 
\eena
In order to calculate the wave front set of $\delta \Delta^\ret$ (and likewise $\delta \Delta^\adv$), we use 
formula~\eqref{41} (and an analogous formula for the advanced propagator), 
as well as the wave front set of the advanced resp. retarded propagator, bounded by
$\cC_{\A/\R}(M, \g)$. The calculus for the wave front set yields
\begin{multline}
\label{ddel}
\WF(\delta \Delta^{\adv/\ret}) = \{(y, p; x_1, k_1; x_2, k_2) \mid y \in J^{-/+}(x_1), x_2 \in J^{-/+}(y);\\
\text{$\exists (y, q_1), (y, q_2)$ such that $(y, q_i) \sim (x_i, -k_i)$, $p = q_1+q_2$}\}
\end{multline}
We now compute the wave front set of $\delta d$ by expressing it in terms of 
$\delta G$ and $\delta \Delta^{\adv /\ret}$ via the $s$-derivative of eq~\eqref{50}, 
and using the wave front set calculus. This gives the bound on the 
wave front set of $\delta d$, thus completing the proof of lemma~6.2.

To complete the proof of eq.~\eqref{wfsc''} for $I_1$, we estimate its wave front set using the 
calculus for the wave front set together with the estimates eq.~\eqref{twfs} for the wave front 
set of $w$, and the estimates on the wave front set of $\delta d$ provided in lemma~6.2. This gives
\begin{multline}
\WF(I_1) \subset \{(y, p; x_1, k_1; \dots; x_n, k_n) \mid \\
\exists (x_1, k_1; \dots; x_n, k_n; z_1, 0; \dots; z_i, q_i; 
\dots; z_j, q_j; \dots z_r, 0) \in \WF(w) \quad \text{such that}\\
(y, p; z_i, q_i; z_j, q_j) \in \WF(\delta d) \quad \text{for some $i,j$}\}\\
\subset \{
(y, p; x_1, k_1; \dots; x_n, k_n) \mid \exists (x_i, q_i) \quad \text{such that}\\
\text{$(x_i, q_i) \sim (y, -p)$ and $(x_1, k_1; \dots; x_i, k_i + q_i;  \dots; x_n, k_n) \in \cC_\T(M, \g)$}\\
\text{or $x_i = x_j = y$ and there exist $q_i, q_j \in T^*_y M$ such that}\\
\text{$p = -q_1-q_2$ and $(x_1, k_1; \dots; y, k_i + q_i;  \dots; y, k_j + q_j; \dots; x_n, k_n) \in 
\cC_\T(M, \g)$} 
\}
\end{multline}
One verifies thereby that 
$I_1$ satisfies the wave front set condition
eq.~\eqref{wfsc''}. The smooth resp. analytic dependence of $I_1$ upon the metric, eq.~\eqref{swfsc''}, 
can be proved in the same way by considering metrics that have in addition a smooth (analytic) dependence
upon a further parameter.

\medskip

\paragraph{Proof of lemma~6.1 for $I_2$:}

We next show that $I_2$ satisfies the wave front set condition
eq.~\eqref{wfsc''}. It was shown in our previous paper~\cite{hw2} that
any distribution that is locally and covariantly constructed from the
metric with a smooth dependence upon the metric and an almost 
homogeneous scaling behavior has a so called 
``scaling expansion''. 
This scaling expansion for $w$ takes the form
\ben
\label{scexp}
w_\g(x_1, \dots, x_{n+r}) = \sum_j (C^{a_1 \dots a_j}_\g \alpha_\g^* u_{a_1 \dots a_j})(x_1, \dots, x_{n+r}) + 
\rho_\g(x_1, \dots, x_{n+r}),  
\een
where $u$ are tensor valued, Lorentz invariant distributions on $(\mr^D)^{n+r-1}$ (we think of 
$\mr^D$ as being identified with the tangent space on $M$ at $x_1$), where $C$ are 
local curvature terms (evaluated at $x_1$) 
that are polynomials in the Riemann tensor and its derivatives, and 
where $\alpha_\g$ is the map
\ben
\alpha: U_{n+1} \owns
(x_1, x_2, \dots, x_{r+n}) \to (e^\mu(x_1, x_2), \dots, e^\mu(x_1, x_{r+n})) \in 
(\mr^D)^{n+r-1},  
\een
where $e^\mu(x, y)$ denotes the Riemannian normal coordinates $y^\mu$ of a point $y$ 
relative to a point $x$. The ``remainder'' $\rho_\g$ is a local, covariant distribution that depends smoothly upon the metric and satisfies 
the additional properties stated in thm.~4.1 of~\cite{hw2}. 
We refer the reader to thm.~4.1 of~\cite{hw2} for the construction and further properties 
of the scaling expansion.
To proceed, we split $I_2 = I_3 + I_4$ further into a contribution $I_3$ arising from the sum 
in our scaling expansion and a contribution $I_4$ arising from the remainder in that expansion. 
We analyze these separately and show that each of them satisfies the wave front set condition
eq.~\eqref{wfsc''}. 

We first analyze $I_3$, given by 
\begin{multline}
I_3(\h, f_1, \dots, f_n) = \sum_{j}
\int \delta(C^{a_1 \dots a_j} \alpha^* u_{a_1 \dots a_j}) (z, y_1, \dots, y_n, x_1, \dots, 
x_{r}) \\
\h(z) \prod_i f_i(y_i) \, \omega\left( : \prod_{j=1}^{r} \varphi(x_j) :_H \right).
\end{multline}
Since the distributions $u$ in the scaling expansion are actually independent of $\g$ (so that $\delta u = 0$), we 
have, dropping the tensor indices, 
\ben
\label{40o}
\WF [\delta (C \alpha^* u) ] \subset
\WF [(\delta C) \alpha^* u ]\cup
\WF [C (\delta \alpha)^* u ]. 
\een
Thus, in order to analyze the wave front set of $\delta (C \alpha^* u)$, we only need 
to analyze the variations $\delta C$ and $\delta \alpha$. 
But $C$ is just a polynomial in the
Riemann tensor and its derivatives, from which one finds
\ben
\WF(\delta C) \subset \{(y, p; x, k) \mid x = y, \quad k = -p\}.
\een
The wave front set of $\delta \alpha$ in turn follows from 
the wave front set of $\delta e^\mu$
(recall that $e^\mu$ is essentially the inverse of the exponential map), 
which in turn can be calculated to be
\bena
\WF(\delta e^\mu) 
&\subset& \{ (y, p; x_1, k_1; x_2, k_2) \mid \text{either of the following holds:}\nonumber\\
&& \text{($y=x_1$ and $p = -k_1$, $k_2 = 0$) or ($y=x_2$ and $p = -k_2$, $k_1 = 0$)}\nonumber\\
&& \text{or ($x_1 = x_2 = y$ and $p = -k_1 -k_2$)}\}. 
\eena
Using the calculus for the wave front set, we find that 
\ben
\label{40}
\WF [\delta (C \alpha^* u) ] \restriction_{\Delta_{n+r+1}} \perp
T(\Delta_{n+r+1}).
\een
Since $\omega$ is a Hadamard state, the distribution $\omega\left( :  \prod \varphi(x_j) :_H \right)$ is actually
a smooth function. Therefore, using again the calculus for the 
front set, and using
the fact that $C \alpha^* u$ has the same support as $t$ [see eq.~\eqref{tsupp}], 
we conclude that $I_3$ is a distribution jointly in $\h, f_1, \dots, f_n$, satisfying the wave front set
condition eq.~\eqref{wfsc''}. The smooth resp. analytic dependence of $I_3$ upon the metric, eq.~\eqref{swfsc''}, 
can be proved in a similar way by considering appropriate families of metrics, instead of the fixed metric, $\g$.

We finally turn our attention to the functional $I_4$, given by
\ben
\label{i4def}
I_4(\h, f_1, \dots, f_n) =
\int \delta \rho(z, y_1, \dots, y_n, x_1, \dots, 
x_r) \h(z) \prod_i f_i(y_i) \,\, \omega\left( : \prod_j \varphi(x_j)  :_H \right). 
\een
We need to show that $I_4$, in fact, defines a distribution 
on $U_{n+1}$, with the wave front property~\eqref{wfsc''}.
Since $\omega\left( : \prod \varphi(x_j) :_H \right)$ is a smooth function, the non-trivial contributions
to the wave front set of $I_4$ arise entirely from $\delta \rho$. 
The wave front set of $\delta \rho$ is analyzed as follows. By construction, 
$\delta \rho$ is already known to be a distribution on $U_{n+r+1}$
away from $\Delta_{n+r+1}$. Let us denote this 
distribution $\delta \rhoo$. It follows from the properties of 
the scaling expansion (cf. thm.~4.1 of~\cite{hw2}) that $\delta \rhoo$ has arbitrary low 
scaling degree at $\Delta_{n+r+1}$ (if the scaling expansion is carried out to sufficiently large order). 
By the arguments given in~\cite{hw2}, this entails that $\delta \rho$ arises from $\delta \rho^0$
by continuing the latter in a unique way to a distribution defined on all of $U_{n+r+1}$, in the sense that
\ben
\label{limeqn}
\delta \rho = \lim_{\lambda \to 0+} \theta_\lambda \delta \rhoo. 
\een
Here, $\theta_\lambda(y, x_1, \dots, x_n) = \theta(\lambda^{-1} S(y; x_1, \dots, x_n))$, where 
$S$ is any smooth function measuring the distance from the total diagonal, and $\theta$ is 
a any smooth, real valued function which vanishes in a neighborhood 
of the origin in $\mr$ and
which is equal to 1 outside a compact set. 
The key point is that we now can derive the
wave front set properties of $\delta \rho$ from the fact that it is the 
unique continuation of $\delta \rho^0$
together with the known properties of $\delta \rho^0$. The relevant 
properties of $\delta \rho^0$ are 
that\footnote{Note that it is essential that we know this property 
for an arbitrary smooth family $\g^{(s)}$ and not just a fixed metric.} 
\ben
\label{keyeqn}
\left\{ \text{closure of wave front set of $\delta \rho_{\g^{(s)}}^0$ in $T^*(\P \times M^{n+r+1})$} \right\}
\Big|_{\P \times \Delta_{n+r+1}} \perp T(\P \times \Delta_{n+r+1})
\een
where $\g^{(s)}$ is any family of metrics depending smoothly upon a parameter $s \in \P$, and
that $\delta \rho^0$ has a certain integral representation (see eqs.~(55)--(57) of \cite{hw2}) which 
can be derived from the fact that it is the remainder in a scaling expansion.
It follows from these properties (by an argument completely
analogous to the one given in the proof of~\cite[prop.~4.1, pp. 336]{hw2}) that 
\ben
\WF (\delta \rho) \restriction_{\Delta_{n+r+1}} \perp T(\Delta_{n+r+1}). 
\een
This estimate can be used to establish equation~\eqref{wfsc''} for $I_4$ by 
applying the wave front set calculus to the defining relation~\eqref{i4def} for $I_4$.
The smooth resp. analytic dependence of $I_4$
 upon the metric, eq.~\eqref{swfsc''},  can be shown by similar methods. This shows that $I_2$ satisfies 
relations eq.~\eqref{wfsc} and~\eqref{swfsc''}, and thereby concludes the proof of lemma~6.1.

\subsubsection{Proof that $D_n$ is symmetric when $\Phi_1 = T_{ab}$}

We now examine the symmetry properties of $D_n$. It is a straightforward consequence of the definition of 
$D_n$ together with the symmetry of the time-ordered 
products, T6, that $D_n(\h; f_1, \dots, f_n)$ is 
symmetric in $f_1, \dots, f_n$ when the fields $\Phi_1, \dots, \Phi_n$ are also exchanged accordingly.
However, the symmetry properties of $D_n$ with regard to exchanges of $\h$ with the $f_i$ are 
not at all manifest from the definition of $D_n$, as $\h$ appears on a completly different footing 
than the $f_i$. We here examine the symmetry properties of $D_n$ under such exchanges, which of course are relevant 
only when one of the fields $f_i \Phi_i$ is equal to the (densitized)
stress energy tensor, say $f_1 \Phi_1 = \eps h_1^{ab} T_{ab}$. 

We claim that the prescription for 
defining time ordered products can be modified (within the allowed freedom) so that the corresponding new $D_n'$ is 
symmetric in the sense that
\ben
\label{D'0}
D_n'(\h_2; \h_1, f_2, \dots, f_n) - D_n'(\h_1; \h_2, f_2, \dots, f_n) = 0. 
\een
We note that 
it is an immediate consequence of this equation, the definition of $D_n$ and the symmetry of the time ordered products, T6, 
that
\ben
\label{symmetryc}
D_n'(\h_1; \dots, \h_i, \dots, \h_j, \dots) = \text{symmetric in $\h_1, \dots, \h_i, \dots, \h_j, \dots$},  
\een
if $f_i \Phi_i = \eps h^{ab}_i T_{ab}, \dots, f_j \Phi_j = \eps h^{ab}_j T_{ab}$, i.e., if any number of the fields
are given by stress energy tensors.

To prove eq.~\eqref{D'0}, let us first consider the simplest case, $n=1$, for which the antisymmetric part of $D_1$ 
is given by\footnote{\label{27}
In this formula, and in other similar formulas below, we are assuming for simplicity that the metric variations
$\h_1$ and $\h_2$ commute, i.e., that $\h_{[1,2]} = \delta_1 \h_2 - \delta_2 \h_1 = 0$. For non-commuting variations, there would
appear the additional term $T_{ab}(\eps h_{[1,2]}^{ab})$ in the formula~\eqref{Edef} for 
$E$ (and corresponding other terms in similar other formulas 
below). An example of non-commuting variations is $\h_1 = \h$ and $\h_2 = \pounds_\xi \g$; in that case $\h_{[1,2]} = - 
\pounds_\xi \h$.} 
\bena
\label{Edef}
E(\h_1, \h_2) &\equiv& D_1(\h_1, \h_2) - D_1(\h_2, \h_1) \nonumber\\
&=& \delta^\ret_1 T_{ab}(\eps h^{ab}_2) - \delta^\ret_2 T_{cd}(\eps h^{cd}_1) + \frac{\i}{2}
[T_{ab}(\eps h^{ab}_1), T_{cd}(\eps h^{cd}_2)].
\eena
We already know, inductively, that $D_1$, and hence $E$, is a c-number distribution that is supported on the total diagonal in 
$M \times M$. Moreover, $E$ is also locally and covariantly constructed out of the metric and scales almost homogeneously 
(with degree $=$ dimension of spacetime) 
under a rescaling of the metric by a constant conformal factor, because $D_1$ has already been 
shown to have these properties. Finally, since $D_1$ satisfies the wave front set properties eqs.~\eqref{wfsc} and~\eqref{swfsc}, 
it follows by the same arguments as in~\cite{hw2} that $D_1$ (and hence $E$) must, in fact, be given by a delta function
times suitable curvature terms of the correct dimension, 
\ben
E(\h_1, \h_2) = \sum_r 
\int \eps[h_1^{ab} (\nabla_{(f_1} \cdots \nabla_{f_r)} h_2^{cd}) C_{abcd}{}^{f_1 \dots f_r} - (1 \leftrightarrow 2)], 
\een
where $C_{abcd}{}^{f_1 \dots f_r}$ are local curvature 
terms of dimension $D-r$. 

We now claim that $E=0$ for any prescription such 
that the quantum stress tensor is conserved, $\nabla^a T_{ab} = 0$. 
To see this, consider the variations 
$h_{ab}$ and $\pounds_\xi g_{ab}$, where $\xi^a$ is an arbitrary smooth, compactly supported
vector field and $h_{ab}$ an arbitrary smooth compactly supported
symmetric tensor field, i.e., choose one of the variations 
to be of pure gauge. 
Using stress energy conservation, and the remark in footnote~\ref{27}, 
one deduces $E(\h, \pounds_\xi \g) = 0$ for any such pair of variations. 
Substituting this into the above expression for $E$, one can show this implies that $E = 0$ by an argument 
similar to that given in the proof of thm.~5.1. 
But it follows from the analysis of section 3.2 above that when $D>2$ 
we can always
adjust our prescription for Wick powers and time ordered products
so as to satisfy $\nabla^a T_{ab} = 0$ in addition to T1--T11a.
Thus, if we take the ``prime'' 
prescription to satisfy conservation of the stress tensor, 
then eq.~\eqref{D'0} follows when $n=1$.

In order to prove eq.~\eqref{D'0} for $n>1$, we use the identity
\bena
\label{d''1}
D_n(\h_1; \h_2, \dots, f_n) - D_n(\h_2; \h_1, \dots, f_n) &=& \frac{2}{\i} (\delta^\ret_1 \delta_2^\ret - 
\delta^\ret_2 \delta_1^\ret) \T \left( \prod_{i=2}^n f_i \Phi_i \right) \nonumber \\
&+& E(\h_1, \h_2) \T \left( \prod_{i=2}^n f_i \Phi_i \right), 
\eena
which follows from our inductive assumptions and the definition of the retarded products by a calculation
similar to those given in the previous subsections. But we have already shown that $E = 0$, and we 
know that $\delta^\ret_1 \delta_2^\ret - \delta^\ret_2 \delta_1^\ret = 0$ from the definition of the retarded 
variation. Hence, the right side of eq.~\eqref{d''1} in fact vanishes, 
thus establishing eq.~\eqref{D'0} for $n>1$.

\paragraph{Remark concerning the cohomological nature of $E=0$, and of T11b:} 
There is an alternative strategy to prove the key identity $E=0$, 
which shows the cohomological nature
of that condition, and therefore---since it is a necessary 
condition for T11b to hold---that there can exist ``cohomological 
obstructions'' to imposing T11b. We define 
\bena
\widehat{\delta} &=& \delta^\ret + \frac{\i}{2}[T_{ab}(\eps h^{ab}), \, \cdot \, ] \nonumber \\
&=& \frac{1}{2} (\delta^\ret + \delta^\adv ) 
\eena
and we view $\widehat{\delta}$ as a ``gauge connection'' on local covariant
fields. Now apply $\widehat{\delta}_3$ to the 
defining relation~\eqref{Edef} for $E(\h_1, \h_2)$, and 
antisymmetrize over the different metric variations $1, 2$ and $3$. We 
obtain
\ben
\widehat{\delta}_{[1} E(\h_2, \h_{3]}) = \widehat{\delta}_{[1} \widehat{\delta}_{2} T_{ab}(\eps h_{3]}^{ab}) = 0, 
\een
where the second equality can be 
verified\footnote{We are assuming that the 
variations 1, 2 and 3 commute, see footnote~\ref{27}. If the variations do not commute, 
there would appear the additional terms $E(\h_1, \h_{[2, 3]}) + 
E(\h_2, \h_{[3, 1]}) + E(\h_3, \h_{[1, 2]})
$ on the left side.
}
by a direct calculation using the Jacobi identity (or alternatively can be viewed as the 
``Bianchi identity'' for the ``connection'' $\widehat{\delta}$, because $E$ is the ``curvature'' of $\widehat{\delta}$). 
Since $E$ is a c-number, the antisymmetrized $\widehat{\delta}$-variation 
of the left side of the equation is actually
equal to $\delta_{[1} E(\h_2, \h_{3]})$, where $\delta$ is the ordinary variation of a functional with respect to the metric. Hence
\ben
\label{curl1}
\delta_{[1} E(\h_2, \h_{3]}) = 0, 
\een
i.e., $E$ has vanishing ``curl''. 

In finite dimensions, every differential form with vanishing 
curl can be written 
as the curl of a form of lower degree, unless there is a 
topological obstruction. In the present case, the key issue is whether
it is possible to
write $E$ as the curl of some $F$, i.e.
\ben
\label{curl}
E(\h_1, \h_2) = \delta_{[1} F(\h_{2]}) 
\een
for some functional 
\ben
F(\h) = \int C^{ab} h_{ab} \eps, 
\een
where $C^{ab}$ is a local curvature term (of the appropriate dimension). 
The point is that, if $E$ could indeed be written in this way, 
and if we could then redefine our prescription for the stress energy 
tensor by ${T'}_{ab} = T_{ab} - C_{ab} \myid$, then the new prescription 
would satisfy $E'=0$ (as well as $\nabla^a {T'}_{ab} = 0$). 
Alternatively, if it is not always possible to write any $E$ 
satisfying eq.~\eqref{curl1} in the form
\eqref{curl}---i.e., if the space of functionals of the metric 
of this type has a non-trivial cohomology
with respect to the differential $\delta$---then if such an $E$ 
arises in eq.~\eqref{Edef} in a quantum field theory, it is clear that
there would be no way consistent with axioms T1-T10 to adjust the 
prescription for defining time-ordered products
so as to make $E$ vanish. Consequently, 
by the arguments given above, it would not be possible to have a conserved
stress-energy tensor in such a quantum field theory, i.e., 
the theory would have a ``gravitational anomaly''. As we have seen above, this 
``cohomological obstruction''
does not occur for the theory of a scalar field, but it could occur 
for quantum field theories containing fields of higher spin. 

In field theories (such as 
scalar field theory) that are invariant under parity, $\eps \to -\eps$, it 
follows that $E$ must transform as\footnote{The minus sign arises simply
because an integration is implicit in the definition of $E$. The integrand
of such an $E$ would be parity invariant.} $E \to -E$. 
We are not aware
of any  $E$ with this transformation property which has vanishing
curl but cannot be written as the curl of some $F$. 
If this could be proven, then this would provide a general proof
that we can satisfy $E=0$ in field theories preserving parity (assuming
that there are no algebraic restrictions on $T_{ab}$). 
However, nontrivial cocycles 
$E$ can occur when parity invariance
is dropped. An example in $D=2$ spacetime dimensions is 
\ben
E_{D=2}(\h_1, \h_2) = \int \bigg[
R \epsilon^{ab} h_{1 \, ac} h_{2 \, b}{}^c + 2\epsilon^{ab} \nabla_c (h_{1 \, a}{}^c 
- \delta_a{}^c h_{1 \, m}{}^m) \nabla_d (h_{2 \, b}{}^d 
- \delta_b{}^d h_{2 \, n}{}^n)
\bigg] \eps \, .
\een
We have checked explicitly that 
$E_{D=2}$ has vanishing curl. However, $E_{D=2}$ is not the curl of 
some $F_{D=2}$, as 
can easily be seen from the fact that, in $D=2$ dimensions, 
the only functional $F_{D=2}$ with the appropriate dimension of length 
is, up to a numerical factor, 
$F_{D=2}(\h) = \int R^{ab} h_{ab} \, \eps$. But $F_{D=2}$ transforms
as $F_{D=2} \to - F_{D=2}$ under parity, while $E_{D=2} \to + E_{D=2}$, so its
antisymmetrized variation cannot be proportional 
to $E_{D=2}$. 

This example explicitly
shows that non-trivial cocycles $E$ can be present, in principle, 
in parity violating theories, at least in $D=2$ dimensions. 
In fact, as we have previously noted, 
gravitational anomalies are known to occur~\cite{aw} for certain parity
violating theories in $D=4k +2$ dimensions.

\subsubsection[Proof that $D_n$ can be absorbed in a redefinition of the \\
time-ordered products]{Proof that $D_n$ can be absorbed in a redefinition of the time ordered products}

We now complete our inductive argument by showing
how to redefine our prescription for the time ordered products so that 
$D'_n = 0$ for the new prescription when $N_\varphi \le k$ factors 
of $\varphi$ are present in 
$f_1 \Phi_1, \dots, f_n \Phi_n$. To do this, we first collect the facts about $D_n$ which we 
have established in the previous subsections, and we summarize the conclusions that can be 
can be drawn from them about the nature of the $D_n$ 

By its very definition, we know that for any choice of the fields $\Phi_i$, 
$D_n(\h; f_1, \dots, f_n)$ is an 
$(n+1)$-times multilinear functional valued in $\W$. 
We showed that the values of this functional are 
actually proportional to the identity operator, allowing us to identify $D_n$ with a functional 
taking values in the complex numbers. This functional is supported only on the total diagonal in $M^{n+1}$, 
i.e., it vanishes if the supports of $\h, f_1, \dots, f_n$ have no common points. 
We also established that the functional $D_n$ depends locally
and covariantly on the metric, and that it has an almost homogeneous scaling behavior under
rescalings of the metric $\g \to \lambda^2 \g$, with $\lambda$ a real constant.
Furthermore, since the multilinear functional $D_n$ has been shown to be in fact a distribution 
(i.e., to be {\em continuous} in the appropriate sense) satisfying the wave front 
set condition~\eqref{wfsc}, it follows that 
\ben\label{25}
D(\h;  
f_1, \dots, f_n) = \sum_{\alpha_1, \dots, \alpha_n} \int_M \eps h_{ab} C^{ab}{}_{\alpha_1 
\dots \alpha_n } \prod_{i=1}^n \eps^{-1} (\nabla)^{\alpha_i} f_i
\cdot \myid, 
\een 
where the $C$ are smooth tensor fields
depending locally and covariantly on the metric, with a suitable almost homogeneous 
scaling behavior. Moreover, since we know that the dependence of $D_n$ on the 
metric is actually smooth resp. analytic in the sense of eq.~\eqref{wfsc'}, it follows by the same
arguments as in~\cite{hw1} that the $C$ have to be a polynomials in 
the metric, its inverse, the Riemann tensor and its derivatives. 
The engineering dimension of the derivatives and curvature monomials in each term in eq.~\eqref{25}
must add up precisely to $d-nD$ where $d$ is the sum of the engineering dimensions of the $\Phi_i$. (Here, it should be noted that
the delta function implicit in eq.~\eqref{25}
has engineering dimension $-nD= n \times \text{engineering
dimension of $\eps^{-1}$ in $D$ spacetime dimensions}$.) 
It follows from the unitarity condition on the time ordered products, T7, together with 
the fact that $\tau^\ret(a^*) = [\tau^\ret(a)]^*$ that the $D_n$ are distributions
satisfying the reality condition $\bar D_n = (-1)^{n+1} D_n$. The $D_n$ also satisfy the symmetry 
condition~\eqref{symmetryc} when one or more of the fields $\Phi_i$ is given by a 
stress energy tensor. Finally, we have shown that,  
when one of the fields $\Phi_i$ is equal to $\varphi$, we automatically have that $D_n = 0$.  

Our proposal for redefining the prescription for time ordered products at the given induction
order is now the following: If $\Phi_1, \dots, \Phi_n$ are fields in $\V$ with 
a total number of $N_\varphi = k$ factors of $\varphi$, then we define 
\ben
\label{cdef}
c\left[
\varphi \nabla_a \nabla_b \varphi \otimes (\otimes_{i=1}^n \Phi_i)
\right] = 2 \i \left( D_{n\,}{}_{ab} - \frac{1}{D-2} g_{ab} D_{n \,}{}^c{}_c \right). 
\een
We also define distributions $c[(\nabla)^r (\varphi \nabla_a \nabla_b \varphi) \otimes (\otimes_i \Phi_i)]$ 
associated with all Leibniz dependent expressions in such a way that eq.~\eqref{leib3} is satisfied. 
and we define $c[\Psi \otimes (\otimes_i \Phi_i)] = 0$ for all $\Psi$ which are ``Leibniz independent'' of 
$\varphi \nabla_a \nabla_b \varphi$ in the sense used in proposition~3.1. 
It is a direct consequence of these definitions that 
\ben
\label{cdef1}
c\left[T_{ab} \otimes (\otimes_{i=1}^n \Phi_i)
\right] = 2 \i D_{n\,}{}_{ab}. 
\een
Because of the symmetry condition~\eqref{symmetryc} satisfied by $D_n$, it follows that 
$c[T_{ab} \otimes \dots  T_{cd} \otimes \dots]$ is symmetric in the respective 
spacetime arguments if one (or more) of the fields $\Phi_i$ is given by a stress tensor. 
Since the $c$ satisfy the Leibniz rule in the first argument by construction, and since
they satisfy the Leibniz rule in the remaining $n$ arguments as a consequence of T10, an analogous
statement also holds by definition for derivatives of the stress tensor. It follows that 
$c$ satisfy the symmetry condition~\eqref{scond}, and the Leibniz condition~\eqref{leib3}.

It is now clear from the properties that we have established 
about the $D_n$ that the so defined coefficients $c$ obey all further restrictions 
that are necessary in order that the new prescrition $\T'$ satisfies T1--T10 and T11a: Since $D_n$ is 
of the form~\eqref{25}, the $c$ are similarly local covariant delta function type distributions with coefficients that are given 
by local curvature terms of the appropriate dimension. The $c$ satisfy the unitarity
constraint eq.~\eqref{ucond} because the $D_n$ satisfy the analogous relation, and the $c$
satisfy the constraint~\eqref{11a}, because we showed that 
$D_n = 0$ when one of the $\Phi_i$ is equal to $\varphi$. 

On account of the formula~\eqref{tabdef} for the free stress tensor $T_{ab}$, the changes 
in the time ordered products corresponding to the $c$ given in eq.~\eqref{cdef} 
via eqs.~\eqref{unique1} and~\eqref{unique2} take the following form
for time ordered products with one factor of $T_{ab}$ and  $n$ factors $\Phi_1, \dots, \Phi_n$ with 
$N_\varphi = k$ factors of $\varphi$:
\ben
\label{chpres}
\T'\left(\eps h_{ab}T^{ab}\prod_{i=1}^n f_i \Phi_i\right) = 
\T \left(\eps h_{ab}T^{ab}\prod_{i=1}^n f_i \Phi_i\right) 
+ 2\i D_n^{ab}(h_{ab}; f_1, \dots, f_n) \cdot \myid
\een
It follows from this relation that the new prescription $\T'$ is 
designed so that $D_n' = 0$ for all $\Phi_1, \dots, \Phi_n$
such that $N_\varphi \le k$. Hence, T11b holds for the new prescription
at the desired order in the induction process. 
This completes the proof that when $D>2$, we can satisfy condition T11b
in addition to conditions T1-T10 and T11a.

\paragraph{Remark:} In $D=2$ spacetime dimensions, 
we cannot define coefficients $c$ by eq.~\eqref{cdef} (because of the factor of $D-2$ in 
the denominator), unless $D_n$ already happens to vanish, in which case there would of course be nothing to show in the
first place. However, $D_n$ is explicitly seen to be nonzero already 
for $n=1$ and $\Phi_1 = \varphi^2$ in $D=2$
spacetime dimensions in the local normal ordering prescription, 
and our previous arguments show that it cannot be made to vanish.

\section{Outlook}

In this paper, we have proposed two new conditions, T10 and T11, that
we argued should be imposed on the definition of Wick polynomials and
time-ordered products in the theory of a quantum scalar field in
curved spacetime. These conditions supplement our previous conditions
T1-T9, and place significant additional restrictions on the definition
of Wick polynomials and time-ordered products that involve derivatives
of the field. We also showed that conditions T1-T10 and T11a can
always be consistently imposed, and in spacetimes of dimension $D>2$,
condition T11b also can be imposed. In addition, we proved that if
these conditions are imposed on the definition of Wick polynomials and
time-ordered products of the free field, then for an {\em arbitrary}
interaction Lagrangian, $L_1$, the perturbatively defined stress-energy
tensor of the interacting field will be conserved. We do not believe
that there are any further natural conditions that should be imposed
on the definition of Wick polynomials and time-ordered products for a
quantum scalar field in curved spacetime. If so, axioms T1-T11
together with the existence proofs and uniqueness analyses of this
paper and our previous papers essentially complete the perturbative
formulation of interacting quantum field theory in curved spacetime
for a scalar field with an arbitrary interaction Lagrangian.

For quantum fermion fields in curved spacetime, one can define a ``canonical
anti-commutation algebra'' in direct analogy to the canonical commutation
algebra ${\mathcal A} (M, \g)$ defined at the beginning of subsection 2.1.
The next step toward the formulation of the theory of interacting fermion 
fields in curved spacetime would be to define the 
fermionic analog of the algebra
$\W$ and to formulate suitable 
fermionic analogs of our axioms T1-T11. We do not
anticipate that any major difficulties would arise in carrying out these
steps, although we have not yet attempted to do so ourselves. We also
would expect it to be possible to prove existence and uniqueness
results for the fermion case
in close parallel to the scalar field case. Indeed, the only
place in our entire analysis where it is clear that differences can arise
is the analysis of obstructions
to the implementation of condition T11b. As previously noted, the analysis
of \cite{aw} establishes that the analog of condition T11b cannot hold
for certain parity violating theories in spacetimes of dimension $4k+2$.

To define the quantum theory of Yang-Mills fields in curved spacetime,
one would presumably start, as in flat spacetime, by ``gauge fixing''
and introducing ``ghost fields''. However, to proceed further in the
spirit of our approach, one would have to formulate the theory
entirely within the algebraic framework, including procedures for
extracting gauge invariant information from the field algebra. Since
many subtleties already arise in the usual treatments of Yang-Mills fields
in flat spacetime due to local gauge invariance, 
we do not anticipate that it will be straightfoward
to extend our analysis to the Yang-Mills case. We expect that it would 
be even less straightforward to extend our analysis to
a perturbative treatment of quantum gravity itself off of an arbitrary
globally hyperbolic, classical solution to Einstein's equation, although 
we also do not see any obvious reasons why this could not be done.

Returning to the case of a scalar field, 
there remain some significant unresolved 
issues even if the renormalization theory as 
presently formulated turns out to be essentially
complete. One such issue concerns the probability interpretation of the
theory. As emphasized, e.g., in \cite{wald}, there is no meaningful notion
of ``particle''---even asymptotically---in a general
curved spacetime. Thus, the only meaningful observables are the smeared
local and covariant quantum fields themselves. Let $\Phi(f) \in \W$ be such 
a field observable for the free scalar field $\varphi$, which is 
``self-adjoint'' in the sense that $\Phi(f)^* = \Phi(f)$. For any state,
$\omega$, the very definition of $\omega$ provides one
with the expectation value,
$\langle\Phi(f)\rangle = \omega(\Phi(f))$, of this observable in 
the state $\omega$. We also can
directly obtain the moments, $\omega([\Phi(f) - \langle\Phi(f)\rangle]^n)$,
of the probability distribution for measurements of $\Phi(f)$ in the 
state $\omega$, since powers of $\Phi(f)$ are also in $\W$. However, to
obtain the probability distribution itself, we need to go to a Hilbert
space representation, such as the GNS representation, where $\omega$
is represented by an ordinary vector in a Hilbert space, and $\Phi(f)$
is represented by an operator $\pi[\Phi(f)]$, so that probabilities can be
calculated by the usual Hilbert space methods. However, a 
potential difficulty arises here.
Although in the GNS representation $\pi[\Phi(f)]$ is automatically
a symmetric operator defined on a dense, invariant domain $\mathcal{D}$,
there does not appear to be any guarantee that $\pi[\Phi(f)]$ will be
essentially self-adjoint on $\mathcal{D}$. If essential self-adjointness
fails, then further input would be needed to obtain a probability 
distribution. Specifically, if $\pi[\Phi(f)]$ has more than
one self-adjoint extension, then additional rules would have to be found
to determine which self-adjoint extension should be used to define the
probability distribution. Worse yet, if $\pi[\Phi(f)]$ does not admit
any self-adjoint extension
at all, it is hard to see how any consistent probability
rules can be given. As far as we are aware, this issue is unresolved for
general observables in $\W$ even for the vacuum state in Minkowski spacetime.

Another issue of interest that has not yet been investigated in depth
concerns whether a useful, non-perturbative, axiomatic characterization of
interacting quantum field theory in curved spacetime can be given. The usual
axiomatic formulations of quantum field theory in Minkowski spacetime,
such as the Wightman axioms~\cite{sw}, make use of properties that are
very special to Minkowski spacetime. It seems clear that a suitable 
replacement for the Minkowski spacetime assumption of covariance of the
quantum fields under Poincare transformations is the condition that
the quantum fields be local and covariant \cite{hw1, bfv}.
It also seems clear that microlocal spectral conditions should provide
a suitable replacement for the usual spectral condition assumptions in 
Minkowski spacetime. However, it is far less clear what should replace
the Minkowski spacetime assumption of the existence of a unique, Poincare
invariant vacuum state, since no analog of this property exists in 
curved spacetime. One possibility for such a replacement might be
suitable assumptions concerning the existence and properties of an
operator product expansion~\cite{h}.

Undoubtedly, the foremost unresolved issue with regard to the
perturbative formulation of quantum field theory in curved spacetime
concerns the meaning and convergence properties of the Bogoliubov
formula, eq.~(\ref{pertseries}), which defines the interacting
field. It is, of course, very well known that ``perturbation theory in
quantum field theory does not converge''. However, as we pointed out
in \cite{hw3}, the usual results and arguments against convergence concern
the calculation of quantities that involve ground states or ``in'' and
``out'' states, and such states would not be expected to have the
required analyticity properties.  We believe that
eq.~(\ref{pertseries}) stands the best chance of making well defined
mathematical sense if it is interpreted as determining the algebraic
relations that hold in the interacting field algebra. The formula does
not, of course, make sense as it stands (except as a formal power
series) since we have not defined a topology on $\W$---so the notion
of ``convergence'' has not even been defined---and, in any case, $\W$
should be ``too small'' to contain the elements of the interacting
field algebra, since $\W$ consists only of {\em polynomial} expressions in
$\varphi$ smeared with appropriate distributions.  However, one could
imagine ``enlarging'' $\W$ by defining a suitable topological
algebra $\bar{\W}$ into which $\W$ is densely embedded. We see no obvious
reason why such a $\bar{\W}$ could not be defined so that 
eq.~(\ref{pertseries}) would define a convergent series in $\bar{\W}$---but,
of course, we also do not see an obvious way of carrying this out! These ideas
appear to be worthy of further investigation.

\medskip

\noindent

{\bf Acknowledgements:} We would like to thank M.~D\"utsch and K.~Fredenhagen for discussions
and for making available to us their manuscript on the Action Ward Identity~\cite{df2} prior
to publication. S. Hollands would like to thank the II.~Institut f\"ur Theoretische Physik, 
Universit\"at Hamburg, for their kind hospitality. 
This work was supported by NFS grant PHY00-90138 to the University of Chicago.

\appendix

\section{Infinitesimal retarded variations}

Let $\g^{(s)}$ be a smooth 1-parameter family of metrics differing 
from $\g\equiv\g^{(0)}$ only within a compact subset $K$.
In this appendix, we show that the retarded variation with respect to the metric defined by 
\ben
\label{a1}
\delta^\ret_\g\left[
\T_\g\left(\prod_{i=1}^n f_i \Phi_i\right)
\right] = 
\frac{\partial}{\partial s}
\tau^\ret_{\g^{(s)}}
\left[ \T_{\g^{(s)}}\left(\prod_{i=1}^n f_i \Phi_i\right) \right] \Bigg|_{s=0}
\een
appearing in our requirement T11b is well-defined and yields an 
element of $\W(M, \g)$. Our proof can easily be generalized to also prove the 
corresponding statement for an infinitesimal retarded variation of the 
potential, T11c, but we shall not treat this case explicitly.

Let $A^{\ret \, (s)}$ be the 
map as defined by eq.~(\ref{a7}) above, let $\omega_2$ be the two-point
function of a Hadamard state on $(M,\g)$ and let
$\omega_2^{(s)}$ be the Hadamard 2-point functions for $(M, \g^{(s)})$,
uniquely specified by the
requirement that $\omega_2^{(s)}$ coincides with $\omega_2$ when both 
arguments are taken
within $M \setminus J^+(K)$. Then it can be verified that $A^{\ret \, (s)}$ 
and $\omega_2^{(s)}$ have a 
smooth dependence upon $s$ in the sense that, when viewed as 
distributions jointly in $s, x_1, x_2$, we have
\begin{multline}
\label{a9}
\WF(A^{\ret \, (s)}) \subset \{(s, \rho; x_1, k_1; x_2, -k_2) \mid 
\text{$\exists y \in M \setminus J^+(K)$ and $(y, p) \in T^*_y M \setminus \{0\}$}\\
\text{such that $(x_1, k_1) \sim (y, p)$ with respect to $\g$ and}\\
\text{such that $(x_2, k_2) \sim (y, p)$ with respect to $\g^{(s)}$}\},
\end{multline}
as well as eq.~\eqref{os}. Since the definition of $\W$ does not 
depend upon the choice of quasifree Hadamard state used in the 
definition of the
generators $W_n$, we can assume without loss of generality that 
the generators $W^{(s)}$ of the algebra
$\W(M, \g^{(s)})$ are defined using the particular 1-parameter family of states 
$\omega^{(s)}$ that we have just described. To compute the action of $\tau^\ret$
on the time ordered products, we recall that the time ordered
products have the following ``(global) Wick expansion,''
\begin{multline}
\label{a9.5}
\T\left( \prod_{i=1}^n f_i \Phi_i \right) = 
\sum_{\alpha_1, \dots, \alpha_n}
\frac{1}{\alpha_1! \cdots \alpha_n!} \int \omega \left[
\T\left(  \prod_{i=1}^n \delta^{\alpha_i}\Phi_i(y_i) \right)
\right] \prod_i f_i(y_i) \cdot \\
\cdot : \prod_i \prod_{j} [(\nabla)^j \varphi(y_i)]^{\alpha_{ij}} :_\omega 
\end{multline}
where we are using the same notation as in the local Wick 
expansion\footnote{Note that
this expansion is entirely analogous to the local Wick 
expansion~\eqref{wickexp}. The only difference is 
that in the local Wick expansion, the time ordered products are 
expanded in terms of the 
{\em local normal ordered products}~\eqref{hndef}, 
while we are using the normal ordered products 
with respect to $\omega_2$ in eq.~\eqref{a10}. The latter are 
globally defined on all of 
$M^n$, whereas the former are only defined in a 
neighborhood $U_n$ of the total diagonal (but, in
contrast to the normal ordered products in eq.~\eqref{a10}, 
the former depend locally and covariantly 
on the metric).} given in~\eqref{wickexp}. Inserting suitable $\delta$-distributions, we can rewrite the Wick expansion
in the form
\bena
\label{a10}
\T \left( \prod f_i \Phi_i \right) &=&
\sum_n \int u_n(y_1, \dots, y_m; x_1, \dots, x_n) \,\, :\prod_i^n 
\varphi(x_i) :_{\omega}
\prod_j f_j(y_j)
\nonumber\\
&=& \sum_n W_n \bigg( u_n(\otimes_i f_i) \bigg),
\eena
where the distributions $u_n$ are defined in terms of 
the $\omega[\T(\prod \delta^{\alpha_i} \Phi_i(y_i))]$ with 
$\sum |\alpha_i| = n$, together
with suitable derivatives of delta functions. On account of 
the delta functions, 
the $u_n$ have the same support as the distributions $w$ in eq.~\eqref{tsupp},
and they  satisfy the same wave front set condition as in eqs.~\eqref{twfs}. 
Furthermore, if we repeat the above steps for our family of metrics $\g^{(s)}$ 
instead of the single metric $\g$ (with $\omega_2$ replaced by $\omega^{(s)}_2$ 
everywhere) then we find that the corresponding distributions $u^{(s)}_n$ 
have a smooth dependence upon $s$, i.e., that they satisfy the same wave front 
set condition as in  eq.~\eqref{tswfs}. By the wave front set calculus, we conclude that, 
for any $n$, and for any fixed choice of smooth compactly supported functions $f_i$, the quantity
$u^{(s)}_n(\otimes_i f_i; x_1, \dots, x_n)$ is indeed a distribution in the variables 
$x_1, \dots, x_n$ belonging to the space $\E'_n(M, \g^{(s)})$. 
Moreover, it follows from the smoothness property in 
$s$ of the $u^{(s)}_n$ that these distributions actually have a smooth dependence upon 
$s$ in the sense that, when viewed as
a distribution jointly in $s, x_1, \dots, x_n$, we have
\begin{multline}
\label{a11}
\WF\bigg( u_n^{(s)}(\otimes_i f_i) \bigg) 
\subset \{(s, \rho; x_1, k_1; \dots; x_n, k_n) \mid \\
(x_1, k_1; \dots; x_n, k_n) \notin [(V^{(s) \, +})^n \cup (V^{(s) \, -})^n] \setminus \{0\} \},
\end{multline}
where $V^{(s) \, +/-}$ are the future/past lightcones associated with the 
metrics $\g^{(s)}$.

Substituting eq.~\eqref{a10} into the the definition of $\tau^\ret_{\g^{(s)}}$, we find
\begin{equation}
\label{a12}
\tau^\ret_{\g^{(s)}} \left[ \T_{\g^{(s)}} \left( \prod_i f_i \Phi_i \right) \right] = 
\sum_n W_n
\bigg(
(A^{\ret \, (s)})^{\otimes n} \left[ u_n^{(s)} (\otimes_i f_i) \right]
\bigg) 
= \sum_n W_n(v_n^{(s)}),
\end{equation}
where the distributions $v_n^{(s)}$ are the elements in the space $\E'_n(M, 
\g)$ defined
by the last equation. It follows from the calculus for wave front sets 
together with the wave front 
set of $A^{\ret \, (s)}$ [see eq.~\eqref{a9}], and the wave front property of 
$u^{(s)}_n$ [see eq.~\eqref{a11}] that $v_n^{(s)}$ are 
distributions depending
smoothly upon $s$ in the sense that, when viewed as distributions jointly in
$s, x_1, \dots, x_n$, we have
\begin{multline}
\label{a13}
\WF( v_n^{(s)} ) 
\subset \{(s, \rho; x_1, k_1; \dots; x_n, k_n) \mid \\
(x_1, k_1; \dots; x_n, k_n) \notin [(V^{(s) \, +})^n \cup (V^{(s) \, -})^n] \setminus \{0\} \}.
\end{multline}
It then
follows that the differentiated functionals $\frac{\partial}{\partial s} 
v^{(s)}_n |_{s=0}$ are in fact well-defined distributions in the class 
$\E'_n(M, \g)$. Thus, expression~\eqref{a1} exists as the well-defined 
algebra element $\sum W_n(\frac{\partial}{\partial s} v^{(s)}_n |_{s=0})$. 
This is what we wanted to show.

\section{Functional derivatives}

In this appendix we define the functional derivatives, $\delta
A/\delta \varphi$ and $\delta A/ \delta g_{ab}$, of any
element $A$ of the space $\F$. We shall elucidate the
calculus of the functional derivative operations and, in particular,
we will derive eqs.~\eqref{DD} and~\eqref{LiefPhi}, which were used in
the proof of theorem 5.1 and in section~6.

Let $A \in \F$. Then $A$ is a $D$-form that is locally constructed out of $\g$,
the curvature, finitely many symmetrized derivatives of the curvature,
$\varphi$, finitely many symmetrized derivatives of $\varphi$, and
test tensor fields $f$ and their symmetrized derivatives. We will
denote these dependences as simply $A=A[\g,\varphi,f]$.  The
functional derivative of $A$ with respect to $\varphi$ is defined by
\ben 
\frac{\partial}{\partial s} A[\g, \varphi + s \psi, f]
\Bigg|_{s=0} = \psi \frac{\delta A}{\delta \varphi} + \d B_\varphi[\g,\varphi,f,
\psi] ,
\label{funder}
\een 
where $B_\varphi$ is a $(D-1)$-form that is similarly locally
constructed out of $\g$, $\varphi$, $f$, and $\psi$. The 
deomposition of the
right side of eq.~(\ref{funder}) into the two terms written there
is uniquely determined by 
the requirements
that (1) no derivatives of $\psi$ appear in the first term and (2) the
second term is exact. The manipulations leading to this decomposition
are the familiar ones that would be
used to derive the Euler-Lagrange equations
if $A$ were a Lagrangian; these manipulations are usually done under an 
integral sign, with the ``boundary term'', $\d B_\varphi$, discarded. 
An explicit formula for $\delta A/\delta \varphi$ was given in
eq.~\eqref{30} above. It is worth noting that if $A$ is an exact form,
i.e., $A = \d C$ for some $C = C[\g,\varphi,f]$, then its functional derivative
vanishes, since clearly eq.~(\ref{funder}) holds with 
$\delta A/\delta \varphi = 0$ and $B_\varphi = 
(\partial/\partial s) C[\g, \varphi + s \psi, f]|_{s=0}$.

Similarly,
the functional derivative of $A$ with respect to $g_{ab}$ is defined by
\ben
\frac{\partial}{\partial s} A[\g + s \h, \varphi, f]
\Bigg|_{s=0} =
h_{ab} \frac{\delta A}{\delta g_{ab}} + \d B_\g [\g,\varphi,f, \h].
\label{funderg}
\een
We can obtain an explicit expression for $\delta A/\delta g_{ab}$
by introducing an arbitrary fixed, background derivative operator,
$\nablao_a$, on $M$, and re-writing $\nabla_a$ and the curvature in terms
of $\nablao_a$ and derivatives of $\g$ with respect to $\nablao_a$.
The resulting explicit formula for $\delta A/\delta g_{ab}$ was given in
eq.~\eqref{delAdelg} above.

Our first result is that functional derivatives with respect to 
$\varphi$ and $\g$ commute modulo exact forms in the sense that
\ben
h_{ab} \frac{\delta}{\delta g_{ab}} \left(\psi \frac{\delta A}{\delta \varphi}
\right) = \psi \frac{\delta}{\delta \varphi} \left(h_{ab} \frac{\delta A}{\delta g_{ab}}\right) + \d B
\label{comder}
\een
for some $(D-1)$-form, $B$, that is locally
constructed out of $\g$, $\varphi$, $f$, $\psi$ and $\h$. 
To prove this, we note that
\bena
\frac{\partial^2}{\partial s \partial t} A[\g + t \h, \varphi + s \psi, f]
\Bigg|_{t=s=0} &=& \frac{\partial}{\partial s} \left(h_{ab} \frac{\delta A}{\delta g_{ab}} + \d B_\g \right) \Bigg|_{s=0} \nonumber \\
&=& \frac{\partial}{\partial s} \left(h_{ab} \frac{\delta A}{\delta g_{ab}}\right) \Bigg|_{s=0} + \d C_\g \nonumber \\
&=& \psi \frac{\delta}{\delta \varphi} \left(h_{ab} \frac{\delta A}{\delta g_{ab}}\right) + \d C_\g + \d B_\varphi
\label{comder2}
\eena
where $C_\g = (\partial/\partial s) B_\g[\g,\phi + s \psi,f,\h] |_{s=0}$.
By equality of mixed partials, we may reverse the order of 
differentiation with respect to $s$ and $t$ on the left side of
eq.~(\ref{comder2}). However,
$\partial^2 A/\partial t \partial s$ is given by a similar expression with
the order of the functional derivatives reversed. This establishes
eq.~(\ref{comder}).

Let us now prove the relation
\ben\label{b-0.5}
(\pounds_\xi g_{ab}) \frac{\delta A}{\delta g_{ab}} +
(\pounds_\xi \varphi) \frac{\delta A}{\delta \varphi} +
(\pounds_\xi f) \frac{\delta A}{\delta f} = \d H.
\een
where $\delta A/\delta f$ is defined by analogy with eqs.~(\ref{funder}) and 
(\ref{funderg}) and is given by an explicit formula analogous to
eq.~\eqref{30}.
This equation is equivalent to eq.~\eqref{LiefPhi} when $A$
depends linearly upon $f$.
Let $F_s$ be the one-parameter family of diffeomorphisms of $M$ generated
by a smooth, compactly supported vector field $\xi^a$.
Since $A$ is locally and covariantly constructed
from $\g, \varphi, f$, we have
\ben
F_s^* A[\g, \varphi, f] = A[F_s^* \g, F^*_s \varphi, F^*_s f],
\een
where $F^*_s$ denotes the pull-back of a tensor field. We differentiate
this equation at $s=0$,
and use the fact that the Lie-derivative of any $D$-form $A$ is given by
$\pounds_\xi A = \d (\xi \cdot A)$,
where $\xi \cdot A$ is the $(D-1)$-form obtained by contracting the index
of the vector field into
the first index of the form. We obtain
\ben
\d( \xi \cdot A[\g, \varphi, f]) = \frac{\partial}{\partial s} A[\g + s
\pounds_\xi \g, \varphi, f] \bigg|_{s=0}
  +\frac{\partial}{\partial s} A[\g, \varphi + s\pounds_\xi \varphi, f]
\bigg|_{s=0}
   +\frac{\partial}{\partial s} A[\g, \varphi, f + s \pounds_\xi f]
\bigg|_{s=0}.
\een
By eq.~\eqref{funderg} (with $h_{ab} = \pounds_\xi g_{ab}$ in that equation),
the first term on the right side is equal to $\pounds_\xi g_{ab} \cdot
\delta A/\delta g_{ab}$ up to some exact form $\d B_\g$. Similarly, the second
term is equal to
$\pounds_\xi \varphi \cdot \delta A/\delta \varphi$, up to some exact from
$\d B_\varphi$. Finally,
the last term is given by $\pounds_\xi f \cdot \delta A/\delta f$ 
plus some $\d B_f$. Thus, we get
eq.~\eqref{b-0.5}, with $H = \xi \cdot A - B_\g - B_\varphi - B_f$.

Finally, we prove the relation
\ben
\label{b0}
(D_\eta D_\xi - D_\xi D_\eta) A = D_{[\xi, \eta]} A + \d C,
\een
for some locally constructed $(D-1)$ form $C$,
where the variational operation $D_\xi$ is defined by
\ben
D_\xi A = \pounds_\xi g_{ab} \cdot \delta A/\delta g_{ab} \,. 
\een
According to
eq.~\eqref{funderg}, we may write
\ben
\label{b1}
D_\xi A[\g] = \frac{\partial}{\partial s} A[\g + s \pounds_\xi \g]
\Bigg|_{s=0} - \d B[\g, \xi]
\een
for some $B$, where we are now omitting reference to the dependence upon
$f, \varphi$ to lighten the notation. Now apply $D_\eta$ to this equation.
\ben\label{b2}
D_\eta D_\xi A[\g] = D_\eta \frac{\partial}{\partial s} A[\g + s
\pounds_\xi \g] \Bigg|_{s=0} - D_\eta \d B[\g, \xi].
\een
The second term on the right side of this equation vanishes, 
since it is the functional derivative of an
exact form. 
Applying eq.~\eqref{b1} to the first term on the right
side of eq.~\eqref{b2}, we get
\ben
D_\eta \frac{\partial}{\partial s} A[\g + s \pounds_\xi \g] \Bigg|_{s=0} =
\frac{\partial^2 }{\partial s \partial t} A[\g + s \pounds_\xi (\g + t
\pounds_\eta \g)] \Bigg|_{s=t=0}
+ \d E[\g, \xi, \eta]
\een
for some $E$. Combining these equations and antisymmetrizing over
$\xi$ and $\eta$, we obtain
\ben
\label{b3}
(D_\eta D_\xi - D_\xi D_\eta) A[\g] =  \frac{\partial^2 }{\partial s
\partial t} A[\g + st (\pounds_\xi \pounds_\eta - \pounds_\eta \pounds_\xi)
\g ] \Bigg|_{s=t=0} + \d K
\een
for some locally constructed $(D-1)$-form $K$.
Applying eq.~\eqref{b1} once more to the first term on the right side of
eq.~\eqref{b3} and using $\pounds_\xi \pounds_\eta - \pounds_\eta \pounds_\xi
= \pounds_{[\xi, \eta]}$,
we obtain the desired relation~\eqref{b0}.


\begin{thebibliography}{99}

\bibitem{aw}
L. Alvarez-Gaume, E. Witten: ``Gravitational Anomalies,''
Nucl.\ Phys.\ B {\bf 234}, 269 (1984).

\bibitem{boas} F. M. Boas: ``Gauge theories in local causal perturbation theory,''
DESY-THESIS 1999-032, (1999) [arXiv: hep-th/0001014]

\bibitem{bpb}
L. Bonora, P. Pasti, M. Bergola: ``Weyl cocycles,'' Class. Quant. Grav. {\bf 3} 635-649
(1986)

\bibitem{bfk} R. Brunetti, K. Fredenhagen and M. K\"ohler: ``The microlocal spectrum 
condition and Wick polynomials on curved spacetimes,'' Commun. Math.
Phys. {\bf 180}, 633-652 (1996)

\bibitem{bf} R. Brunetti and K. Fredenhagen: ``Microlocal Analysis and 
Interacting Quantum Field Theories: 
Renormalization on physical backgrounds,''  
Commun. Math. Phys. {\bf 208}, 623-661 (2000)

\bibitem{bfv}
R.~Brunetti, K.~Fredenhagen and R.~Verch,
``The generally covariant locality principle: A new paradigm for local  quantum physics,''
Commun.\ Math.\ Phys.\  {\bf 237}, 31 (2003), 
[math-ph/0112041]; see also K.~Fredenhagen,
``Locally covariant quantum field theory,'' [arXiv:hep-th/0403007].

\bibitem{df} 
M.~D\"utsch and K.~Fredenhagen:
``A local (perturbative) construction of observables in gauge theories:
the example of QED'' Commun. Math. Phys. {\bf 203}, 71 (1999)

\bibitem{fd} 
M.~D\"utsch and K.~Fredenhagen:
``Algebraic quantum field theory, perturbation theory, 
and the loop  expansion,'' Commun. Math. Phys. {\bf 219}, 5 (2002) [arXiv: hep-th/0001129]; 
``Perturbative algebraic field theory, and deformation quantization,''
[arXiv: hep-th/0101079]

\bibitem{mwi} 
M.~D\"utsch and K.~Fredenhagen:
``The Master Ward Identity and Generalized Schwinger-Dyson Equation in
Classical Field Theory'', Commun.\ Math.\ Phys.\  {\bf 243}, 275 (2003), 
[hep-th/0211242]

\bibitem{df2}
M. D\"utsch and K.~Fredenhagen: ``Causal Perturbation Theory in Terms of Retarded Products, 
and a Proof of the Action Ward Identity,'' [hep-th/0403213]

\bibitem{h} S. Hollands: ``PCT Theorem for the Operator Product Expansion in Curved Spacetime,''
Commun. Math. Phys.~{\bf 244}, 209--244 (2004), [gr-qc/0212028]

\bibitem{hr} S.~Hollands and W.~Ruan,
``The state space of perturbative quantum field theory in curved
space-times,''
Annales Henri Poincare {\bf 3}, 635 (2002)
[arXiv:gr-qc/0108032]. 

\bibitem{hw1} S. Hollands and R. M. Wald: ``Local Wick Polynomials and Time Ordered Products
of Quantum Fields in Curved Space,'' Commun. Math. Phys. {\bf 223}, 289-326 (2001), [gr-qc/0103074]

\bibitem{hw2}
S.~Hollands and R.~M.~Wald:
``Existence of local covariant time-ordered-products of quantum fields in  curved spacetime,'' Commun. Math. Phys. {\bf 231}, 309-345 (2002), [gr-qc/0111108]

\bibitem{hw3}
S.~Hollands and R.~M.~Wald:
``On the Renormalization Group in Curved Spacetime,''
Commun. Math. Phys. {\bf 237}, 123-160 (2003), [gr-qc/0209029]

\bibitem{horm}
L. H\"ormander: {\it The Analysis of Linear Partial Differential Operators I,} Berlin, Springer-Verlag 1983

\bibitem{mo}
V.~Moretti: ``Comments on the stress-energy tensor operator in curved spacetime,''
Commun.\ Math.\ Phys.\  {\bf 232}, 189 (2003)
[arXiv:gr-qc/01090]

\bibitem{stora} 
R.~Stora: ``Pedagogical Experiments in Renormalized Perturbation Theory,'' contribution 
to the conference ``Theory of Renormalization and Regularization,'' in Hesselberg, Germany
(2002), {\tt http://wwwthep.physik.uni-mainz.de/scheck/Hessbg02.html} 

\bibitem{sw} R. F. Streater and A. A. Wightman: {\it PCT, Spin and Statistics and All That,} New York, Benjamin 1964

\bibitem{wald} R. M. Wald: {\it Quantum Field Theory on Curved Spacetimes and Black Hole Thermodynamics,}
The University of Chicago Press, Chicago (1990)

\bibitem{wz}
J.~Wess and B.~Zumino:
``Consequences Of Anomalous Ward Identities,''
Phys.\ Lett.\ B {\bf 37}, 95 (1971).

\end{thebibliography}
\end{document}